\documentclass[twocolumn]{aastex63}
\usepackage[utf8]{inputenc}

\usepackage{longtable}
\usepackage{color}
\usepackage{lineno}

\usepackage{url}
\usepackage{hyperref}
\usepackage{amsmath, amssymb}
\usepackage{natbib}
\usepackage{ulem}
\usepackage{float}

\newcommand{\UCSC}{Department of Astronomy and Astrophysics, University of California, Santa Cruz, CA 92064, USA}
\newcommand{\Einstein}{NASA Einstein Fellow}

\newcommand{\Duke}{Department of Physics, Duke University, Durham North Carolina 27708, USA}
\newcommand{\Moore}{Gordon and Betty Moore Foundation, 1661 Page Mill Road, Palo Alto, CA 94304, USA}
\newcommand{\Harvard}{Harvard-Smithsonian Center for Astrophysics, 60 Garden Street, Cambridge, MA 02138, USA}

\newcommand{\JHU}{Department of Physics and Astronomy, The Johns Hopkins University, Baltimore, MD 21218.}
\newcommand{\STScI}{Space Telescope Science Institute, Baltimore, MD 21218.}

\newcommand{\Chicago}{Department of Astronomy \& Astrophysics \& Kavli Institute for Cosmological Physics, University of Chicago, 5640 South Ellis Avenue, Chicago, IL 60637, USA}
\newcommand{\Cambridge}{Institute of Astronomy and Kavli Institute for Cosmology, Madingley Road, Cambridge, CB3 0HA, UK}

\newcommand{\NCUG}{Graduate Institute of Astronomy, National Central University, 300 Zhongda Road, Zhongli, Taoyuan 32001, Taiwan}
\newcommand{\Carnegie}{Observatories of the Carnegie Institute for Science, 813 Santa Barbara St., Pasadena, CA 91101, USA}

\newcommand{\CarnegieWashington}{Carnegie Institution of Washington, Las Campanas Observatory, Casilla 601, Chile}
\newcommand{\UCSD}{1 Center for Astrophysics and Space Science, University of California San Diego, La Jolla, CA 92093, USA}
\newcommand{\FSU}{Department of Physics, Florida State University, 77 Chieftain Way, Tallahassee, FL 32306, USA}
\newcommand{\Portsmouth}{Institute of Cosmology and Gravitation, University of Portsmouth, Portsmouth, PO1 3FX, UK}
\newcommand{\Pitt}{Pittsburgh Particle Physics, Astrophysics, and Cosmology Center. Physics and Astronomy
Department, University of Pittsburgh, Pittsburgh, PA 15260, USA}
\newcommand{\LBNL}{Lawrence Berkeley National Laboratory, 1 Cyclotron Road, Berkeley, CA, 94720, USA}
\newcommand{\Berkeley}{Department of Astronomy, University of California, Berkeley, 94720, USA}
\newcommand{\Penn}{Department of Physics and Astronomy, University of Pennsylvania, Philadelphia, PA 19104, USA}
\newcommand{\Southhampton}{School of Physics and Astronomy, University of Southampton,  Southampton, SO17 1BJ, UK}
\newcommand{\Spain}{Institute of Space Sciences, Campus UAB, Carrer de Can Magrans, s/n, E-08193 Barcelona, Spain}
\newcommand{\SpainTwo}{Institut d’Estudis Espacials de Catalunya (IEEC), E-08034 Barcelona, Spain}
\newcommand{\utaustin}{University of Texas at Austin, 1 University Station C1400, Austin, TX 78712-0259, USA}

\newcommand{\massstepstretch}{$0.072 \pm 0.041$}
\newcommand{\w}{$-1.17 \pm 0.12$}
\newcommand{\onepw}{$-0.17 \pm 0.12$}
\newcommand{\Om}{$0.258 \pm 0.090$}

\begin{document}

\title{Cosmological Results from the RAISIN Survey: Using Type Ia Supernovae in the Near Infrared as a Novel Path to Measure the Dark Energy Equation of State}

\suppressAffiliations

\author[0000-0002-6230-0151]{D. O. Jones}
\affiliation{\UCSC}
\affiliation{\Einstein}

\author[0000-0001-9846-4417]{K. S. Mandel}
\affiliation{\Cambridge}

\author[0000-0002-1966-3942]{R. P. Kirshner}
\affiliation{\Moore}
\affiliation{\Harvard}

\author{S.~Thorp}
\affiliation{\Cambridge}

\author{P. M. Challis}
\affiliation{\Harvard}

\author{A.~Avelino}
\affiliation{\Harvard}

\author{D.~Brout}
\affiliation{\Harvard}
\affiliation{\Einstein}

\author{C.~Burns}
\affiliation{\Carnegie}

\author{R.~J.~Foley}
\affiliation{\UCSC}

\author{Y.-C.~Pan}
\affiliation{\NCUG}

\author{D.~M.~Scolnic}
\affiliation{\Duke}

\author{M.~R.~Siebert}
\affiliation{\UCSC}

\author[0000-0002-7706-5668]{R.~Chornock}
\affiliation{\Berkeley}

\author[0000-0003-3431-9135]{W.~L.~Freedman}
\affiliation{\Chicago}

\author{A.~Friedman}
\affiliation{\UCSD}

\author{J.~Frieman}
\affiliation{\Chicago}

\author[0000-0002-1296-6887]{L.~Galbany}
\affiliation{\Spain}
\affiliation{\SpainTwo}

\author[0000-0003-1039-2928]{E.~Hsiao}
\affiliation{\FSU}

\author[0000-0003-0313-0487]{L.~Kelsey}
\affiliation{\Portsmouth}

\author{G.~H.~Marion}
\affiliation{\utaustin}

\author{R.~C.~Nichol}
\affiliation{\Portsmouth}

\author{P.~E.~Nugent}
\affiliation{\LBNL}
\affiliation{\Berkeley}

\author{M.~M.~Phillips}
\affiliation{\CarnegieWashington}

\author{A.~Rest}
\affiliation{\JHU}
\affiliation{\STScI}

\author{A.~G.~Riess}
\affiliation{\JHU}
\affiliation{\STScI}

\author{M. Sako}
\affiliation{\Penn}

\author{M. Smith}
\affiliation{Institut de Physique des Deux Infinis, Universit\'{e} de Lyon 1, CNRS/IN2P3, F-69622 Villeurbanne Cedex, France}

\author[0000-0002-3073-1512]{P.~Wiseman}
\affiliation{\Southhampton}

\author[0000-0001-7113-1233]{W.~M.~Wood-Vasey}
\affiliation{\Pitt}

\correspondingauthor{D.~O.~Jones}
\email{david.jones@ucsc.edu}

\begin{abstract}

Type Ia supernovae (SNe\,Ia) are more precise standardizable candles when measured in the near-infrared (NIR) than in the optical. With this motivation, from 2012-2017 we embarked on the RAISIN program with the Hubble Space Telescope ({\it HST}) to obtain rest-frame NIR light curves for a cosmologically distant sample of 37 SN\,Ia ($0.2\lesssim z \lesssim 0.6$) discovered by Pan-STARRS and the Dark Energy Survey. By comparing higher-$z$ {\it HST} data with 42 SN\,Ia at $z<0.1$ observed in the NIR by the Carnegie Supernova Project, we construct a Hubble diagram from NIR observations (with only time of maximum light and some selection cuts from optical photometry) to pursue a unique avenue to constrain the dark energy equation of state parameter, $w$.  We analyze the dependence of the full set of Hubble residuals on the SN\,Ia host galaxy mass and find Hubble residual steps of size $\sim$0.06-0.1~mag with 1.5- to 2.5-$\sigma$ significance depending on the method and step location used. Combining our NIR sample with Cosmic Microwave Background (CMB) constraints, we find $1+w=-0.17\pm0.12$ (statistical$+$systematic errors).  The largest systematic errors are the redshift-dependent SN selection biases and the properties of the NIR mass step.  We also use these data to measure ${\rm H}_0=75.9\pm 2.2~{\rm km~s^{-1}~Mpc^{-1}}$ from stars with geometric distance calibration in the hosts of 8 SNe\,Ia observed in the NIR versus ${\rm H}_0=71.2\pm3.8~{\rm km~s^{-1}~Mpc^{-1}}$ using an inverse distance ladder approach tied to {\it Planck}. Using optical data we find $1+w=-0.10\pm0.09$ and with optical and NIR data combined, we find $1+w=-0.06\pm0.07$; these shifts of up to $\sim$0.11 in $w$ could point to inconsistency in the optical versus NIR SN models. There will be many opportunities to improve this NIR measurement and better understand systematic uncertainties through larger low-$z$ samples, new light-curve models, calibration improvements, and eventually by building high-$z$ samples from the {\it Roman Space Telescope}.

\end{abstract}

\keywords{cosmology: observations -- cosmology: dark energy -- supernovae: general}

\section{Introduction}

After five decades of development as cosmological distance probes \citep{Kirshner10}, Type Ia supernovae (SNe\,Ia) are now mature tools for understanding cosmic acceleration \citep{Riess98,Perlmutter99} and systematic errors will soon become the largest source of uncertainty in the parameters of cosmic expansion \citep{Betoule14,Riess16,Scolnic18,Brout19,Jones19,Brout22}. In the optical bands where most SN\,Ia data have been obtained, substantial deviations from the behavior of a standard candle are reduced by correcting for the observed relations between light curve shape and luminosity \citep{Pskovskii77,Phillips93,Hamuy96,Riess96,Guy07,Jha07,Conley08,Mandel11,Burns14}, reddening by dust \citep{Riess96,Phillips99,Burns14,Thorp21,Mandel22}, and the intrinsically redder color of less luminous SNe\,Ia \citep{Tripp98,Mandel17}.

However, SNe\,Ia have been found to be more nearly standard candles in the near-infrared (NIR; wavelengths covered by the rest-frame $zYJHK$ bands) and their light is also less sensitive to the effects of dust extinction at these wavelengths \citep{Krisciunas04,Krisciunas07,WoodVasey08,Mandel09,Mandel11,Barone-Nugent12,Phillips12,Avelino19,Mandel22}.  This paper reports an attempt to realize these advantages in a cosmologically distant SN Ia sample.

At low redshift, significant samples of SNe\,Ia have been observed in the NIR by the Carnegie Supernova Project \citep[CSP-I;][]{Contreras10,Stritzinger11,Krisciunas17} at Las Campanas Observatory in Chile and the CfA Supernova Survey \citep{WoodVasey08,Friedman15} using PARITEL at the F.L.\ Whipple Observatory in Arizona.
NIR SN\,Ia light curves from  SweetSpot \citep{Weyant18} using the WIYN telescope at Kitt Peak National Observatory in Arizona have also been assembled, and additional data from CSP-II \citep{Hsiao19,Phillips19}, the VISTA Extragalactic Infrared Legacy Survey (VEILS) \footnote{\url{https://www.eso.org/sci/publications/announcements/sciann17237.html}.}, UKIRT \citep{Konchady21}, and {\it HST} (the SIRAH program; {\it HST}-GO 15889, PI: Jha) are forthcoming. 

Measurements of the dark energy equation of state parameter, $w$, require comparing low-$z$ SNe to those at larger cosmological distances.
Measurements of $z > 0.1$ SNe in the rest-frame NIR have only been obtained by \citet{Stanishev18} (four SNe) and \citet{Freedman09}.  The only previous measurement of the dark energy equation of state using NIR data was the pioneering study of \citet{Freedman09}, who found $w = -1.05 \pm 0.13\ ({\rm stat}) \pm 0.09\ ({\rm sys})$ by comparing 21 low-$z$ SNe to 35 high-$z$ SNe using templates as red as the rest-frame $I$ band with observations in $YJ$ from the Magellan Baade telescope.  This analysis was unable to apply a number of corrections that are now commonly included in cosmology analyses but were less well studied or entirely unknown at the time that \citet{Freedman09} was published, including corrections for distance (Malmquist) biases and the dependence of Hubble residuals on host galaxy mass \citep{Kelly10,Lampeitl10,Sullivan10}.

Though high-$z$ NIR SN\,Ia observations are rare, the value of NIR data is clear from low-$z$ analyses.  \citet{Avelino19} used CSP-I and CfA data to show in head-to-head comparisons of the same SNe that the scatter in the distances as measured in the infrared is smaller by 35\% than distances determined from optical light curves.  Studies such as \citet{WoodVasey08,Mandel09,Kattner12,Mandel22} have also shown evidence that SNe at NIR wavelengths  yield more precise distance measurements.
  
  Low-$z$ NIR data have recently been used to demonstrate a significant relationship between SN\,Ia distance residuals and the mass of the SN\,Ia host galaxy \citep{Ponder20,Uddin20}, which was originally discovered at optical wavelengths \citep{Kelly10,Lampeitl10,Sullivan10}, although \citet{Johansson21} found mildly conflicting results with no evidence for a mass-step in the $JH$ bands and $\sim$2-$\sigma$ evidence for a step in the $Y$ band.
\citet{Burns18} and \citet{Dhawan18} used NIR data to measure the Hubble constant, H$_0$, using this new SN\,Ia wavelength range to provide additional evidence for tension between Cepheid$+$SN distance ladder measurements of H$_0$ \citep[e.g.,][]{Riess21b} and the cosmic microwave background \citep[CMB;][]{Planck18}.

With the advent of the {\it Nancy Grace Roman Space Telescope}, developing the optimal ways to measure NIR distances at cosmological redshifts is now urgent.  Although {\it Roman} will observe SNe\,Ia primarily at rest-frame optical wavelengths at maximum redshifts up to $\gtrsim$2.5 \citep{Hounsell18}, {\it Roman} SN\,Ia observations will be in the rest-frame NIR in the important regime at $z \lesssim 0.7$ where Baryon Acoustic Oscillation constraints will be limited by the cosmic volume \citep{Weinberg13}.

Here, we present NIR cosmological parameter measurements from the RAISIN survey.
RAISIN (an anagram for “SNIA in the IR”) used 23 SNe\,Ia discovered in the Medium Deep Survey (MDS) of Pan-STARRS \citep{Chambers16} in Hawaii and another 23 discovered by the Dark Energy Survey (DES) at Cerro Tololo in Chile \citep{DES05}.  We triggered {\it Hubble Space Telescope} observations of those objects with {\it WFC3-IR} using the $F125W$ and $F160W$ filters.  After applying a number of well-motivated cuts to the data, we use 42 low-$z$ SNe\,Ia from CSP-I (hereafter CSP) and 37 high-$z$ ($z > 0.2$) SNe\,Ia from RAISIN to measure $w$.

In this work, we pursue a ``NIR-only" cosmological analysis that uses a different wavelength range than previous cosmological analyses with SNe and will have reduced systematic uncertainties due to dust.  An optimal combination of optical$+$NIR data would yield the most precise cosmological constraints, but we show that an NIR-only analysis offers additional independent distance information.
Throughout this paper we enumerate areas that will be improved to shrink these errors in future NIR studies and compare our results to a combined optical$+$NIR measurement.

In Section \ref{sec:data} we describe the RAISIN data, including survey
properties, classifications, and photometric measurements.  In Section \ref{sec:analysis} we describe our analysis method.  In Section \ref{sec:results} we present our baseline cosmological results and in Section \ref{sec:sysdiscussion} we discuss additional analysis variants and measure the correlation of Hubble residuals with host galaxy mass. In Section \ref{sec:discussion} we discuss the implications of our results for future missions such as the {\it Roman Space Telescope}.  In Section \ref{sec:conclusions} we conclude.

\begin{figure*}
    \centering
    \includegraphics[width=7in]{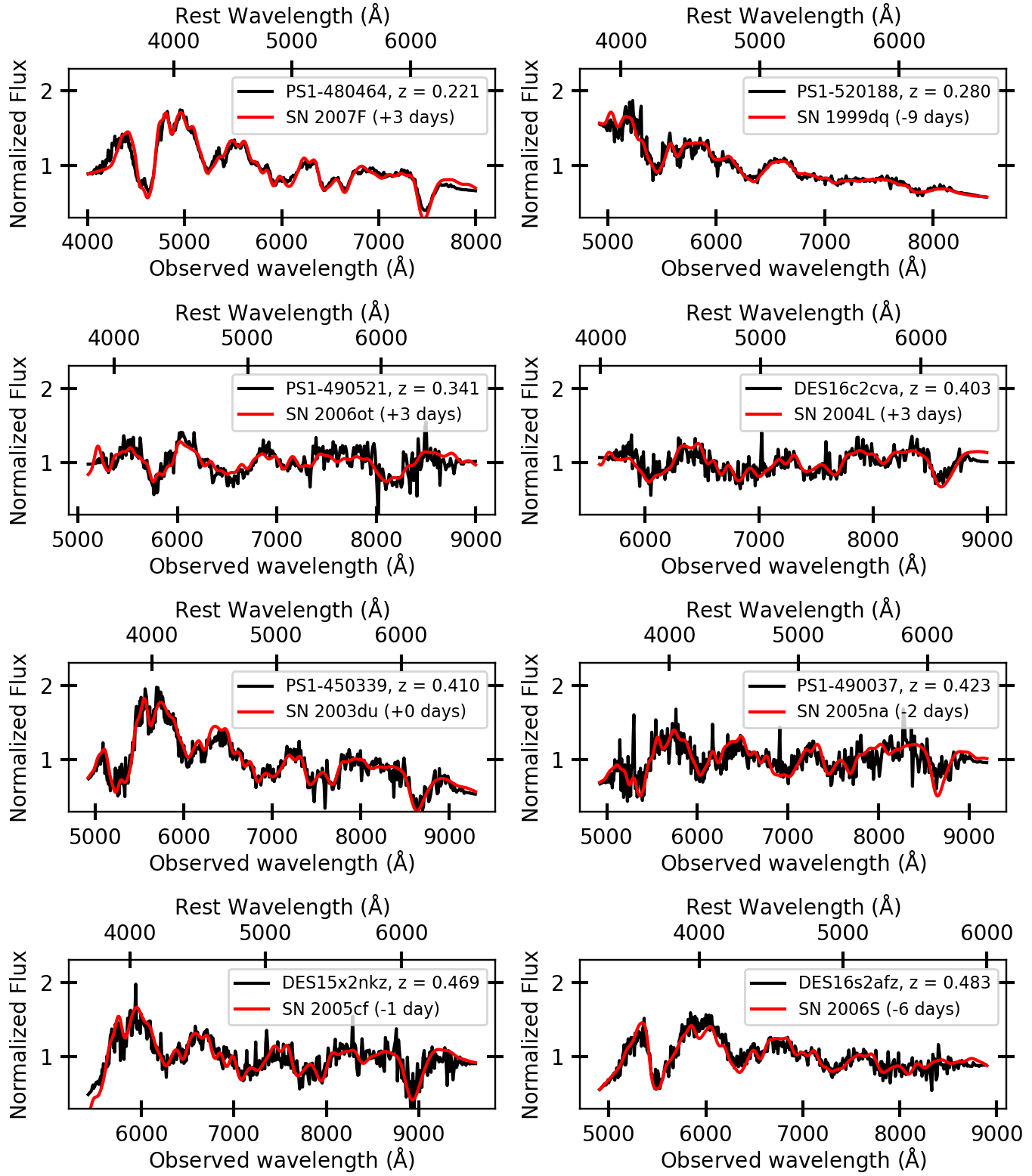}
    \caption{Representative classification spectra for RAISIN SNe (black) with the best template matches from SNID (red).  Though the template match to PS1-490521 (SN~2006ot) is a peculiar Ia, the SNID matches to normal SNe\,Ia are also excellent.}
    \label{fig:spectra}
\end{figure*}

\section{The RAISIN Sample}
\label{sec:data}

\begin{figure*}
\centering
  \includegraphics[width=6in]{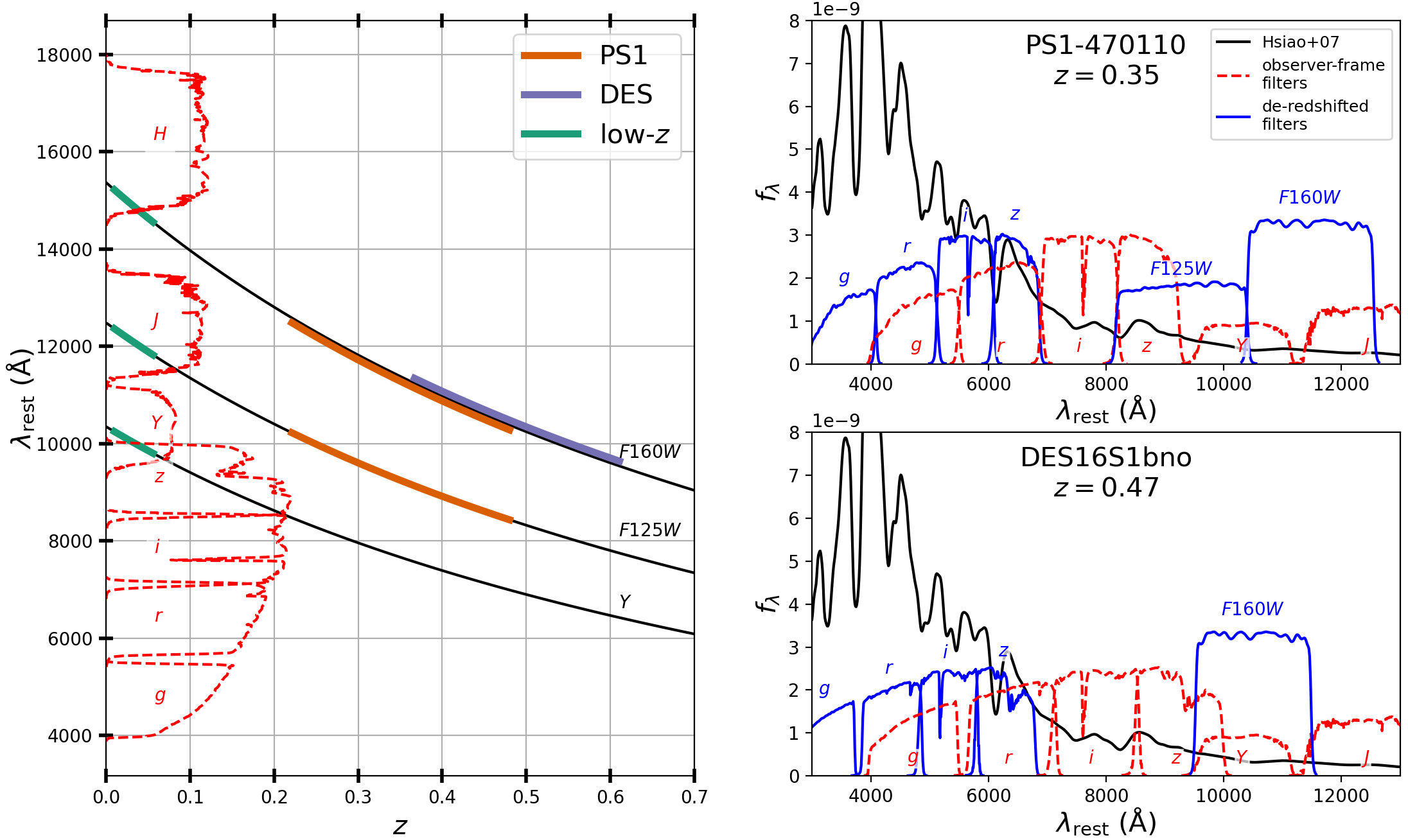}
  \caption{Rest-frame wavelengths probed by RAISIN observations as a function of redshift. Left: Rest-frame wavelengths corresponding to the observer-frame $Y$, $F125W$ ({\it HST}'s $J$-band) and $F160W$ ({\it HST}'s $H$-band) as function of redshift, with the low-$z$, RAISIN1 (PS1) and RAISIN2 (DES) redshift ranges highlighted.  The shapes of $grizYJH$ filter throughputs (red; arbitrary units) are shown for illustration. Right: de-redshifted (solid blue) and observer-frame (dashed red) filters for the median-redshift RAISIN1 (top) and RAISIN2 (bottom) SNe.  The \citet{Hsiao07} model (black) is shown for illustration.}
  \label{fig:raisinwave}
\end{figure*}

\begin{figure*}
  \includegraphics[width=7in]{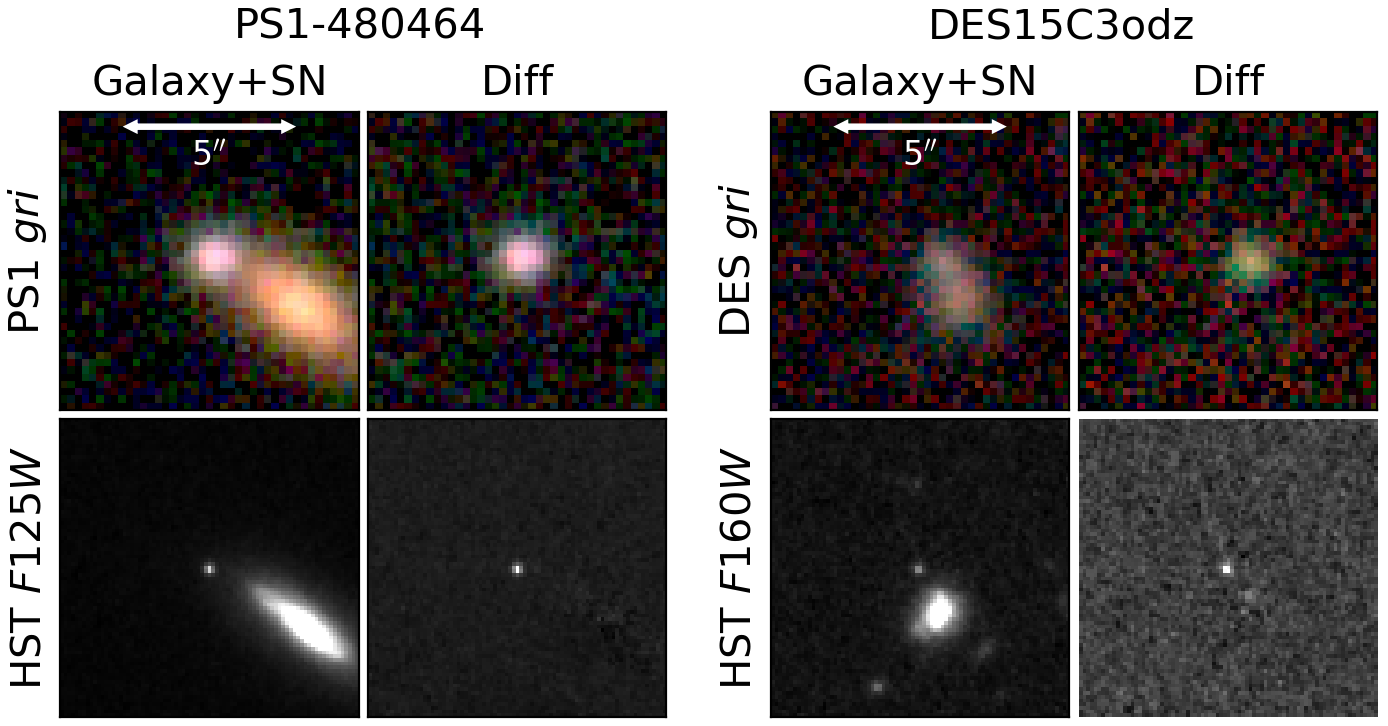}
  \caption{Stamp images of RAISIN SNe from Pan-STARRS (left) and DES (right) with
    $gri$ images (top) and RAISIN images from {\it HST} (bottom).  The SN
    location is centered in each image.}
  \label{fig:stamps}
\end{figure*}

\subsection{Overview and Strategy}

The RAISIN program was carried out in cycle 20 through {\it HST}-GO 13046 (hereafter RAISIN1; PI: Kirshner) and cycle 23 through {\it HST}-GO 14216 (hereafter RAISIN2; PI: Kirshner).  RAISIN1 followed 23 spectroscopically classified SNe from the Pan-STARRS MDS \citep{Chambers16} and RAISIN2 followed 23 spectroscopically classified SNe from DES \citep{DES05}.
 RAISIN1 observed SNe in the redshift range of $0.22 \leq z \leq 0.50$ and RAISIN2 observed SNe at  $0.35 \leq z \leq 0.61$ to observe at redshifts where the available {\it HST} filters overlap with the rest-frame $YJH$ filters to minimize $K$-correction uncertainties.
The median redshifts of PS1 and DES SNe aligned well with these targeted redshift ranges.

Due to occasional poor weather during PS1 and DES observing seasons and the need for {\it HST} template imaging after each SN had faded, RAISIN observations for each program extended over a period of 1.5-2 years; RAISIN1 observations were taken from 2012 October 29 to 2014 June 17 and RAISIN2 observations occurred from 2015 September 28 to 2017 November 21st.

To select RAISIN candidates, we identified Pan-STARRS and DES SNe that were likely to be SNe\,Ia discovered before maximum light.  These candidates were then classified spectroscopically; classifications for MDS and DES SNe were carried out either by the RAISIN team in collaboration with the survey teams or by the MDS or DES teams themselves.  For RAISIN1, we selected candidates that had been discovered by the Pan-STARRS team and were consistent with having a light curve phase approximately 5-10 days before maximum light.  For RAISIN2, we similarly selected candidates based on their apparent light curve phase and additionally required that the photometric redshift of their host galaxies were consistent with the target redshift range of $\sim 0.5 \pm 0.1$.

Most classifications were from Magellan (16 SNe) and Gemini South (13 SNe), with additional classifications from Gemini North (four SNe), the MMT (six SNe), the AAT (three SNe), and Keck (three SNe).  Each classification was determined using the SN IDentification software (SNID; \citealp{Blondin07}), which uses cross-correlation matching to template SNe to yield SN types, light curve phases, and redshifts.  
Representative spectra for RAISIN targets are shown in Figure \ref{fig:spectra}.
Spectroscopically classified SNe\,Ia with rising light curves within the target redshift range were submitted to STScI for scheduling as non-disruptive Targets of Opportunity on {\it HST}. To improve the distance measurement, two additional epochs were obtained for each SN after the initial {\it HST} observation, spaced by approximately five rest-frame days.  Template images for each SN were taken at least six months after the initial observations, with the exception of a single SN for which no template was needed because it was located in a region with no apparent galaxy light.

\subsection{Ground-based Data and Photometry}

\subsubsection{The Pan-STARRS Medium Deep Survey}
The Pan-STARRS MDS, the source of RAISIN1 SNe, observed 70 square degrees in $griz$ filters over approximately 4 years, with 
$gr$, $i$, and then $z$ observations on successive nights.  MDS also observed in the $y$ filter, primarily during bright time, but the S/N of $y$-band observations for objects in our targeted redshift range was too low to be of use for SN\,Ia cosmology.  Approximately 5200 SNe were discovered across the four years of the MDS, with over 3000 host galaxy redshifts measured \citep{Jones17}. Five hundred twenty SNe were spectroscopically classified during the survey, with $\sim$350 of these being SNe\,Ia and having a median redshift of $\sim$0.35 \citep{Scolnic18}.  Cosmological parameter measurements from these data are presented in \citet{Scolnic18} for the spectroscopically classified sample and \citet{Jones18} for the full sample.  Further details regarding the MDS are given in \citet{Chambers16}.

The optical photometry for RAISIN1 comes from the Pantheon cosmological analysis \citep{Scolnic18}, with the exception of five SNe that were not included in \citet{Scolnic18}\footnote{SN 520107 has a large shape uncertainty, while SN 470240 has an unusually blue optical color but passes the NIR-based cuts in Table \ref{table:raisin_cuts}.  The other three SNe, 480794, 540087, and 540118 appear to pass standard cosmology cuts and were used for the cosmological analysis in \citet{Jones18}.}.  Light curves of those five SNe were published by \citet{Villar20} and are included in the data release that accompanies this paper.  Pan-STARRS SN photometry is from {\tt Photpipe} \citep{Rest05}, with details specific to the MDS given in \citet{Rest14} and \citet{Scolnic18}.  In brief, MDS images are astrometrically aligned and their zeropoints are measured using the PS1 catalog. PS1 zeropoints have been calibrated to a 3~mmag relative precision across the sky \citep{Schlafly12,Scolnic15,Brout21b}.  Single-season, inverse variance-weighted template image stacks are then convolved and subtracted from the nightly images, and SNe are discovered and their light curves subsequently measured by performing DAOPHOT forced photometry on the resulting difference images \citep{Stetson87}.

\subsubsection{The Dark Energy Survey}

DES observed in $griz$ filters over 27 square degrees with a cadence of approximately once per week.  These 27 square degrees were split into eight 2.7 deg$^2$ ``shallow" fields, with depths of $\sim$23.5~mag, and two ``deep" fields, with depths of approximately 24.5~mag.   DES spectroscopically classified 251 SNe\,Ia at redshifts $0.02 < z < 0.85$ \citep{Abbott18}.  After sample cuts, their three-year spectroscopic sample includes 207 SNe\,Ia at a median redshift of $z = 0.36$.  Cosmological parameter measurements from the DES spectroscopic sample are given in \citet{Abbott18} with additional publications describing the calibration \citep{Burke18,Lasker19}, primarily based on observations of the CALSPEC standard star C26202, bias corrections \citep{Kessler19}, photometry \citep{Brout19b}, spectroscopic classification \citep{Smith20b}, and systematic uncertainties \citep{Brout19} of these data.

For RAISIN2, DES SN discovery uses a difference imaging procedure similar to the MDS, but final optical photometry is carried out using the scene-modeling algorithm presented in \citet{Brout19b}.  Scene modeling \citep{Holtzman08} uses imaging data to build a pixel-based model of the galaxy$+$SN that is convolved with each night's measured point spread function (PSF) model.  The amplitude of the SN is allowed to vary epoch-to-epoch while the galaxy brightness is fixed.  The robustness of the algorithm has been tested with artificial sources and recovers fluxes to an accuracy of 3~mmag.

\subsubsection{The Carnegie Supernova Project}

There are two sources of well-sampled low-$z$ NIR SN\,Ia data, the CfA and CSP samples \citep{Friedman15,Krisciunas17}.  These were combined to yield a sample of 89 low-$z$, NIR-observed SNe in \citet{Avelino19}.  However, most CfA SNe are either at redshifts where the effect of peculiar flows dominates the distance uncertainty ($z \lesssim 0.01$) or lack data prior to maximum light, which makes the determination of the time of maximum light uncertain and increases distance uncertainties.  Therefore, we restrict ourselves to the CSP data for the low-$z$ SN sample used in this work, although the CfA data remain extremely useful for training SN standardization models in the NIR.  The release of CSP-II data in the near future will add an additional 125~SNe with NIR light curves and 90~SNe with NIR spectra \citep{Hsiao19,Phillips19}.

CSP uses DAOPHOT \citep{Stetson87} to measure natural-system photometry on SN images after a template image, taken once the SN light has faded, has been subtracted. CSP images are calibrated with respect to primary standards BD $+17^{\circ}4708$ (optical bandpasses) and Vega (NIR), with additional details given in \citet{Krisciunas17}.

\subsubsection{Spectroscopic Classifications and Redshifts}
\label{sec:class}

Spectroscopic classifications and SN redshifts for each RAISIN SN were determined using SNID \citep{Blondin07}.  SNID determines a classification quality through a combination of the overlap between the observed spectrum and a template spectrum ($lap$), and the height of the cross-correlation peak ($r$). We ensure that the best template match has an $rlap > 5$, indicating a good match, and that the top three spectroscopic matches are all normal SNe\,Ia.  For SN~PS1-520107, the classification was unclear so we remove this SN from our sample.  For SN~PS1-490521 the best-matched spectrum was to the peculiar SN\,Ia 2006ot but, as discussed in Section \ref{sec:selection}, we choose to include 06bt- and 06ot-like SNe in our cosmology analysis as they cannot be reliably classified (or ruled out) from noisier high-$z$ spectra.

The PS1 and DES teams measured host galaxy redshifts for these objects using cross-correlation matches to galaxy templates.  For PS1, redshifts were estimated using the {\tt rvsao} package \citep{Kurtz98}, and for DES the {\tt Marz} package was used \citep{Hinton16}.

For CSP classifications and redshifts, see \citet{Krisciunas17} and references therein.  SN~classifications were measured from SNID, as well as SN redshifts.  As discussed in \citet{Krisciunas17}, all CSP SNe have host galaxy redshifts, and nearly all of these are measured from the NASA/IPAC extragalactic database (NED\footnote{The NASA/IPAC Extragalactic Database (NED) is funded by the National Aeronautics and Space Administration and operated by the California Institute of Technology.}).  We also use updated group redshifts from the Pantheon$+$ analysis \citet{Scolnic21}  for SN~2005al and 2009ab, which differ significantly from the values in the latest CSP data release.

\subsection{HST Data and Photometry}
\label{sec:hstphot}

The {\it HST} data for RAISIN1 SNe were taken with three epochs each of $F125W$ (approximately the $J$ band) and $F160W$ (approximately the $H$ band).  For RAISIN2 SNe, only the $F160W$ band was used for most SNe; five RAISIN2 SNe included data in both filters.  Figure \ref{fig:raisinwave} illustrates the rest-frame wavelengths probed by the {\it HST} data as a function of redshift.

The first {\it HST} observations occurred at a median of 12.2 observer-frame days relative to $B$-band maximum light for RAISIN1 ($+$9.2 rest-frame days) and 13.9 observer-frame days ($+$9.5 rest-frame days) for RAISIN2 due to the time required for SN discovery, classification, and the latency in the {\it HST} scheduling.  The median S/N is 12.5 for RAISIN1 and 12.9 for RAISIN2 with template observations taken at an average of 189~days after the first SN detection. See Appendix \ref{app:hstphot} for a discussion of how we correct for a small amount of late-time SN flux in some template images.

Examples of RAISIN observations are shown in Figure \ref{fig:stamps}.  Optical photometry and spectra of these SNe are provided in the online data release accompanying this paper to allow these measurements to be of use in future NIR analyses such as SIRAH ({\it HST}-GO 15889).

Our team measured photometry from the RAISIN images according to the following steps:

\begin{enumerate}
\item For each epoch, FLT images, which are images that have been bias-subtracted, dark-subtracted, and flat-fielded, are drizzled\footnote{ ``Drizzling" refers to the process of linearly reconstructing an image from under-sampled, dithered data \citep{Fruchter02}.} together and the final (template) epoch is subtracted.
\item We then measured SN centroids from the difference images and performed 0.4\arcsec\ aperture photometry on the difference images with aperture corrections from {\it HST}.  We verified these aperture corrections using drizzled images of the CALSPEC standard star P330E.
\item Using artificial stars injected into the data frames, we corrected the resulting measurements and uncertainties for host galaxy noise.
\item We calibrated the data using publicly available zeropoints that were measured from observations of four white dwarf standards (GD153, G191B2B, GD71, GRW+70D5824), and the G-type star P330E\footnote{\url{https://www.stsci.edu/hst/instrumentation/wfc3/data-analysis/photometric-calibration/ir-photometric-calibration}.}.
\end{enumerate}

\noindent We describe these steps in additional detail in Appendix \ref{app:hstphot}.  Coordinates of the RAISIN SNe and the final RAISIN photometry are given in Appendix \ref{sec:app_phot}.

\section{Analysis}
\label{sec:analysis}

In this section, we describe the NIR light curve models we used (Section \ref{sec:lcmodels}), present our method for distance measurements (Section \ref{sec:distmeas}), create a dispersion model to measure distance-dependent biases (Section \ref{sec:dispmodel}), apply sample selection cuts (Section \ref{sec:selection}), measure the dependence of distance on host galaxy mass (Section \ref{sec:host}), and estimate each source of systematic uncertainty in our NIR measurement of $w$ (Section \ref{sec:analysis_systematics}).  Our cosmological parameter measurements are then presented in Section \ref{sec:results}.

\subsection{NIR Light Curve Models}
\label{sec:lcmodels}

\begin{figure*}
  \includegraphics[width=7in]{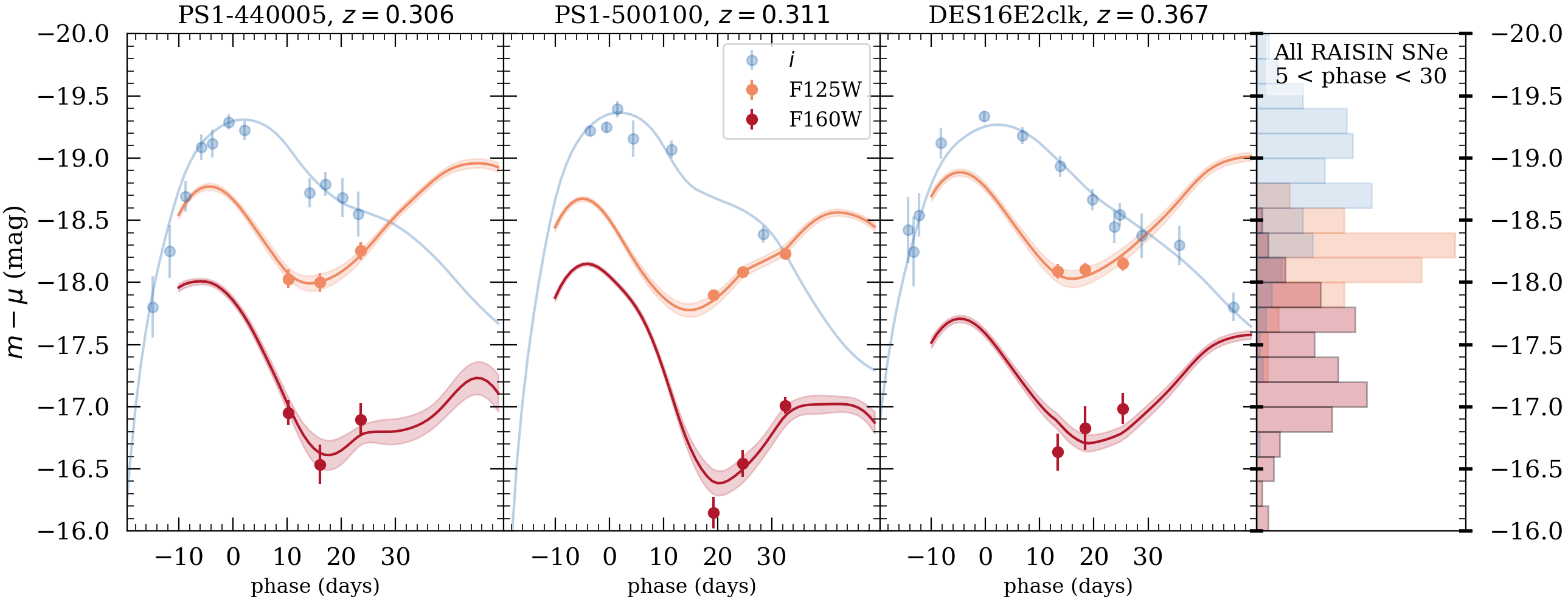}
  \caption{Examples of RAISIN SN light curves.  Apparent magnitude minus distance modulus (a proxy for absolute magnitude) versus phase is shown for three RAISIN
  SNe with observations in $i$ (PS1 or DES), $F125W$, and $F160W$.  Solid lines show the SNooPy model fit to the data.  The optical bands, including $i$, are used to estimate the time of maximum light, while the {\it HST} NIR bands are used to measure the distance (we allow $s_{BV}$ fitting to show the best-fit NIR model in this example).  Histograms with approximate absolute magnitude distributions for all RAISIN SNe are shown on the right after applying the sample selection cuts discussed in Section \ref{sec:selection} below.}
  \label{fig:lcexamples}
\end{figure*}

SN\,Ia standardization models rarely make use of the NIR due to the small number of well-calibrated, high-cadence NIR light curves.  In the last decade, only the SALT2 \citep{Guy07,Guy10} and SiFTO \citep{Conley08} light curve fitters have been used for measurements of $w$, with SALT2 being by far the most common model.  SALT2, however, extends only to a central filter wavelength of approximately 7000~\AA, with the recently-developed SALT3 model \citep{Kenworthy21} now extending this wavelength to 8700~\AA\ to include the rest-frame $z$ band.

However, there are now two well-vetted light-curve models extending to the NIR: the SNooPy model \citep{Burns11,Burns14} and the BayeSN model \citep{Mandel22,Thorp21}.  SNooPy benefits from over a decade of use by the community and includes light curve templates for the $uBVgriYJH$ bands that have been trained using well-calibrated CSP light curves.  
BayeSN is a hierarchical Bayesian Spectral Energy Distribution (SED) model for SNe\,Ia (BayeSN-SED) that models time-dependent optical-NIR (to 1.8 $\mu$m) SN Ia SEDs as a combination of physically-distinct intrinsic spectral components and host galaxy dust effects. By having a SED-based model continuous in both time and wavelength, BayeSN removes the need to compare high-redshift photometric data to an approximate rest-frame filter and effectively allows a $K$-correction that varies with light-curve stretch.
At the same time, BayeSN has yet to be integrated in publicly available tools like SNANA \citep{Kessler10}, so the simulation of distance-dependent biases for a cosmology analysis is not yet feasible.

Though we incorporate both SNooPy and BayeSN in our analysis, we use SNooPy as our baseline method due to the greater ease of determining distance biases and because BayeSN was under development during the time when much of this analysis was carried out. The advantages of BayeSN's statistical framework are somewhat mitigated due to the limited number of NIR observational epochs in the RAISIN data. Still, we find good consistency between the distance measurements from these two models, as demonstrated in Section \ref{sec:bayesn}; we plan to use BayeSN in a future optical$+$NIR analysis of these data.  
 
 \subsubsection{SNooPy Model Philosophy}
 \label{sec:snoopy}
 
The SNooPy model philosophy assumes that the observed variation in SN colors is caused by extinction in the SN host galaxy, but allows the selective-to-total extinction ratio $R_V$ to be specified by the user --- the nominal method in this analysis --- or fit by the data.  We also allow SNooPy to fit negative extinctions for a more agnostic treatment of SN color; SNooPy assumes the luminosity versus color relationship to be linear with no minimum value (the same philosophy as the Tripp relation; \citealp{Tripp98}).  This effectively means that $R_V$ should be interpreted as a luminosity versus color trend in this work, rather than a physical dust law.

SNooPy also includes options to correct for the dependence of SN luminosity on light curve shape.  We adopted the ``EBV\_model2", that uses the $s_{BV}$ parameter to determine the relation between SN light curve shape and luminosity \citep{Burns18}.  The $s_{BV}$ parameter is a stretch parameter that is insensitive to extinction because it takes advantage of the fact that the delay between time of maximum light and the time of reddest color is insensitive to reddening \citep{Lira98}. To enable easier determinations of distance biases and systematic uncertainties in later stages of the analysis, we incorporated this SNooPy model into the SNANA software \citep[Section \ref{sec:simulations};][]{Kessler10} and used the SNooPy $iYJH$ templates generated from low-$z$ CSP data to fit our NIR data and measure distances.  Example RAISIN light curves with SNooPy fits are shown in Figure \ref{fig:lcexamples}.

\subsection{Distance Measurement Method}
\label{sec:distmeas}

Because RAISIN NIR light curves include just three post-maximum light epochs per SN, we are unable to make robust shape and color corrections or determine the time of maximum light from NIR data alone; we find a Hubble residual RMS of 0.24~mag when fitting $A_V$ and 0.28~mag when fitting $s_{BV}$ from NIR-only RAISIN data (in Appendix \ref{app:sims}, we will use simulations to confirm that larger RMS values are expected given the small number of post-maximum light epochs per SN).  

Therefore, our nominal distance measurement approach is to use the SNooPy model to fit {\it only} the NIR SN amplitude; we use optical data to determine the time of maximum light and fix the SN shape and color to a nominal value\footnote{We adopt arbitrary values of $s_{BV}=1$ and $A_V=0$ for simplicity, as the bias correction stage will adjust the derived distances for the average $s_{BV}$ and $A_V$ of the data themselves.  We have confirmed that adjusting the nominal value of these parameters in the distance fits does not change the scatter by more than $\sim$0.01~mag.}.  This approach yields a Hubble residual RMS of 0.18~mag.

We relax either the NIR-only aspect of this analysis or the nominal approach of fitting only the amplitude of the NIR data --- i.e., not fitting $s_{BV}$ or $A_V$ --- in several places in this analysis as described below:

\begin{enumerate}
    \item {\bf Selection cuts}.  Optical data are used to make sample selection cuts to remove SNe with high values of $E(B-V)$, which cannot be estimated from NIR data alone.  This is discussed in Section \ref{sec:selection}.
    \item {\bf Time of maximum light}.  Because we do not have NIR data near maximum light in the high-$z$ sample, we must use optical data to constrain the phase of the SN observations.
    \item {\bf Scatter model}.  SNe\,Ia were discovered and spectroscopically classified at optical wavelengths, so we determine the consequence of these selection effects on our sample.  To estimate distance-dependent selection effects for the SNooPy model, we use optical$+$NIR CSP data to estimate the covariance matrix of the distance residuals after estimating and correcting for $s_{BV}$ and $A_V$.  This is described below in Section \ref{sec:dispmodel}.
    \item {\bf Bias corrections}: To determine and correct for the redshift dependence of the intrinsic populations of $s_{BV}$, we fit $s_{BV}$ using the NIR data.  Though the resulting Hubble residual scatter increases, as discussed above, this is a useful tool to compare data distributions of $s_{BV}$ to simulated distributions of $s_{BV}$.  This is described in Appendix \ref{app:sims}.
    \item {\bf Wavelength-dependence of the mass step}. In Section \ref{sec:mass_step_results}, we compare estimates of the host galaxy mass step between optical and NIR data in a self-consistent manner that accounts for the correlation between $s_{BV}$ and galaxy mass.  Therefore, in this section we will correct the NIR Hubble residuals for their dependence on the optical$+$NIR-measured $s_{BV}$ parameter.  
    We use both the SNooPy derived $s_{BV}$-NIR luminosity correlation and the directly measured dependence of our NIR Hubble residuals on $s_{BV}$.  A $s_{BV}$-corrected mass step is not used for cosmological parameter estimation but only to investigate the physics and wavelength-dependence of the mass step.
    \item {\bf Comparing to optical and optical$+$NIR measurements of $w$}.  In Section \ref{sec:cosmo}, we compare the baseline results to results that fit the SNooPy model to optical and optical$+$NIR data.
\end{enumerate}

\subsection{Dispersion Model and Distance Biases}
\label{sec:dispmodel}

\begin{figure}
  \includegraphics[width=3.5in]{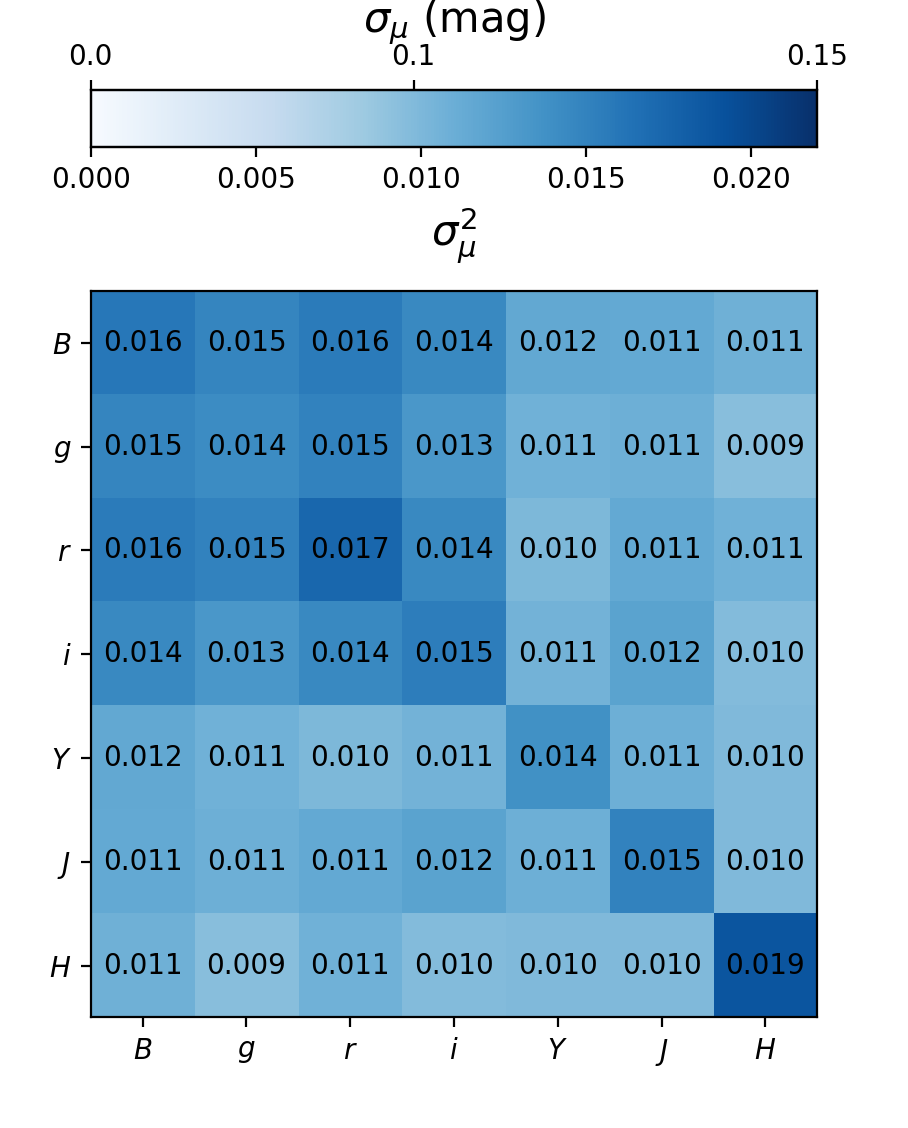}
  \caption{Optical-to-NIR covariance matrix for the SNooPy model computed from CSP data (similar to Fig.\ 6 of \citealp{Mandel11}), with
    covariance values labeled and the top axis of the colorbar indicating
    the uncertainties on the diagonal in each band.  After shape and color-correction,
    we found the maximum dispersion to be $\sim$0.15 mag in the
    optical bands and the minimum to be $\sim$0.12 mag in the $Y$ band.}
  \label{fig:covmat}
\end{figure}

Because we compare SNe\,Ia across a large span of redshift, we must determine distance-dependent biases, that are caused by the increased likelihood of discovering higher-luminosity SNe at larger redshifts in magnitude-limited surveys; these can be up to $\sim$5\% in distance for the PS1 and DES samples \citep{Scolnic18,Brout19}.  To determine these biases, we build a dispersion covariance matrix from the optical to NIR to understand the ways in which optical SN\,Ia selection effects $-$ which are caused by PS1 and DES selection criteria $-$ affect distances measured in the NIR.

Because there has been no measurement of the covariance in magnitude between different SNooPy bands, we estimated this model from the low-$z$ CSP data in our sample.  
To generate this model, we use low-$z$ SNe at $0.015 < z < 0.1$ to ensure that the results are minimally dependent on cosmological parameters while also mitigating the effect of peculiar velocity uncertainties at $z < 0.015$ \citep{Peterson21}.

We then simultaneously fit the SNooPy model to all the optical and NIR bands in the low-$z$ CSP sample to estimate the time of maximum light, $s_{BV}$, and $A_V$ for each SN.  Keeping these parameters fixed, we then fit the low-$z$ data with the SNooPy model for each of the $BVgriYJH$ bands individually. We use the resulting Hubble residuals in each band to generate a band-to-band covariance matrix that can be used to generate large Monte Carlo simulations and determine wavelength-dependent distance biases.  
We note that the additional distance scatter predicted from a peculiar velocity uncertainty of 250 km s$^{-1}$ \citep{Scolnic18} is smaller than the scatter observed in each band at the median redshift $z = 0.023$, and we find that the observed optical-to-NIR covariances are very similar when restricting the sample to $z > 0.025$ or $z > 0.03$.

The resulting covariance matrix is shown in Figure \ref{fig:covmat} and is now included in the public SNANA software\footnote{\url{https://snana.uchicago.edu/}.}.   We find similar dispersion measurements in most bands, with values ranging from $\sim$0.116 to 0.136~mag; the maximum dispersion is in the $H$ band and the minimum is $Y$ but the band-to-band differences are not highly significant in this sample. 
Covariance is on the order of $\sim$0.015$\text{ mag}^2$ between neighboring optical bands (correlations of $\gtrsim$0.95) and $\sim$0.010$\text{ mag}^2$ between NIR-to-NIR and optical-to-NIR bands (correlations of $\sim$0.78-0.86).  These correlations can be understood as the correlation between Hubble residuals measured using one band at a time.  
Slightly higher correlations are found when we examine the \citet{Uddin20} data, which might be due to the exclusion of fast-declining, red, and low-$z$ SNe in the present analysis.
NIR-derived distances in our sample are well-correlated with optical distances --- implying a significant wavelength-independent component of SN Hubble residual scatter --- but are less correlated with optical bands than optical bands are with each other, implying that NIR data contain additional information useful for constraining SN\,Ia distances.

\subsection{Sample Selection Cuts}
\label{sec:selection}

\begin{table}
    \centering
    \caption{RAISIN Cuts}
    \begin{tabular}{lrrrrrr}
      \hline \hline
      &\multicolumn{2}{c}{CSP DR3}&\multicolumn{2}{c}{RAISIN1}&\multicolumn{2}{c}{RAISIN2}\\
      &N$_{\mathrm{cut}}$&N$_{\mathrm{rem}}$&N$_{\mathrm{cut}}$&N$_{\mathrm{rem}}$&N$_{\mathrm{cut}}$&N$_{\mathrm{rem}}$ \\
      \hline

&\nodata&134&\nodata&23&\nodata&22\\
$\sigma_{host} < 0.1$ mag&0&134&2&21&3&19\\
spec. SN\,Ia&0&133&1&20&0&19\\
$z > 0.01$&21&116&0&20&0&19\\
pre-max&22&92&0&20&0&19\\
pre-max, $\chi^2$ cut&35&57&0&20&0&19\\
NIR data&8&49&0&20&0&19\\
SNooPy fitting&0&49&0&20&0&19\\
$E(B-V) < 0.3$&1&49&0&20&0&19\\
$0.75 < s_{BV} < 1.18$&4&45&1&19&0&19\\
$\sigma(s_{BV}) < 0.2$&2&43&0&19&0&19\\
Chauvenet&1&42&0&19&1&18\\
      \hline
      \hline\\[-1.5ex]
      \multicolumn{7}{p{.45\textwidth}}{
      \begin{minipage}{3.4in}
      
{\bf Note.}  The effect of selection cuts on the low-$z$ and RAISIN datasets.  These include requiring (1) photometric uncertainty from bright hosts $<$ 0.1 mag, including systematic uncertainty due to subtraction residuals, (2) spectroscopic confirmation that the SN is Type Ia, (3) redshift $>$ 0.01 to limit peculiar velocity errors, (4) data taken before maximum light, both before removing points with $\chi^2 > 10$ from the light curve fit and (5) after removing those points (see text), (6) existence of NIR data, (7) successful convergence of the fit to the SNooPy light-curve model, (8) $E(B-V) < 0.3$, (9) values for the $s_{BV}$ parameter that indicate a normal SN\,Ia light curve, and (10) two remaining Hubble diagram outliers (2006bt, DES15C1nhv) that fails Chauvenet's criterion (see text).
\end{minipage}
}
\end{tabular}
\label{table:raisin_cuts}
\end{table}

Next, we apply selection cuts to our sample to ensure that each SN has well-measured photometry and distances.  First, as discussed in Appendix \ref{sec:hostnoise}, we remove five SNe with host galaxy noise contributions that add greater than 0.1~mag to the photometric errors, based on an injected star analysis, as these are indications of subtraction residuals.  Due to PSF variation caused by the breathing of {\it HST}, it is difficult to have clean subtractions on galaxies that are bright relative to the SN; this is because even few-percent variations in the PSF when multiplied by a bright galaxy core can cause large residuals relative to the brightness of the SN.  This PSF variation will unavoidably bias our sample against SNe in high surface brightness environments, making it even more important to better characterize the SN-host galaxy environment correlation.  We visually inspect each difference image to ensure that the remaining SNe appear to have high-quality subtractions.  We also remove SN~PS1-520107, which has an unclear spectroscopic classification.  We note that SN~2006bt is peculiar \citep{Foley10}, but we chose to include it for consistency across the redshift range as similar events would be impossible to identify and exclude in our high-$z$ sample given the noisier spectra at those redshifts.  However, 2006bt subsequently fails Chauvenet's criterion (see below).  We further remove SNe that lack either optical or NIR data before maximum light as this makes it impossible to accurately determine the phase of a given observation.  We also ensure that pre-maximum light data exist both before and after removing $>$10-$\sigma$ SNooPy light-curve fit outliers from the data; if a SN had a mis-identified time of maximum light we found that a SN could appear to have pre-maximum light data when in reality it did not, but this issue was fixed once we removed outlying data points from the fit.

Next, we remove SNe measured to have $E(B-V) > 0.3$~mag from optical data.  Though only a single SN in our sample fails this cut, these SNe could in principle cause large outliers on our NIR-only Hubble diagram
given that we perform light-curve fits with fixed $A_V$.  High-$A_V$ SNe could also have some sensitivity to differences in $R_V$ that are difficult to measure due to the extrinsic/intrinsic degeneracy in SN colors \citep[e.g.,][]{Mandel22}.   We also remove SNe with extreme values of the $s_{BV}$ parameter that are not well-represented in the SNooPy training sample.  These include $s_{BV} > 1.18$  and $s_{BV} < 0.75$, values for which SNooPy is not well trained.   Cutting at $s_{BV} > 0.75$ also excludes 91bg-like and transitional SNe\,Ia (see \citealp{Burns14,Gall18}), which removes SNe that would be poorly standardized by our baseline method that does not fit for $s_{BV}$.  At $s_{BV} > 1.18$ in particular, only a single high-stretch SN is included in the SNooPy model training. 
SNe with $s_{BV}$ measurement uncertainties $>$0.2 are also removed, as we cannot reliably estimate whether they pass cuts.
 Finally, two SNe --- SN~2006bt and DES15C1nhv --- fail Chauvenet's criterion as 2.8- and 3.1-$\sigma$ outliers on the Hubble diagram.  A sample of 38 objects (the high-$z$ sample) is expected to have just 0.07 3.1-$\sigma$ (or greater) outliers on average, and a sample with 44 objects (the CSP sample) is expected to have just 0.2 $>$2.8-$\sigma$ outliers; both numbers are below the Chauvenet threshold of 0.5.  We treat the low- and high-$z$ distributions differently in this calculation as they have substantially different photometric data coverage and scatter (Section \ref{sec:optical}).

The full set of distances, light curve parameter measurements (using optical data), and cuts on those data for both the low-$z$ and RAISIN samples are given in Appendix \ref{sec:app_dist}. For each subsample, selection cuts are summarized in Table \ref{table:raisin_cuts}.

\subsection{Simulations and Bias Corrections}
\label{sec:simulations}

\begin{figure}
    \centering
    \includegraphics[width=3.5in]{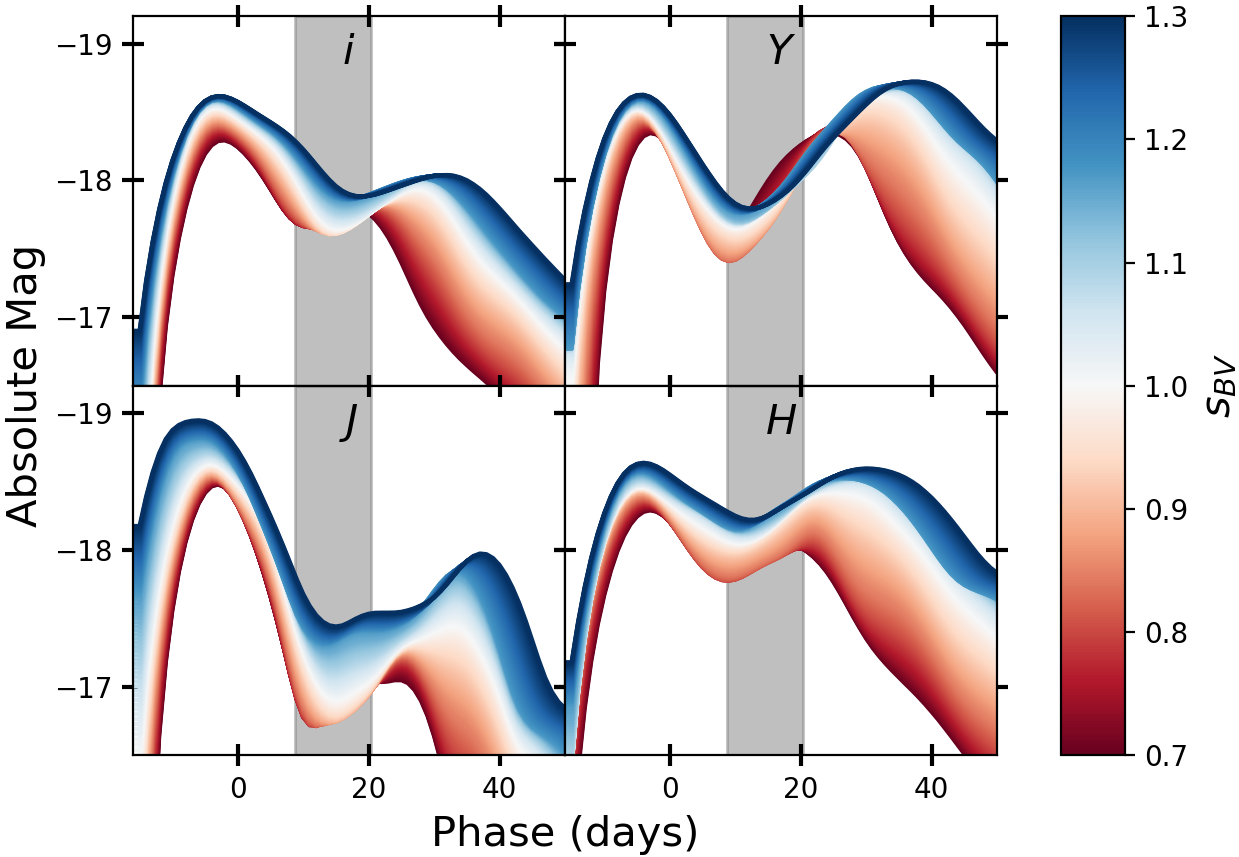}
    \caption{SNooPy model $iYJH$ light curves as implemented in SNANA and as a function of the color-stretch parameter, $s_{BV}$.  The rest-frame $Y$ band model in particular has very little sensitivity to $s_{BV}$ at the epochs of the RAISIN data.  The phase range within which 68\% of RAISIN observations fall are shown in grey bands (for CSP the 68\% range covers phases from $\sim -3$-40~days).}
    \label{fig:snoopy}
\end{figure}

\begin{figure*}
  \includegraphics[width=7in]{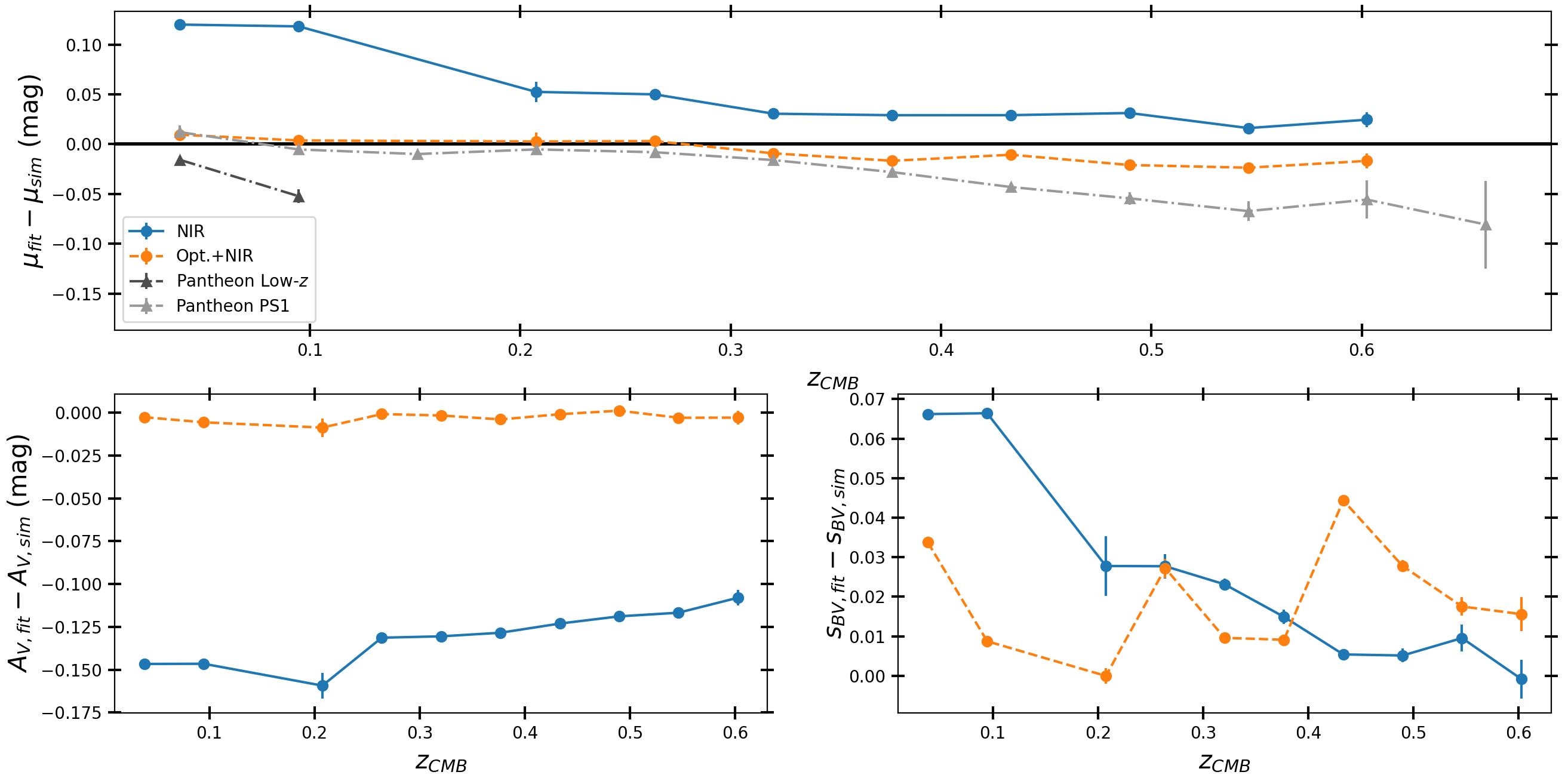}
  \caption{Distance biases (top), $A_V$ biases (bottom left) and $s_{BV}$ biases (bottom right) as a function of redshift.  Results are shown for both the NIR-only light curve fits (blue), which assume zero reddening and $s_{BV} = 1.0$, and the optical$+$NIR fits (orange), which include $A_V$ and $s_{BV}$ as free parameters.  Predicted low-$z$ (black triangles) and PS1 (grey triangles) distance biases from Pantheon are also shown in the top panel.  Some bin-to-bin fluctuation in the bias corrections as a function of redshift is caused by simulating SNe at only the exact redshifts of RAISIN objects for a smaller, more computationally efficient simulation.}
  \label{fig:biascor}
\end{figure*}

Using the dispersion model from Section \ref{sec:dispmodel}, we use the SNANA \citep{Kessler10} simulation framework to generate realizations of each subsample in this analysis (CSP, RAISIN1, and RAISIN2).  SNANA is the ``industry standard'' tool for accurately simulating large SN\,Ia samples for cosmological parameter measurements, and for this reason both the MDS and DES teams have already developed sophisticated SNANA simulations of their surveys that yield accurate realizations of their survey data.  SNANA includes sky noise, survey zeropoints, survey cadences, host galaxy properties, realistic shape and color distributions for the SN population, and the quantitative criteria that were used to trigger followup observations.

These simulations are presented in Appendix \ref{app:sims}.  In brief, we based our simulations on pre-existing CSP \citep{Kessler19}, PS1 \citep{Scolnic18}, and DES \citep{Kessler19} simulations, but we applied them to a SNooPy model that we implemented within the SNANA framework (Figure \ref{fig:snoopy}).  We then estimated intrinsic population distributions of the $s_{BV}$ values using the aggregate NIR data and, although NIR data are only weakly dependent on $A_V$, we simulate a nominal exponential $A_V$ scale of $\tau = 0.2$~mag and vary this scale length in our systematic error budget.  This exponential scale is well matched to our measurements from optical observations, from which we estimate $\tau \simeq 0.15-0.18$~mag; preliminary optical$+$NIR analysis of the RAISIN sample from BayeSN also gives $\tau=0.21 \pm 0.04$.

The simulations then give the distance bias: the average difference between simulated and measured distance as a function of redshift, shown in Figure \ref{fig:biascor}.  We correct the measured distances for the average bias at each SN redshift.  The predicted optical distance biases have similar sizes
to bias corrections generated for the Pantheon analysis \citep{Scolnic18} using the \citet{Guy10} optical scatter model and the SALT2 standardization model.  The NIR bias corrections are dominated by the difference in mean stretch and $A_V$ between the low- and high-$z$ samples.  The uncertainties on the differences in these distributions are used to determine systematic uncertainties on the bias corrections.

\subsection{Host Galaxy Mass Dependence}
\label{sec:host}

\begin{figure}
    \centering
    \includegraphics[width=3.5in]{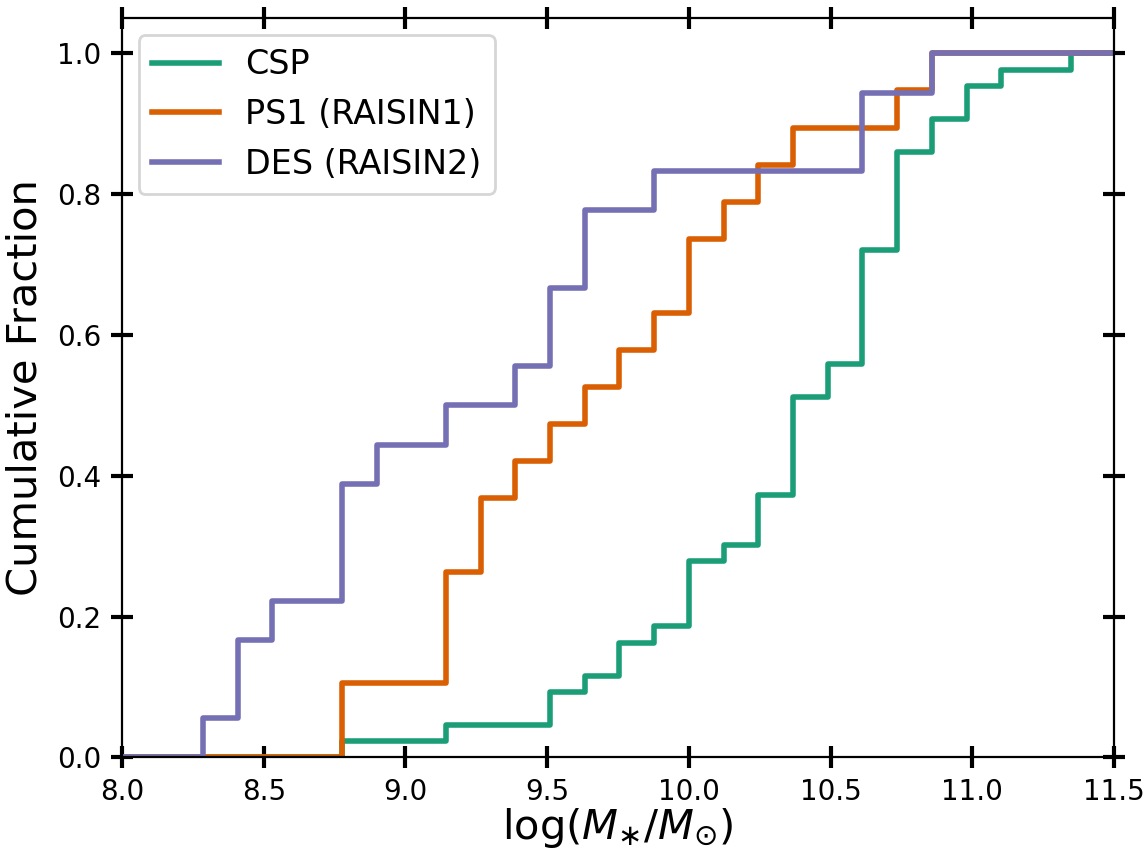}
    \caption{Cumulative fraction of SNe as a function of host galaxy mass for CSP and RAISIN SNe.  CSP SNe are found in significantly more massive host galaxies.}
    \label{fig:mass_hist}
\end{figure}

The step-like dependence of SN Hubble residuals on host galaxy mass has been well established \citep{Kelly10,Lampeitl10,Sullivan10} and recently there have been 
reported --- but somewhat conflicting --- detections of the host mass step in the NIR \citep{Ponder20,Uddin20,Johansson21}.  Constraining the wavelength dependence of the mass step could help to distinguish among theories for the origin of the step, e.g., dust properties \citep{Brout21} versus progenitor metallicity \citep{Rose21}.  We attempt to measure and correct for a host galaxy mass step in the RAISIN data.
We measure host galaxy masses ourselves to ensure consistency across the sample using a procedure described in Appendix \ref{app:hostmass}.  We simultaneously estimate the mass step and the average Hubble residual in three redshift bins to ensure our measurement is independent of cosmological parameters.  The resulting masses are shown in Figure \ref{fig:mass_hist}.

\subsection{Systematic Uncertainties}
\label{sec:analysis_systematics}

\begin{figure*}
  \includegraphics[width=7in]{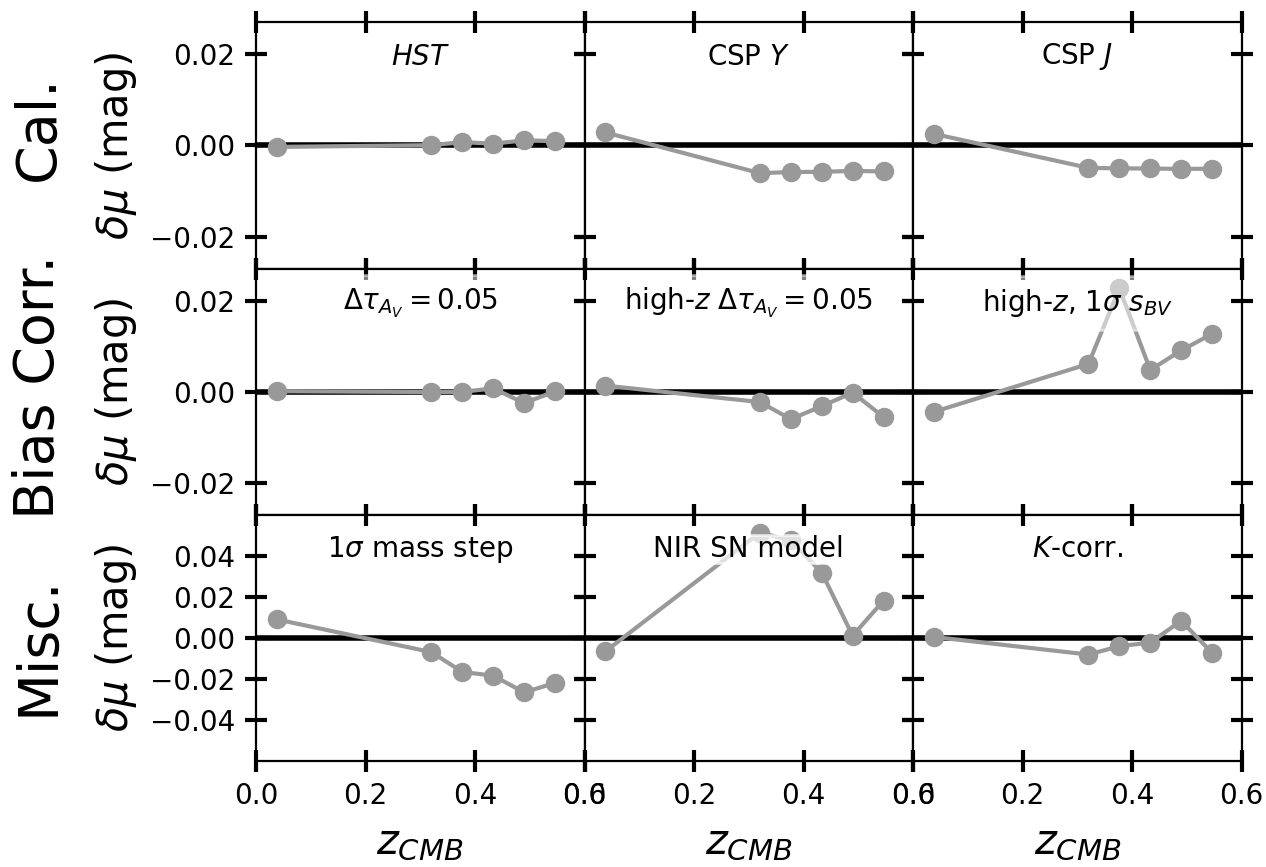}
  \caption{Selected systematic shifts in distance modulus as a function of redshift for this analysis, grouped into calibration systematics (top), bias correction systematics (middle) and other systematics (bottom).  
  The systematics shown in the middle panel are a shift in the scale length of the $A_V$ distribution (left), a shift in the scale length of the $A_V$ distribution for the high-$z$ data only (center), and 1-$\sigma$ shifts in the high-$z$ stretch distribution parameters (right).  The systematics shown in the bottom panel are due to the uncertainty in the size of the mass step (left), the SNooPy model (center), and $K$-correction uncertainties (right).  The weighted average of each distance vector is subtracted to show relative low- to high-$z$ differences, even if the systematic only changes the distances in one part of the redshift range.  For visualization purposes, only bins with at least 3 SNe are shown to reduce noise while illustrating $z$-dependent trends.}
  \label{fig:syserr}
\end{figure*}

We treat systematic uncertainties in much the same way as previous cosmological analyses
(e.g., \citealp{Scolnic18,Brout19,Jones19}): we determine 1-$\sigma$ uncertainties
in data and analysis parameters $-$ e.g., peculiar velocities, Milky Way reddening, calibration, and bias corrections $-$ apply each systematic to the data or analysis, and construct a covariance matrix of the distances resulting from each analysis choice or data modification.  The systematic uncertainty covariance matrix is then added to the (diagonal-only) statistical covariance matrix.  The systematic uncertainty covariance matrix is given by:

\begin{equation}
C_{sys}^{jk} = \sum_{n=1}^{N}\frac{\partial f(z_j)}{\partial S_n}\frac{\partial f(z_k)}{\partial S_n}\sigma^2(S_n)
\label{eqn:syscov}
\end{equation}

\noindent for a given set of systematics $S_n$ applied to each light curve or distance measurement. For the $n$th systematic, $\frac{\partial f(z_j)}{\partial S_n}$ is the change
in distance after applying systematic $S_n$ to a SN at redshift $z_j$.  The size of each systematic uncertainty is given by $\sigma(S_n)$.

Details about individual systematic uncertainties are given below.  Compared to previous analyses, our choice of the SNooPy model and the limited sample size allows us to omit a few second-order systematic uncertainties, including redshift evolution in nuisance parameters and redshift evolution of the host galaxy mass step.  We also do not include the choice of SN\,Ia dispersion model as a systematic uncertainty because we have constrained this directly from the low-$z$ Hubble diagram itself.  We are therefore not subject to the G10/C11 scatter model degeneracy that causes significant uncertainties in optical-only analyses \citep[e.g.,][]{Brout19}.  The redshift dependence of a subset of our systematic uncertainties is shown in Figure \ref{fig:syserr}.

\subsubsection{Calibration}

This analysis is limited to two photometric systems, {\it HST} and CSP (the Swope telescope).  The {\it HST} CALSPEC calibration affects both the RAISIN and low-$z$ datasets, while the CSP calibration to CALSPEC affects only the low-$z$ data.  The CALSPEC calibration was updated in \citet{Bohlin20} with the revised uncertainty in the $F125W$ and $F160W$ bands now at the level of 0.5\%.  The CSP calibration is from \citet{Krisciunas17}; the $JH$ calibration is tied to Vega while the optical calibration is tied to observations of BD$+$17$^{\circ}$4708.  We have updated the optical CSP zeropoints using the latest CALSPEC spectra of BD$+$17$^{\circ}$4708\footnote{Available at \url{https://archive.stsci.edu/hlsps/reference-atlases/cdbs/current_calspec/}.}.  The CSP $JH$ calibration is uncertain at the level of approximately 2\%, and we apply independent 2\% $JH$ systematic errors to the CSP magnitudes as a conservative estimate.  We also assume a larger, 3\% systematic error for the $Y$-band calibration as it was calibrated to \citet{Castelli03} atmospheric models rather than Persson standards \citep{Persson98,Krisciunas17}.  Better calibration of low-$z$ samples is a promising area for improvement to enable improved future ground-based, NIR measurements.

\subsubsection{Bias Corrections}

The uncertainty in the host galaxy bias corrections is dominated by uncertainty in the intrinsic distributions of $s_{BV}$ and $A_V$ in our samples.  The method of \citet{Scolnic16} gives 1$\sigma$ uncertainties on the derived population parameters and we re-compute bias corrections after varying the $s_{BV}$ distribution parameters for the low-$z$ and high-$z$ data by their 1-$\sigma$ uncertainties.  The population parameters after 1-$\sigma$ variations are shown in Appendix \ref{app:sims}.  We include two additional extinction variants: first, we reduce the exponential scale length by 0.03 in $E(B-V)$ (from 0.13 to 0.1) and second, we reduce the exponential scale length of the extinction distribution by 0.03 in $E(B-V)$ for {\it only} the CSP sample.  These $E(B-V)$ shifts are equivalent to a shift in mean $A_V$ of 0.05 for our default $R_V = 1.52$ or 0.1 for $R_V = 3.1$; they are greater than the measured difference in mean $A_V$ between the low- and high-$z$ samples when optical$+$NIR data are used in the fitting.

\subsubsection{NIR SN Standardization Model}
\label{sec:nir_model_sys}

We substitute BayeSN distances for SNooPy distances to estimate the systematic error in the SNooPy NIR SN\,Ia standardization model.  Though BayeSN has not yet been implemented in SNANA, e.g., bias corrections cannot yet be estimated, we can still compare the un-corrected distances after fitting the RAISIN and low-$z$ data with BayeSN.  To ensure that the BayeSN distances are fit in the same way as the SNooPy distances, we keep the $A_V$ and shape parameter fixed to values equivalent to the SNooPy values of $A_V = 0$ and $s_{BV} = 1$ (BayeSN $A_V = 0$, $\theta \simeq -1$), and we apply the SNooPy bias corrections to the BayeSN distances, making the assumption that the SNooPy bias corrections remain valid for BayeSN measurements.  SNooPy and BayeSN distances are well-correlated (see further discussion in Section \ref{sec:bayesn}) and the average Hubble residual between the low- and high-$z$ samples changes by just 2~mmag when using BayeSN instead of SNooPy.

\subsubsection{Mass Step}
\label{sec:massstepsys}

We adopt two variants of the host galaxy mass step.  In the first, we shift the mass step by its 1-$\sigma$ uncertainties from the maximum likelihood approach described in Appendix \ref{app:hostmass}.  In the second, we adopt the NIR mass step location suggested by \citet{Ponder20} of 10.44~dex: we measure the mass step at a location of 10.44~dex, recompute the step, and re-apply the step to the data.

\subsubsection{$K$-corrections}

Because the \citet{Hsiao07} model is used to generate SNooPy's $K$-corrections, use of the SNooPy model requires an understanding of the uncertainties in $K$-corrections due to the limited number of SNe that were used to construct the \citet{Hsiao07} model.  To create a model for $K$-corrections with realistic uncertainties, we create composite spectra following the method of \citet{Siebert19} using a limited number of individual NIR spectra from the Infrared Telescope Facility (\citealp{Marion09}, with data from G.~H.~Marion, private communication) that were used to create the original \citet{Hsiao07} template.  Composite spectra are created at phases of approximately $-9,-3, +7, +15, +39$~days, epochs that match the phases of the individual spectra themselves, and then bootstrap-resampled 50 times to create estimated uncertainties.  The $+$15-day composite spectrum is within a few days of the epoch of most RAISIN NIR data\footnote{We note that this is also near the epoch in which the $H$-band spectral break occurs \citep{Hsiao13}, but this would only affect our low-$z$ data where $K$-corrections are small.}.  We then use the shape of the \citet{Hsiao07} template light curve at each wavelength to interpolate between the phases covered by the composite spectra and use a Savitzky-Golay algorithm to smooth over the telluric regions.  We then assume those uncertainties are representative of the uncertainties on the \citet{Hsiao07} model and shift the model by the standard deviation of the bootstrap-resampled spectra.  This will result in a systematic change in the measured distances caused by the statistical uncertainty in the \citet{Hsiao07} model, particularly at high redshift.  

We note some small slope differences between the \citet{Hsiao07} template and our composite spectra, but find good consistency overall.  
We also inspect the 2011fe NIR spectra from \citet{Hsiao13} $-$ a SN with somewhat narrower than average stretch $-$ and see some broadband color differences that could be attributed to relative calibration errors in the spectra or perhaps intrinsic variation between SNe.  Methods such BayeSN, which effectively has a SN stretch-dependent $K$-correction, is capable of modeling this intrinsic variation.  Future, well-sampled, public NIR spectral databases will reduce this systematic uncertainty and are an important area to improve future analyses.

\begin{figure*}
  \includegraphics[width=7in]{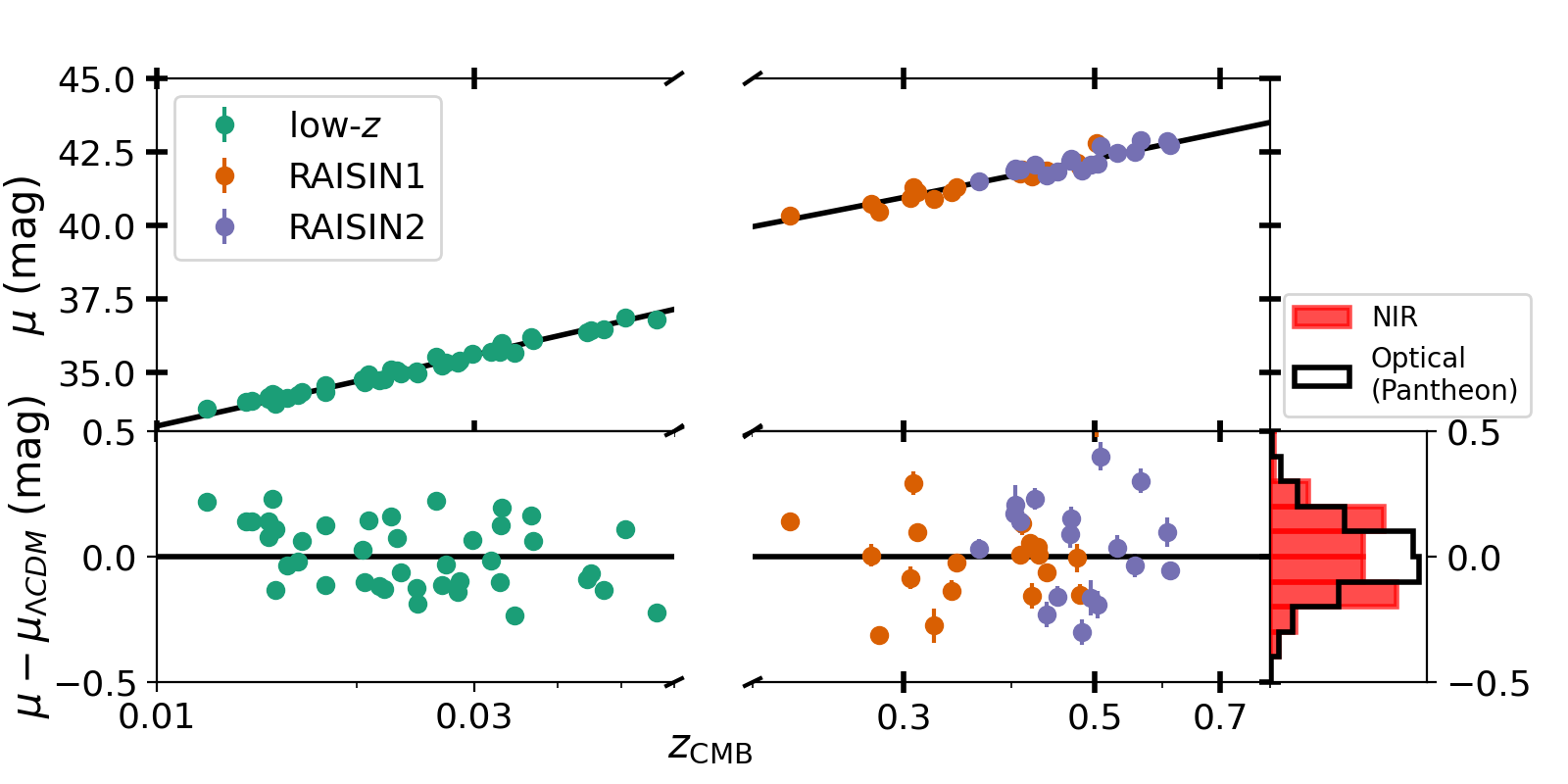}
  \caption{The RAISIN Hubble diagram, with low-$z$ SNe from CSP (green), and high-$z$ samples from RAISIN1 (orange) and RAISIN2 (purple).  
  We show photometric uncertainties only, neglecting the contribution of intrinsic scatter for visual clarity.}
  \label{fig:hubble}
\end{figure*}

\subsubsection{Peculiar Velocities and MW $E(B-V)$}

We apply peculiar velocity and Milky Way (MW) reddening systematics following \citet{Scolnic18,Jones19}.  For MW $E(B-V)$, we reduce the \citet{Schlafly11} values by 5\% according to the systematic uncertainty in the dust temperature correction.  For peculiar velocities, we conservatively vary the mass-to-light bias parameter $\beta$ in the 2M$++$ cosmic flow maps by its 5$\sigma$ statistical uncertainty \citep{Lavaux11}.

\subsection{Cosmological Parameter Measurements}

We use a combination of CosmoMC \citep{Lewis02} and CosmoSIS\footnote{\url{https://bitbucket.org/joezuntz/cosmosis/wiki/Home}.} \citep{Zuntz15} with likelihood chains from \citet{Planck18}, including temperature, polarization, and lensing, to measure cosmological parameters from the vector of SN distances and the systematic uncertainty covariance matrix.  To measure SN-only constraints we use CosmoMC to sample the likelihood, while for combining SNe with {\it Planck}, we use the CosmoSIS importance sampler to combine SN constraints with results from the {\it Planck} chains.  Luminosity distances $d_L$ from the $wCDM$ model, in Mpc, are given by:

\begin{equation}
  \begin{split}
    d_L(z,w,\Omega_m,\Omega_{\Lambda},\Omega_k) = (1 + z)\frac{c}{H_0}\int_0^z\frac{dz}{E(z)},\\
    E(z) = [\Omega_M(1 + z)^3 + \Omega_k(1 + z)^2 + \Omega_{\Lambda}(1 + z)^{3(1+w)}]^{1/2}.
  \end{split}
  \label{eqn:dl}
\end{equation}

\noindent Here $\Omega_m$, $\Omega_{\Lambda}$, and $\Omega_k$ are the cosmic matter density, dark energy density and spatial curvature, respectively.  We use the following function to estimate cosmological parameters:

\begin{equation}
    \chi^2 = (\hat{\mu} + \Delta\mathcal{M} - \mu_{w{\rm CDM}})C^{-1}(\hat{\mu} + \Delta\mathcal{M} - \mu_{w {\rm CDM}}),
\end{equation}

\noindent where $C$ is the covariance matrix from the combined statistical and systematic uncertainties, $\hat{\mu}$ is the distance modulus measured from the data as described in Section \ref{sec:distmeas} and including the bias corrections and host mass step as discussed in Sections \ref{sec:simulations} and \ref{sec:host}.  The model distance modulus $\mu_{wCDM} = 5\log(d_L)-5$.  $\Delta\mathcal{M}$ is the offset between the assumed SNooPy absolute magnitude of a SN\,Ia, which is also degenerate with the assumed value of H$_0$, and the best-fit global value.

\section{Results}
\label{sec:results}

\begin{table}[]
    \centering
    \caption{Summary of Systematic Uncertainties on $w$ from NIR Data Alone}
    \begin{tabular}{lccc}
    \hline \hline
Error&$\Delta w$&$\Delta\sigma_w$&$\sigma_{w,sys}/\sigma_{\textrm{stat}}$\\
\hline
All Sys.&-0.040&0.082&0.954\\
\hline
Bias Corr.&-0.040&0.070&0.809\\
$-$ $s_{BV}$&-0.040&0.069&0.805\\
$-$ $A_V$&0.000&0.009&0.099\\
Mass Step&-0.012&0.036&0.425\\
Phot. Cal.&0.006&0.026&0.306\\
$-$ Low-$z$&0.005&0.026&0.305\\
$-$ $HST$&0.000&0.002&0.025\\
Pec. Vel.&-0.017&0.010&0.110\\
NIR SN Model&0.008&0.008&0.097\\
Template Flux&0.008&0.008&0.097\\
MW E(B-V)&0.000&0.003&0.041\\
$k$-corr.&-0.002&0.002&0.026\\
\hline\\
    \multicolumn{4}{l}{
  \begin{minipage}{2.75in}
    {\bf Note.} Summary of systematic uncertainties on the $w$CDM model with constraints from SNe$+${\it Planck}.  $\Delta w$ is the shift in $w$ resulting from applying each systematic uncertainty and $\Delta\sigma_w$ is the additional uncertainty that would need to be added in quadrature to the statistical error to yield the uncertainty on $w$ from a given variant.  Note that some variants shift the value of $w$ but do not increase its uncertainties.
\end{minipage}}
    \end{tabular}
    \label{table:w_sys}
\end{table}

\begin{table*}[]
    \centering
    \caption{Comparing Optical and NIR Measurements of $w$}
    \begin{tabular}{lcccc}
    \hline \hline
&$w_{stat}$&$w_{stat+sys}$&$\sigma_{w,sys}$&$\Delta\sigma_w/\sigma_{\textrm{stat}}$\\
Opt. only&$-1.099\pm0.074$&$-1.102\pm0.092$&0.055&$0.739$\\
NIR only&$-1.128\pm0.086$&$-1.168\pm0.119$&0.082&$0.956$\\
Opt.$+$NIR&$-1.043\pm0.056$&$-1.059\pm0.074$&0.048&$0.864$\\
\hline\\
    \end{tabular}
    \label{table:w_optnir}
\end{table*}

After performing the analysis described in the previous section, we produce the RAISIN NIR-only Hubble diagram shown in Figure \ref{fig:hubble}.  The high-$z$ distance measurements are presented in Appendix \ref{sec:app_dist}.  The total root mean square (RMS) of the Hubble residuals is 0.167~mag, 18\% higher than the Pantheon Hubble residuals.
However, the CSP scatter for those SNe with NIR data near maximum light is 0.136 mag, slightly {\it lower} than the Pantheon scatter of 0.141~mag, showing the value of well-sampled NIR light curves near maximum light.  Future samples with additional epochs near maximum light and a revised NIR model will reduce the scatter well below that achievable from optical data \citep{Avelino19}.  Additional data in the gap between the low-$z$ and RAISIN redshifts would be particularly beneficial in creating an extended NIR Hubble diagram.

\begin{figure}
  \includegraphics[width=3.5in]{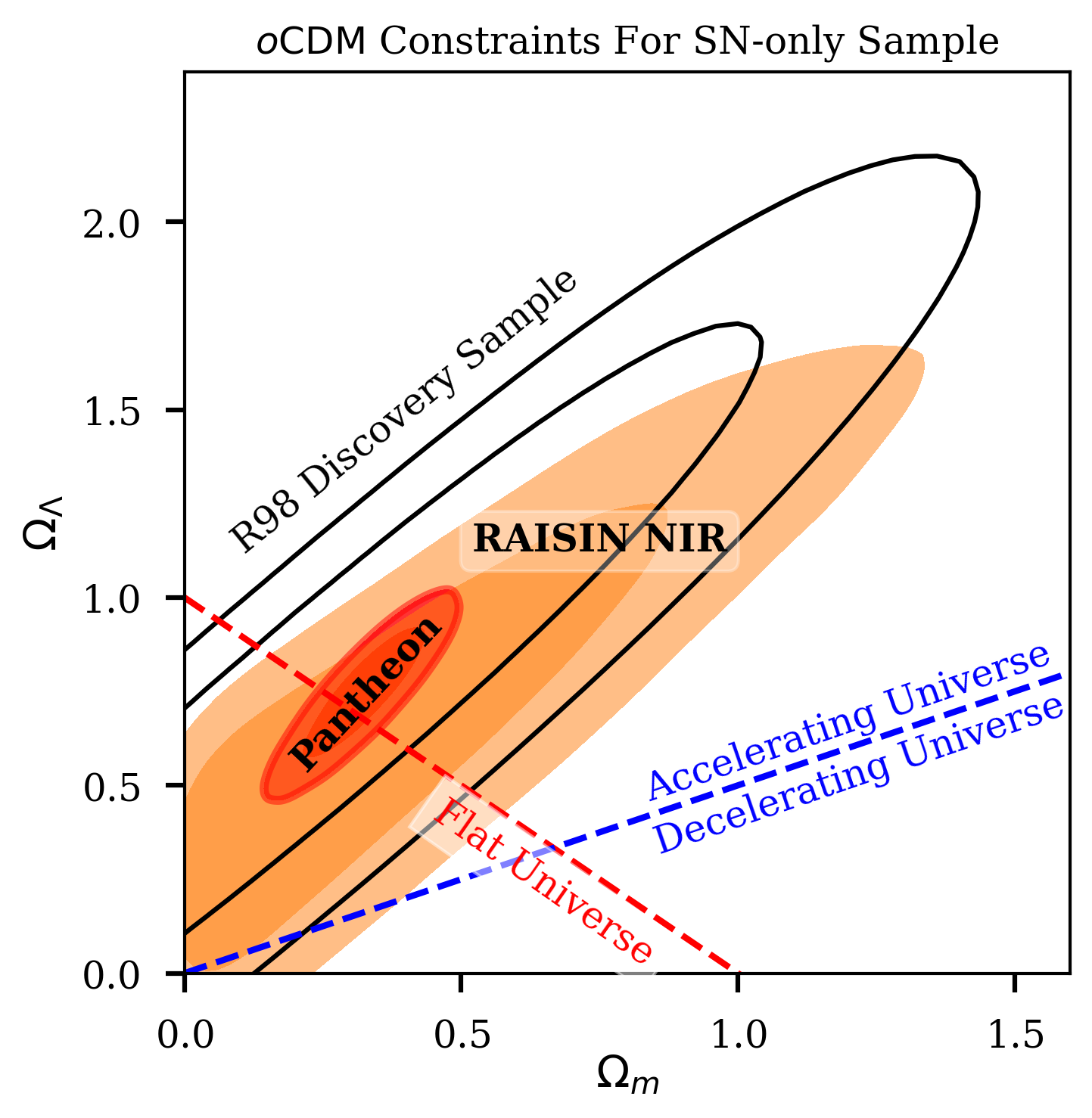}
  \caption{Cosmological parameter measurements from $o$CDM (a CDM model allowing non-zero curvature) with SNe alone.  Open contours show the \citet{Riess98} discovery sample, red contours show the Pantheon constraints from \citet{Scolnic18}, and the results from RAISIN SNe are in orange.  All contours show the 68\% and 95\% confidence intervals.}
  \label{fig:ocdm}
\end{figure}

\begin{figure}
  \includegraphics[width=3.5in]{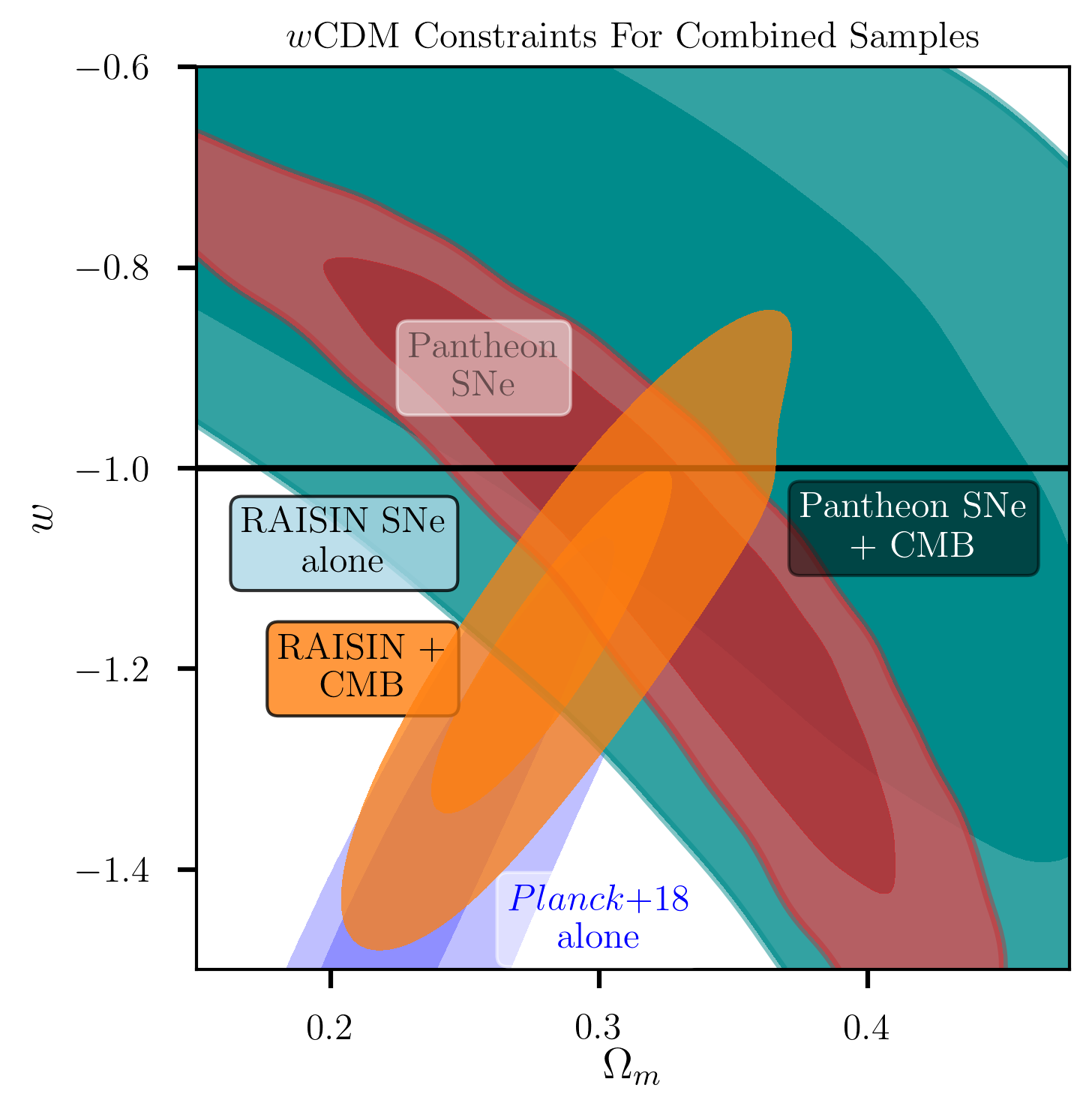}
  \caption{Cosmological parameter measurements from $w$CDM.  RAISIN$+$CMB cosmological constraints (orange) are consistent with those from the Pantheon sample (black; \citealp{Scolnic18}), with a statistically insignificant shift towards phantom dark energy ($w < -1$).  All contours show the 68\% and 95\% confidence intervals.}
  \label{fig:wcdm}
\end{figure}

\begin{figure}
    \centering
    \includegraphics[width=3.5in]{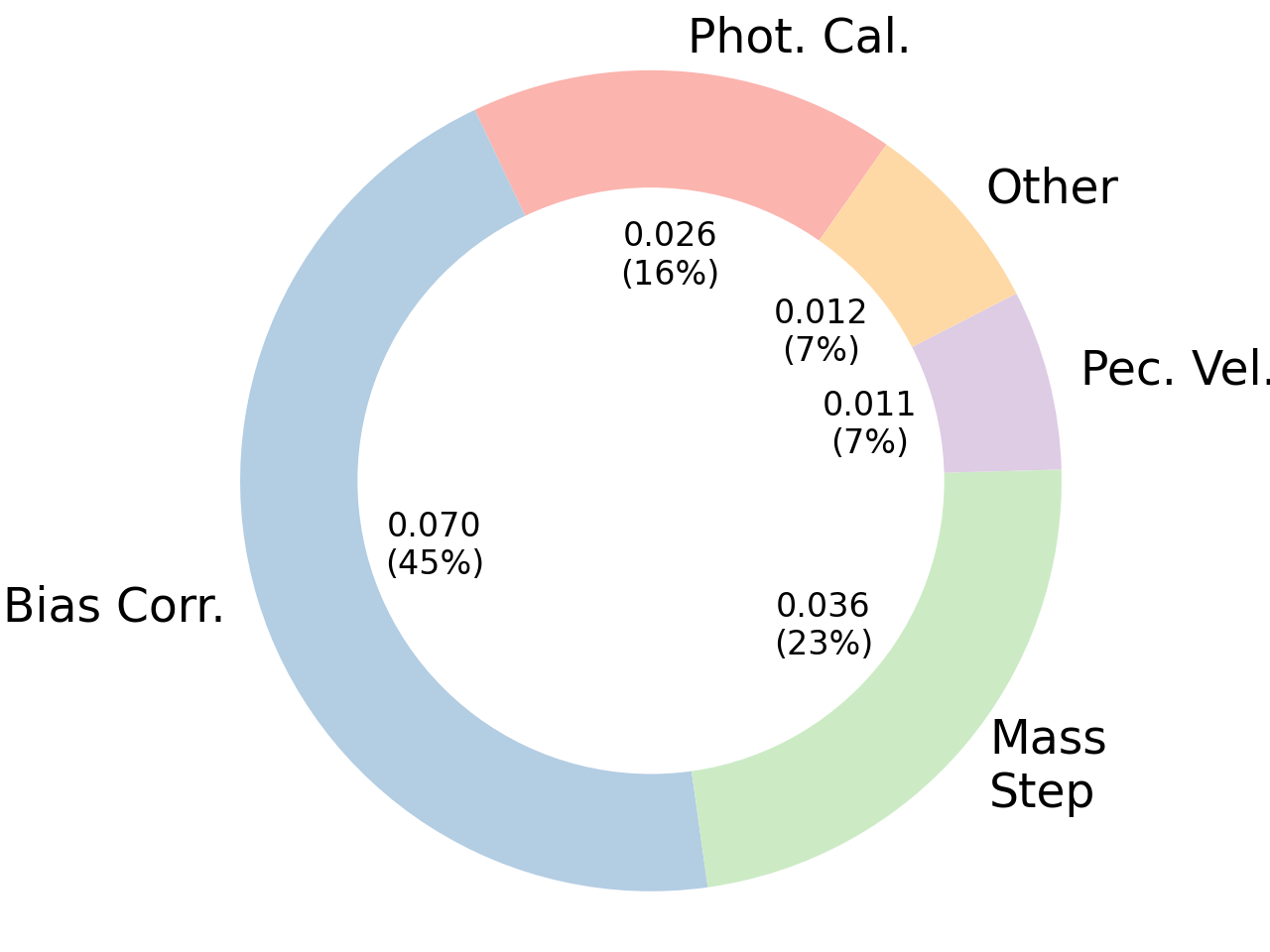}
    \caption{Summary of systematic uncertainties on $w$ for the RAISIN cosmological analysis (Table \ref{table:w_sys}).  Bias correction and mass step uncertainties dominate the error budget, with photometric calibration and uncertainties in the peculiar velocities contributing significantly.  Other systematic uncertainties are added in quadrature to give the ``Other" contribution.}
    \label{fig:raisin_sys}
\end{figure}

\subsection{$\Omega_m$ and $w$}
\label{sec:cosmo}

By using CosmoMC to constrain cosmological parameters from SNe alone, Figure \ref{fig:ocdm} shows the constraints on non-flat CDM relative to Pantheon and the cosmic acceleration discovery sample from \citet{Riess98}.  These data thereby confirm cosmic acceleration with high significance, using a new method and wavelength range that has less sensitivity to dust extinction uncertainties.   
In a flat universe, from RAISIN SNe alone, we find $\Omega_m = $ \Om.

With CMB constraints from \citet{Planck18} we find a dark energy equation of state of $1+w = $~\onepw\ (Figure \ref{fig:wcdm}).  This measurement is consistent with the previous NIR measurement from \citet{Freedman09} of $1+w = -0.05 \pm 0.13~{\rm (stat)}~\pm 0.09~{\rm (sys)}$ but with smaller uncertainties.

Statistical and systematic uncertainties in this analysis are approximately equal.  The dominant systematic uncertainty in this measurement is the bias correction --- predominantly due to uncertainty in the intrinsic $s_{BV}$ distribution --- which will improve with larger sample sizes and improved light curve fitting methods.  The second-largest systematic uncertainty is the size of the NIR mass step (Section \ref{sec:mass_step_results}), which will similarly improve with larger sample sizes.
The full list of systematic uncertainties on the $w$CDM model are shown in Table \ref{table:w_sys} and visually in Figure \ref{fig:raisin_sys}.  Though we have attempted to be comprehensive in our accounting of systematic uncertainties, we note that there may be substantial additional uncertainties in this novel measurement that will become more apparent only with larger sample sizes.

\subsection{The Hubble Constant}
\label{res:h0}

In addition to measuring $w$, we measure H$_0$ with NIR SN data alone.  We use Cepheid distances with CSP and RAISIN SN data to measure the Hubble constant from the distance ladder approach and use SN$+$CMB data to constrain the H$_0$ parameter with an ``inverse distance ladder" approach \citep[e.g.,][]{Cuesta15} that combines the angular size of the CMB sound horizon with $w$CDM constraints from CSP, RAISIN and {\it Planck}.

To measure H$_0$ from the distance ladder, we use publicly available Cepheid or tip of the red giant branch (TRGB) distances for all ten SNe with CSP photometry and with independent TRGB or Cepheid measurements to calibrate the NIR luminosity of SNe\,Ia.  Nine of these SNe --- SNe~2006D, 2007af, 2007on, 2007sr, 2009Y, 2011iv, 2012fr, 2012ht, and 2015F\footnote{Photometry for SN~2012fr is published in \citet{Contreras18}, photometry for SNe~2012ht and 2015F are published in \citet{Burns18}, and photometry for SN~2011iv are published in \citet{Gall18}.}--- have redshifts of $z < 0.01$ and therefore were not included in our baseline sample for measuring $w$.  One SN, 2007A, was already included in our sample as it has a redshift of $z = 0.017$.  

We use a combination of available Cepheid and TRGB distance calibrations \citep{Freedman19,Anand21, Riess21b} and take the agnostic approach of giving all available distances as originally provided equal weight.  We use Cepheid distances from \citet{Riess21b} for five SNe: 2006D, 2007A, 2009Y, 2012ht, and 2015F.  We use the mean of available Cepheid and TRGB distances for three SNe: 2007af \citep{Freedman19,Riess21b}, 2007sr \citep{Freedman19,Anand21,Riess21b}, and 2012fr \citep{Freedman19,Anand21,Riess21b}. Finally, we use the mean TRGB distances for two SNe from \citet{Freedman19} and \citet{Anand21}, SNe~2007on and 2011iv, which are in the same host galaxy of NGC 1404.  These two SNe are fast decliners (\citealp{Gall18} find $s_{BV} = 0.57 \pm 0.01$ and $0.64 \pm 0.01$ for 2007on and 2011iv, respectively) with decline rates below our nominal $s_{BV}$ cuts so we provide results with and without them. 

We use these Cepheid and TRGB distances to calibrate the SN luminosity of the CSP and RAISIN samples following the method of \citet{Riess21b}.
If we include SNe~2007on and 2011iv in our $H_0$ measurement, we measure $H_0 = 77.5 \pm 2.1~{\rm km~s^{-1}~Mpc^{-1}}$ from Cepheid$+$TRGB and H$_0 = 76.6 \pm 2.6~{\rm km~s^{-1}~Mpc^{-1}}$ from TRGB alone.  Excluding the two fast-declining SNe in NGC~1404, however, we measure H$_0 = 75.9 \pm 2.2~{\rm km~s^{-1}~Mpc^{-1}}$ from Cepheid$+$TRGB and H$_0 = 72.9 \pm 3.0~{\rm km~s^{-1}~Mpc^{-1}}$ from TRGB distances alone. 
We note that the statistical significance of this TRGB measurement with a reduced sample is limited as we have just three TRGB calibrator galaxies and the measurement is statistically consistent with other recent TRGB-based measurements \citep{Freedman19,Anand21}.  The majority of the uncertainty derives from the error in the mean of the modest samples of NIR SN calibrators.

Because our distance measurement method does not correct for any dependence of SN luminosity on $s_{BV}$ and our Hubble flow sample excludes such fast-declining objects (we require $s_{BV} > 0.75$) it is possible these fast-decliners, which make up a large fraction of the calibrator sample, could bias our results; therefore, our baseline result excludes the fast-decliners.

\begin{figure}
    \centering
    \includegraphics[width=3.5in]{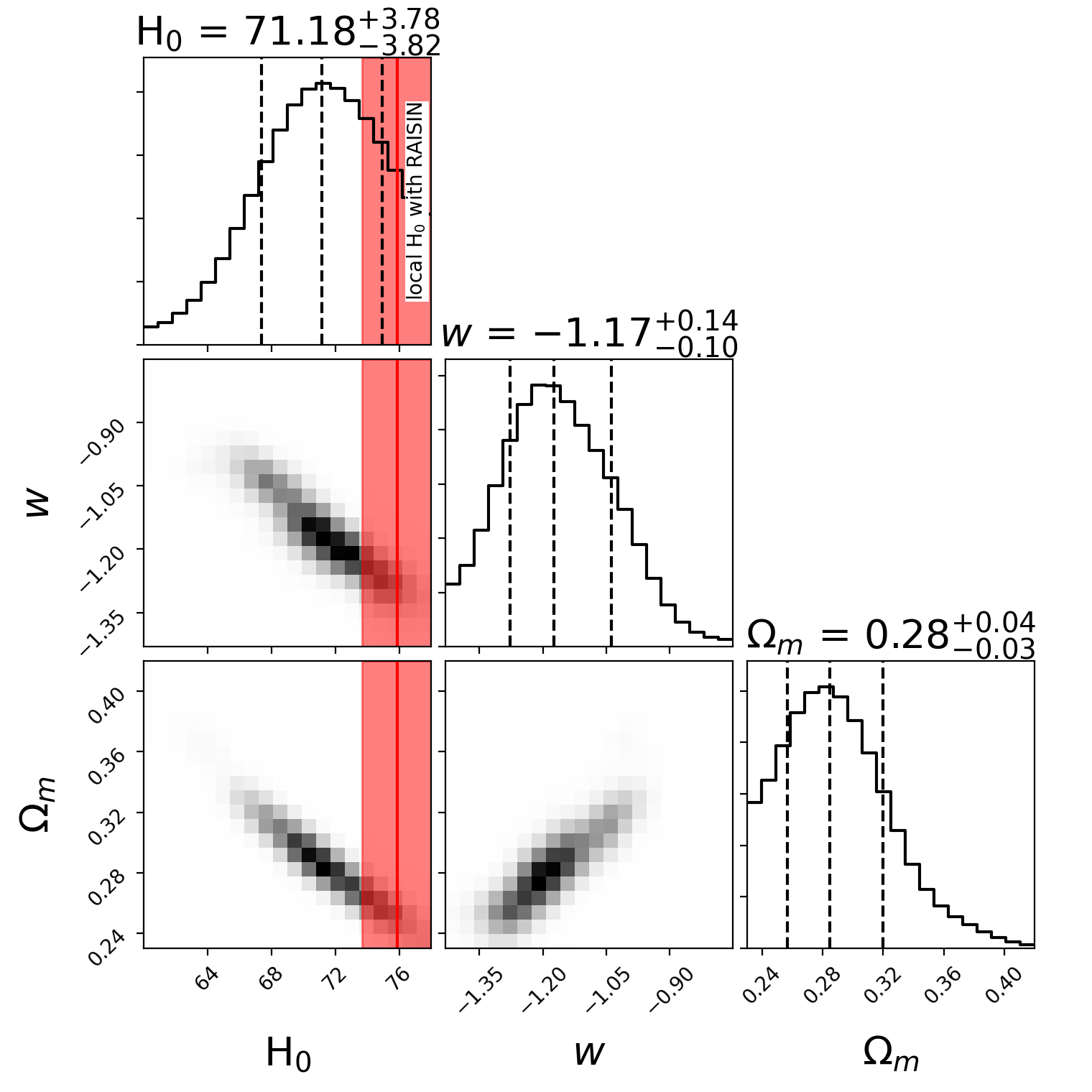}
    \caption{Constraints on H$_0$, $w$, and $\Omega_m$ from RAISIN and CMB data combined using an inverse distance ladder method.  This figure was generated with the {\tt corner} package \citep{corner}. }
    \label{fig:corner}
\end{figure}

From the inverse distance ladder in a $w$CDM model, we find a model parameter H$_0 = 71.2\pm3.8~{\rm km~s^{-1}~Mpc^{-1}}$ (Figure \ref{fig:corner}).  The value of H$_0$ is consistent with the conventional result of $67.4 \pm 0.5$ assuming $\Lambda$CDM with {\it Planck} data alone.
Independent approaches to measuring SN distances at cosmological redshifts such as this one demonstrate the ways in which future measurements can help determine what role, if any, the present dark energy could play in the tension between local (e.g., \citealp{Riess16,Pesce20,Huang20,Wong20,Freedman21,Riess21b}) and CMB H$_0$ measurements (e.g., \citealp{Dhawan20}).

Though the uncertainties on these measurements are large, we find that moving to NIR wavelengths does not appear to change the size of the difference between CMB and local H$_0$ measurements (see also \citealp{Burns18,Dhawan18}).

\subsection{Comparing to Optical and Optical$+$NIR Measurements}
\label{sec:optnirresults}

Finally, Table \ref{table:w_optnir} compares our NIR-only measurement of $w$ to measurements using optical data alone and optical$+$NIR data from SNooPy.  In computing the optical and optical$+$NIR measurements, we apply the same base set of systematic uncertainties but add calibration uncertainties at the level of 3~mmag for each PS1 band, 5~mmag for each DES band, and 1\% for each optical CSP band \citep{Scolnic18,Burke18,Krisciunas17}.  

Interestingly, we find substantial shifts when considering optical or optical$+$NIR data versus NIR data alone.  In the NIR-only case versus the optical$+$NIR case in which NIR data primarily have the effect of constraining the SN color, $w$ shifts by up to 0.11.  Though the statistical significance of this shift is not entirely clear given the correlated data and some correlation in the systematic uncertainties, this shift could point to the need for revisions to the NIR or optical SN standardization model in future work, particularly given forthcoming surveys such as the {\it Roman Space Telescope} SN survey that will observe in the NIR \citep{Rose21b}.

Although the sensitivity of the NIR-only measurement to bias correction uncertainties yields a result with significantly higher systematic errors in our baseline analysis, larger sample sizes would result in more precise bias estimations from a better constraint on the redshift-dependent stretch distribution of SNe.  However, given that NIR-only approaches will always have difficulty constraining the potential evolution of dust extinction with redshift, we believe that the optimal approach in future work will be an optical$+$NIR measurement; Table \ref{table:w_optnir} shows that this measurement has the lowest total systematic uncertainty budget of our three wavelength regimes.  Though the overall improvement in the precision of $w$ is small, we note that alternative distance measurement methods such as BayeSN \citep{Mandel22} appear to more optimally weight the NIR versus optical data and yield significantly reduced distance uncertainties when NIR data are included.

\section{Exploring Alternative Analysis Methodologies and Constraining the Mass Step}
\label{sec:sysdiscussion}

\begin{figure*}
    \centering
    \includegraphics[width=7in]{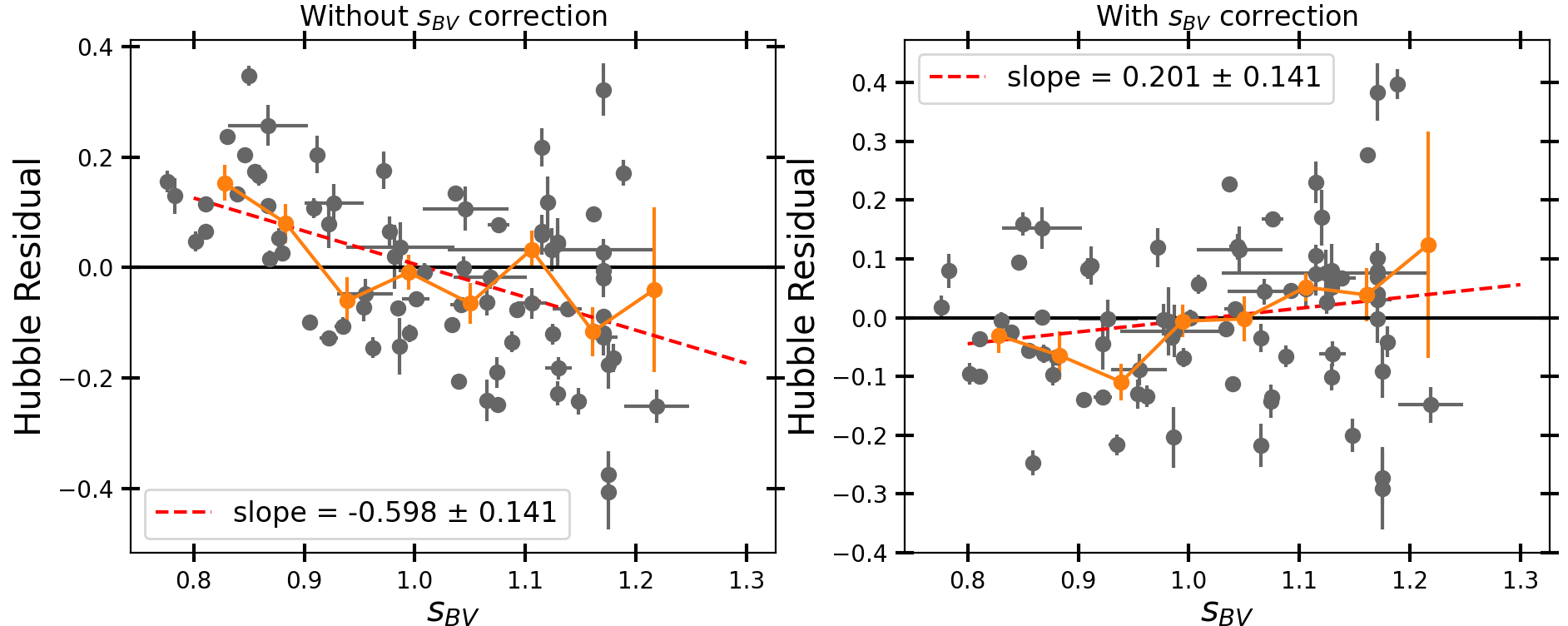}
    \caption{NIR Hubble residuals as a function of $s_{BV}$, before the Hubble residuals are corrected for $s_{BV}$ (left) and after (right).  Correcting for the marginally significant $\sim$2-$\sigma$ slope in the right-hand panel would shift our high-$z$ distances by just $\sim$0.015~mag.}
    \label{fig:sbv}
\end{figure*}

\begin{figure*}
  \includegraphics[width=7in]{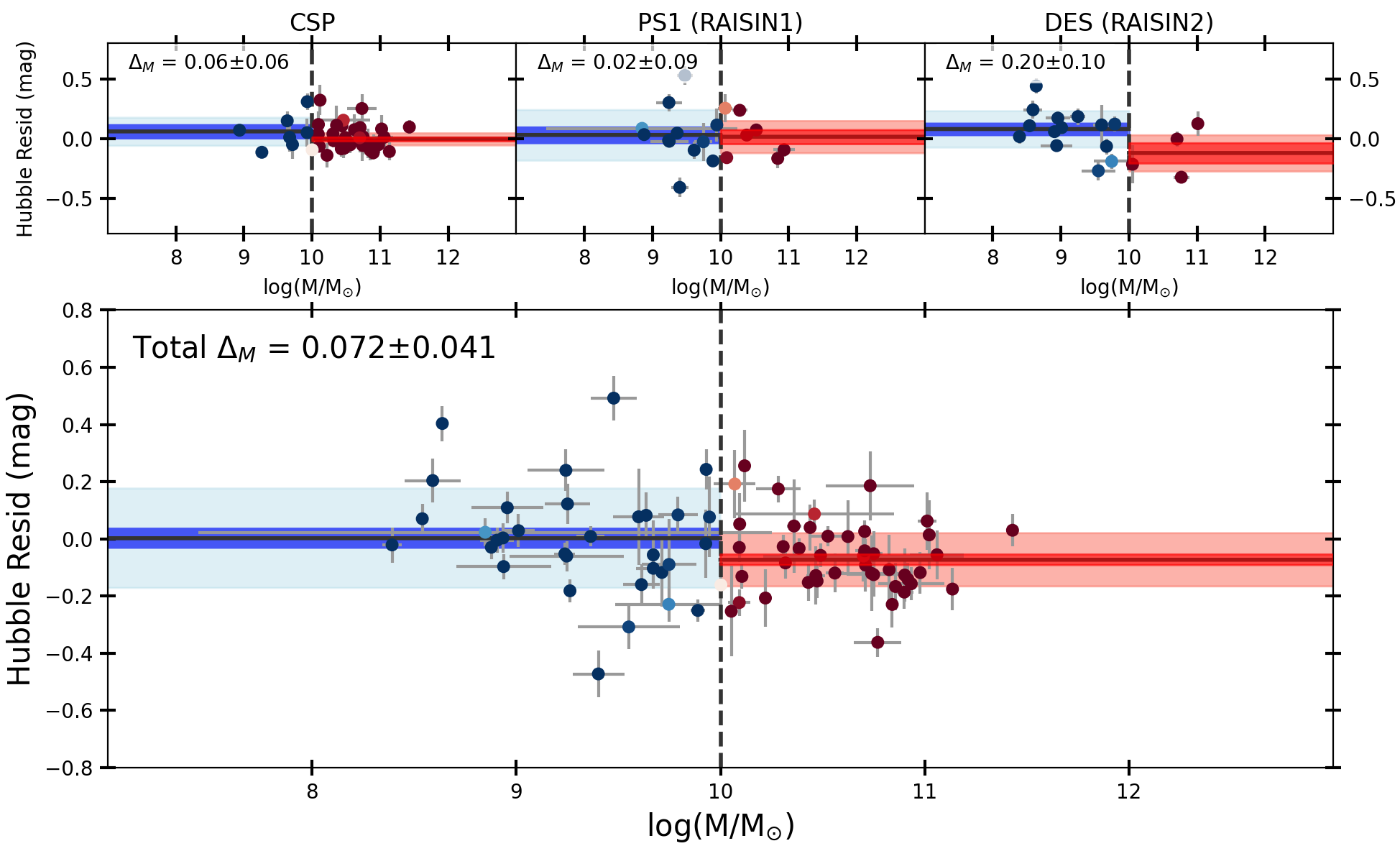}
  \caption{Maximum likelihood, NIR mass step measurements with CSP (top left), RAISIN1 (top center), RAISIN2 (top right) and all data combined (bottom).  Dark shading indicates the uncertainty on the mean for high-mass (red) and low-mass (blue) hosts, while light shading indicates the dispersion of each population.  The shading of the points indicates the probability that a SN is in a low-mass host (blue) or high-mass host (red).  In this figure (unlike in Figure \ref{fig:hubble}), we correct individual distances for $s_{BV}$ to derive a measurement that is directly comparable to optical analyses.  We also simultaneously fit for three $z$-dependent distance bins to remove the effect of cosmological parameter values on the mass step size.}
  \label{fig:massstep}
\end{figure*}

In this section, we explore variations on our nominal analysis and examine potential systematic uncertainties in the cosmological parameter measurement.

\subsection{The Mass Step}
\label{sec:mass_step_results}

Using our nominal distance measurement method we find a maximum likelihood size of the mass step of $\Delta_M = 0.022\pm0.047$~mag at a step location of 10~dex and $0.026 \pm 0.040$~mag at the \citet{Ponder20} location of 10.44~dex (the step location is fixed during each fit).  These results are used to generate the distance measurements in Section \ref{sec:results} above.
However, the differences between the NIR and optical measurements are affected by the fact that stretch is correlated with host galaxy mass; our nominal distance method does not fit for stretch and there is a median difference of $\Delta s_{BV} = -0.099$ between the high- and low-mass subsets of our sample.
Correcting for this stretch differential, as done in an optical-only analysis, would have the effect of making the mass step larger.

To measure the mass step in a way that is more comparable to the way optical mass steps are measured, we test the impact of applying an $s_{BV}$ correction to the NIR distances using the best-fit $s_{BV}$ from optical$+$NIR data.  We see a strong trend between Hubble residual and optical$+$NIR-derived $s_{BV}$ as shown in Figure \ref{fig:sbv} and find that the scatter in our Hubble residuals is reduced by 10\% by making this correction.  We do not see a significant change in the dependence of Hubble residual on $s_{BV}$ in the low-$z$ data (slope $-0.622 \pm 0.189$) compared to the high-$z$ data (slope $-0.702 \pm 0.244$).

After $s_{BV}$ fitting we measure a larger mass step, as we would predict, of \massstepstretch.  This result is shown in Figure \ref{fig:massstep}.  If instead we use the location of the mass step from \citet{Ponder20} of 10.44~dex, we find a mass step of 0.057$\pm$0.035~mag.  If we adopt the approximate linear, empirical correction between $s_{BV}$ and Hubble residual shown in Figure \ref{fig:sbv}, instead of using the default SNooPy $s_{BV}$-luminosity relation, we find consistent steps of $0.086\pm0.043$~mag and $0.094\pm0.037$~mag for 10 and 10.44~dex step locations, respectively.  We conclude there is $\sim$2-$\sigma$ evidence for a mass step in these data even if we do not {\it a priori} assume that the SNooPy luminosity-$s_{BV}$ relation is correct.

Using the high-$z$ data alone, we measure a mass step of $0.114\pm0.081$~mag at a step location of 10~dex and a step of $0.122\pm0.091$~mag at 10.44~dex.  Unfortunately, the sample is too small to constrain the size of the step with these high-$z$ data alone, which yield a statistically insignificant 1-$\sigma$ evidence for the NIR mass step.

If we use the optical data to measure the mass step with the same SNe, we find a step of $0.11\pm0.03$~mag from SNooPy and $0.08\pm0.03$~mag from SALT3 (the same step size is measured from SALT2).  These are consistent with optical mass steps in the literature, which can range from $0.04\pm0.019$~mag from DES \citep{Smith20} to $0.119\pm0.032$ from the Nearby Supernova Factory \citep{Aldering02,Rigault18}; the largest dataset (Pantheon; \citealp{Scolnic18}) yields a step of $0.072\pm0.01$\footnote{This value is from the analysis variant in which \citet{Scolnic18} do not apply bias corrections to the SN shape and color parameters and is the analysis version that is most comparable to our method here.}.

These mass step results qualitatively support $-$ albeit with limited significance $-$ the findings of \citet{Ponder20} and \citet{Uddin20}, who also found evidence that the NIR mass step was of order the same size as the optical mass step.

\subsection{Comparing to Alternative Distance Measurement Methods}
\label{sec:altdist}

\begin{figure}
    \centering
    \includegraphics[width=3.5in]{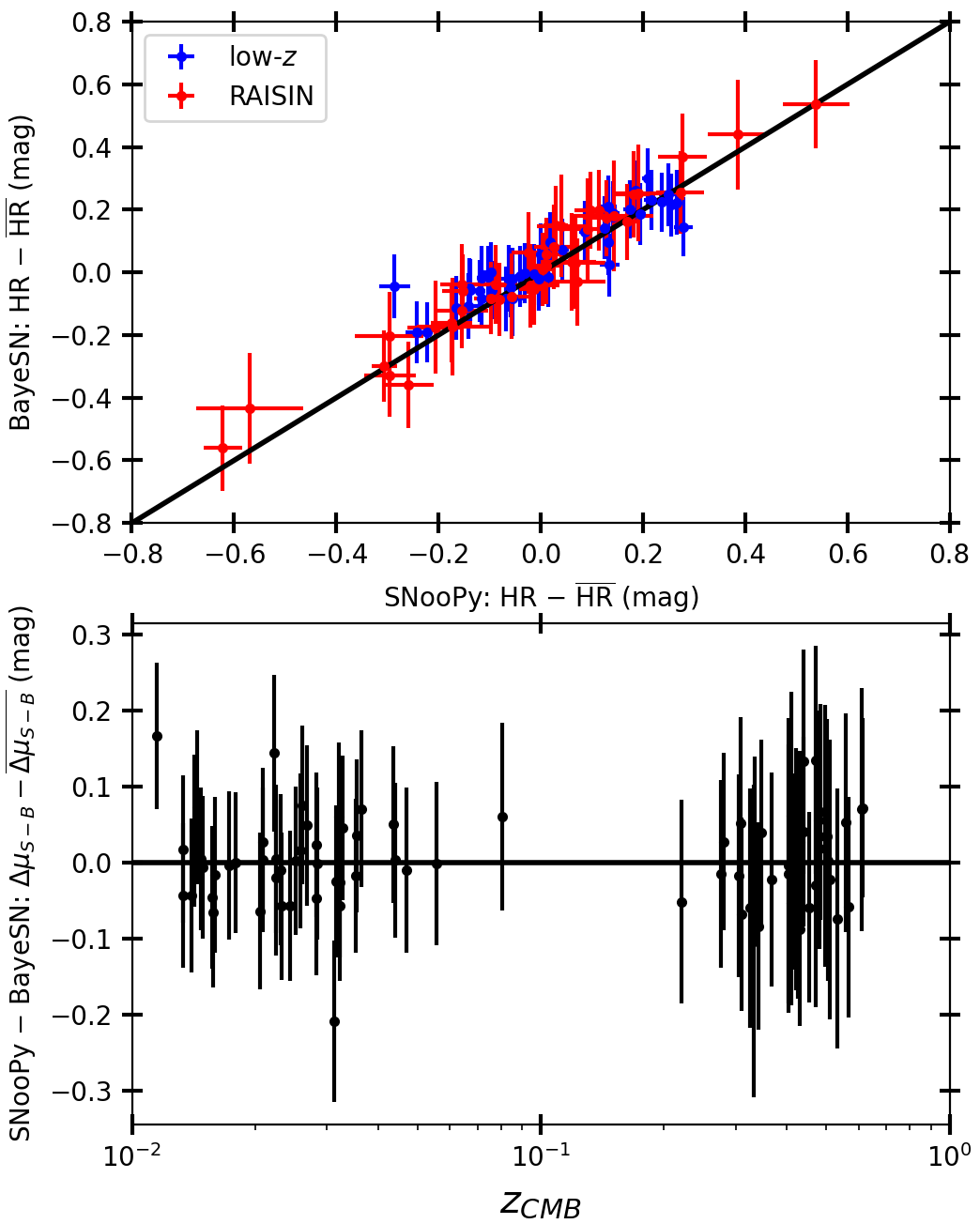}
    \caption{Top: SNooPy versus BayeSN Hubble residuals for the low-$z$ (blue) and RAISIN (red) samples.  Bottom: SNoopy distance moduli minus BayeSN distance moduli as a function of redshift with an average redshift-independent offset removed (cosmological parameter measurements are not affected by a global offset).  The average value of the SNooPy distance modulus minus the BayeSN distance modulus changes by only 2~mmag between the low-$z$ and RAISIN samples.}
    \label{fig:bayesn}
\end{figure}

\subsubsection{NIR Distances from BayeSN}
\label{sec:bayesn}

Although we have not yet simulated distance biases with the BayeSN SED model, we did compare our SNooPy distances before bias correction to raw BayeSN distances that assume a fixed stretch and zero host dust extinction $A_V$, analogous to the approach we adopted for SNooPy.  We fix the BayeSN primary intrinsic component $\theta = -1$, approximately equivalent to our chosen SNooPy $s_{BV} = 1$.  We use the same time of $B$-band maximum light for both SNooPy and BayeSN and fit only the NIR light curve data to obtain photometric distance estimates.

As demonstrated in Figure \ref{fig:bayesn}, these distances are consistent across the redshift range of our sample, with the average difference between BayeSN and SNooPy distances changing by just 2~mmag from the low-$z$ to high-$z$ samples.  As SNooPy has rarely been used at high redshift, this consistency between the two models gives us confidence that the raw distances that we measure with SNooPy are robust.

\subsubsection{Optical Distances from SALT3 and SNooPy}
\label{sec:optical}

In Table \ref{table:optnir}, we explore the differences between several methods of measuring NIR, optical, and optical$+$NIR distances.  We compare the baseline NIR-only distances to optical$+$NIR distances with both the baseline $R_V = 1.5$ and a Milky Way-like $R_V = 3.1$.  We also compare to optical distances from SALT3 and finally, we use an approach where optical$+$NIR data are first used to determine $s_{BV}$ and then $s_{BV}$ is fixed to that value in an NIR-only fit that measures the distance.  This essentially allows the NIR distances to include the anticipated $s_{BV}$-luminosity relation.

We find Hubble residual differences of $\sim$1-2\% between our baseline NIR method and methods using optical data and the default SNooPy value of $R_V = 1.5$ \citep{Folatelli10}.  With $R_V = 3.1$, the difference in distances is much larger, at 0.065 mag between low- and high-redshift, though we note that using this value of $R_V$ increases the Hubble residual scatter by 28\%.  In a NIR-only analysis, however, using $R_V = 3.1$ changes the distances by just $\sim$0.02-0.04 mag due to the lower sensitivity of the NIR to the value of $R_V$. Because we allow negative extinction in this analysis and because SNooPy does not model the intrinsic correlation between color and luminosity, the $R_V$ parameter is more analogous to the nuisance parameter $\beta$ in the Tripp relation than to a physical dust law.

The difference between our baseline distances and SALT3 distances is just $0.005\pm0.038~{\rm mag}$.
Although this difference in distance is small,
the bias corrections shown in Figure \ref{fig:biascor} are $\sim$0.05~mag smaller in the NIR-only distances, increasing the potential discrepancy between SALT3 and the baseline NIR measurement.
However, given the large uncertainties on the distance in measurement, a SALT3-based (or SALT2-based) analysis would still yield a consistent measurement of $w$.

Finally, the procedure of adjusting the NIR distances by a value of $s_{BV}$ measured in the optical results in a statistically insignificant $-0.016 \pm 0.027$~mag change.  The lack of a small $s_{BV}$ correction in our baseline analysis therefore does not appear to have a significant impact on the resulting cosmological parameter estimation.  We also explore altering the semi-arbitrary $s_{BV}$ cuts in our baseline distance measurements, particularly given that the $s_{BV}$ values tend to cluster near $\sim$1.18, perhaps due to a limited high-$s_{BV}$ training sample in SNooPy.  Fortunately, we find that our analysis is relatively insensitive to the minimum and maximum $s_{BV}$; both a ``loose" cut of $0.6 < s_{BV} < 1.3$ and a ``tight" cut of $0.8 < s_{BV} < 1.15$ change the average Hubble residual by just 7~mmag when comparing SNe at $z < 0.1$ to those at $z > 0.1$.  The mass step measurement is changed by less than 0.01~mag.

\begin{figure*}
    \centering
    \includegraphics[width=7in]{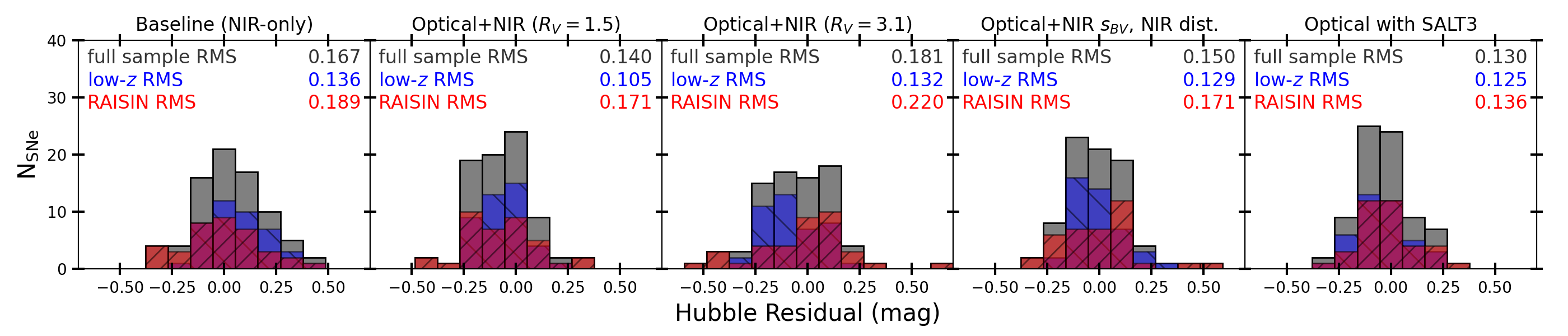}
    \caption{Hubble residual scatter for different analysis variants.   From left to right, we show our baseline NIR-only Hubble residuals, optical$+$NIR SNooPy distances with $R_V = 1.52$ \citep{Folatelli10}, optical$+$NIR SNooPy Hubble residuals with $R_V = 3.1$, NIR SNooPy Hubble residuals after applying the $s_{BV}$ parameters measured from optical$+$NIR data, and optical SALT3 Hubble residuals.  The SALT3 panel removes two CSP SNe with colors that are too red ($c > 0.3$) to pass standard SALT cosmology cuts; with these SNe included, scatter increases negligibly to 0.132~mag in the CSP sample.}
    \label{fig:optical}
\end{figure*}

In Figure \ref{fig:optical}, we examine the dispersion of Hubble residuals for different methods that use optical data.  We find that the Hubble residual dispersion is lowest when using the optical$+$NIR with $R_V = 1.5$ ($0.140$~mag) or the SALT3 model ($0.130$~mag).  SALT3 appears to standardize SNe\,Ia more precisely than SNooPy at high redshift, while SNooPy distances are more precise than SALT3 distances for the CSP data on which SNooPy was trained. The SALT3 results are nearly the same as those from SALT2, which was used for most previous optical-only analyses.

If an optical versus NIR difference becomes statistically significant in future work, it could point to unrecognized systematic uncertainties in analyses with optical data, NIR data, or both.  In optical analyses, the G10 scatter model is a source of significant uncertainty in correcting for distance-dependent biases \citep{Scolnic18,Brout19} and proposed alternative scatter models could shift $w$ by 4\% \citep{Brout21}.  In the NIR, differences in stretch-luminosity correlations between \citet{Burns14} and \citet{Burns18}, for example, may point to systematic uncertainties in SNooPy NIR standardizations.  These results show that NIR observations constitute an extremely useful consistency check on optical analyses.

\begin{table}[]
    \centering
    \caption{Optical$+$NIR Distance Comparisons}
    \begin{tabular}{lr}
    \hline \hline
&$\Delta \mu$ (mag)\\\cline{2-2}
Optical+NIR ($R_V = 1.5$)&$0.011\pm0.030$\\
Optical+NIR ($R_V = 3.1$)&$-0.065\pm0.032$\\
Optical with SALT3&$0.005\pm0.038$\\
Optical+NIR $s_{BV}$, NIR dist.&$-0.016\pm0.027$\\
\hline\\
    \multicolumn{2}{l}{
  \begin{minipage}{3in}
    {\bf Note.} For optical$+$NIR analysis methods without bias corrections, the difference between low- and high-$z$ distances relative to the baseline NIR-only method: 
    \begin{align}
    \Delta\mu = [{\bar{\mu}(z > 0.1)} - {\bar{\mu}(z < 0.1)}]_{opt+NIR} \nonumber\\
    - [{\bar{\mu}(z > 0.1)} - {\bar{\mu}(z < 0.1)}]_{NIR-only}.\nonumber
    \end{align}
\end{minipage}}
    \end{tabular}
    \label{table:optnir}
\end{table}

\subsection{The Late-Time NIR Model}
\label{sec:latetimenir}

\begin{figure}
    \centering
    \includegraphics[width=3.5in]{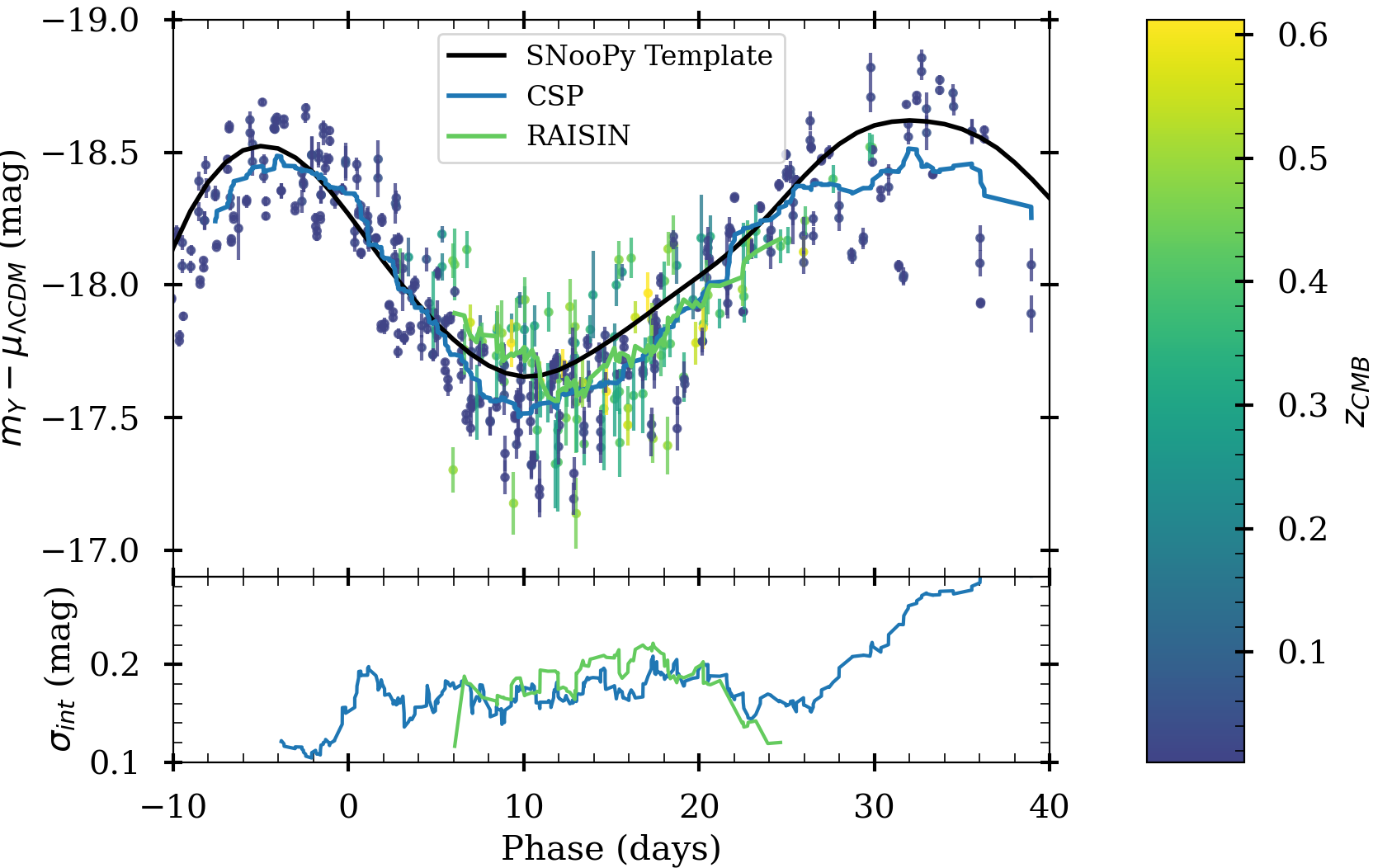}
    \caption{
    The SNooPy model (black) compared to  CSP  (blue) and RAISIN (red) measurements after using the SNooPy $K$-corrections and the cosmological distance to convert the observations to rest-frame absolute magnitude.  Solid lines in the top panel show the rolling means of the CSP and RAISIN absolute magnitudes and the rolling standard deviation, with the average peculiar velocity error and photometric noise subtracted in quadrature, is shown in the bottom panel.  NIR SN\,Ia observations appear to have the lowest RMS near maximum light and slightly higher RMS from $\sim$10-25 days.}
    \label{fig:nirphase}
\end{figure}

Our Hubble diagram compares SNe observed near maximum light in the low-$z$ CSP sample to SNe observed only at later phases in the higher-$z$ RAISIN sample; the initial NIR epoch in RAISIN occurred at $+9$ rest-frame days on average.  
This could introduce systematic uncertainty if the SN standardization model is not as accurate at late times as it is near maximum light where most previous NIR data exist.  Although we account for this systematic uncertainty by using two SN\,Ia standardization models, both use much of the same training data.

To investigate potential biases due to the phase at which each SN was observed in the NIR, we investigated whether there is a trend in Hubble residuals as a function of the median phase at which each SN was observed.  By comparing the Hubble residuals of SNe observed at a median phase of $<+15$~days to those observed at a median phase of $>+15$~days ($+15$~days is approximately the sample median), we detected a ``step" in the NIR Hubble residuals at 3-$\sigma$ significance.  This step appears to show that RAISIN SNe\,Ia observed at a phase of $>+15$~days are fainter by $0.164\pm0.051$~mag than RAISIN SNe\,Ia observed at a median phase of $<+15$~days.  Though this trend is potentially concerning, we note that the difference in stretch between the early versus late-observed samples is 0.12, which could explain $\sim$0.05~mag of the $\sim$0.16-mag difference (Figure \ref{fig:sbv}).  
This selection bias could be caused by the fact that low-stretch SNe are intrinsically fainter, so it may have been harder to classify them and subsequently trigger {\it HST} observations prior to maximum light.

To inspect potential biases in the SNooPy model as a function of phase, in Figure \ref{fig:nirphase} we show the SNooPy $Y$-band template alongside the CSP and RAISIN observations after the observed data have been converted to the rest frame.  Both the high- and low-$z$ data agree relatively well with the SNooPy model, albeit with slight offsets at $+10<{\rm phase}<+20$~days of $\sim 0.05$~mag.  These offsets suggest that the late-phase NIR SNooPy model should be adjusted as more data become available, but would only shift the mean high-$z$ distance measured here by approximately 0.02~mag.  The variance of the NIR data also appears to increase slightly near the phase at which most of the later RAISIN observations occur, while observations near maximum light have the lowest scatter.  The slightly higher scatter of the RAISIN magnitudes compared to CSP might also be an indication of $K$-correction uncertainties in the high-$z$ distances.  We observe similar trends when comparing the $J$-band SNooPy model to our data.

The apparent dependence of Hubble residual on the phase at which a SN was observed could also be due to subtle errors in the relationship between SN luminosity and $s_{BV}$ or that the NIR SN color-luminosity dependence differs significantly from the relation that would be expected from dust extinction alone.  Although retraining the SNooPy model is beyond the scope of this work, and will likely require significant additional rest-frame NIR data to be robustly re-determined, additional low-$z$ data and model retraining are important ways in which future NIR analyses could be improved.

\section{Discussion}
\label{sec:discussion}

Although this study reaffirms the existence of cosmic acceleration and establishes the value of NIR SN\,Ia data for cosmological measurements, much of the value of this dataset will be in preparing for NIR SN\,Ia samples from the {\it Roman Space Telescope}.  Although the SN survey strategy is not yet finalized, {\it Roman} will observe in the rest-frame NIR to $z \approx 0.7$, the regime in which cosmological parameter constraints from SNe\,Ia will likely be more precise than volume-limited Baryon Acoustic Oscillations \citep{Weinberg13}.  Measuring the most
precise cosmological constraints from these data will be the path to obtaining the best possible constraints on cosmological parameters.

With {\it Roman} on the horizon, the systematic error budget presented here suggests a path forward for improved NIR cosmological constraints.  Here's what we need:

\begin{enumerate}
\item Larger photometric NIR samples with coverage near maximum light across multiple photometric systems to improve calibration systematics.
\item Larger photometric and spectroscopic NIR samples to improve training of models,
  and optical-to-NIR measurements of the SN\,Ia dispersion to constrain any redshift-dependence of the value of $R_V$.
\item Optical-to-NIR SN Ia SED models and standardization methods that separate SN\,Ia intrinsic dispersion from the effects of host galaxy dust as a function of wavelength, and improve the modeling of high-stretch (slowly declining) SNe.
\item Improved constraints on the host galaxy mass step and other host galaxy dependences as a function of wavelength to break degeneracies with dust properties.
\item An untargeted NIR low-$z$ sample with well-understood selection effects.

\end{enumerate}

{\it These improvements would improve the accuracy of the cosmological measurement presented here even without obtaining additional high-$z$ data}.  A public release of CSP-II NIR spectra and photometry will address several of these key points \citep{Hsiao19,Phillips19}.  As we enter the {\it Roman} era, new data will also reduce bias correction systematics by constraining the distribution of SN shape and color as a function of redshift.  We elaborate on a number of these potential improvements below.

\subsection{NIR Dispersion Model}

The NIR dispersion model measured both in this work and in the BayeSN SED model show that NIR distances offer significant independent leverage to constrain distance measurements compared to those obtained only with optical data.  
Further, NIR data, in combination with optical, provide additional leverage to improve constraints on both $R_V$ and $A_V$, as demonstrated in recent analyses that use NIR data to break the degeneracy between host galaxy dust and intrinsic SN physics \citep{Mandel22,Thorp21}.

In this study, we see that even without shape and color corrections, just three epochs of NIR data at $\sim$10-20 days after maximum light are sufficient to constrain dark energy properties.  It is possible that even fewer epochs or filters would still yield competitive cosmological constraints, though additional epochs and filters give valuable ability to avoid data outliers and to constrain dust properties.  Much of the systematic uncertainty in this analysis stems from the fact that these SNe were selected for spectroscopic followup in the optical.  In large, untargeted NIR samples like those from {\it Roman}, this source of systematic uncertainty will vanish.

\subsection{Building a New, Well-Calibrated NIR SED Model}

SN\,Ia analyses and standardization models in the NIR are currently limited by the sample sizes of well-calibrated datasets, both photometric and spectroscopic, which introduce substantial calibration systematics and uncertainties in the precision of the standardization model training itself. New, well-calibrated NIR samples from CSP-II, SIRAH, VISTA, UKIRT \citep{Konchady21}, and eventually {\it Roman} will be needed for improved model training.  Simultaneous NIR spectroscopy will also be important for a better understanding of $K$-corrections.  The SIRAH sample in particular, with extremely well-calibrated photometry and grism spectroscopy tied to ground-based NIR observations, will be a powerful link between precise NIR cosmological analyses from space and the ground.  

This analysis shows that the NIR$+$optical measurement of $w$ has similar precision to the optical measurement alone.  Although we find that systematic uncertainties on $w$ are lowest when we use NIR data in combination with optical, there is no improvement in the precision of cosmological parameter measurements when adding NIR data in this analysis. This is likely because our light curve fitting approaches are not yet optimally weighting the SN data at different wavelengths, a problem that previous studies have found can be addressed with improved uncertainty treatments in SN standardization models \citep{Mandel11,Mandel22}.

\subsection{The Mass Step}

The underlying cause of the host galaxy mass step is a subject of debate, with numerous explorations of alternative global host galaxy dependencies \citep{Hayden13,Pan14} or relationships between Hubble residuals and local host galaxy properties near the SN location \citep{Rigault13,Rigault15,Jones15b,Jones18,Kim18,Rigault18,Roman18,Kelsey21}.  The degeneracy between physical effects, potential data reduction artifacts \citep{Solak21}, and dust has made measuring the wavelength dependence of the host galaxy mass step, and other host galaxy dependencies, extremely important for next-generation cosmological measurements.  Measurement of whether a step-like dependence of NIR distance measurement on global or local galaxy color, e.g., \citet{Roman18,Jones18,Kelsey21}, would be particularly interesting to understand the effects of dust or progenitor ages on SN distance measurements.  Previous measurements of the mass step in the NIR have been limited by sparse data \citep{Ponder20} and the need to include optical data with the NIR observations to constrain the best-fit parameters such as $s_{BV}$ and $A_V$ \citep{Uddin20}.  Flexible simulation-based approaches, e.g. \citet{Pierel21}, will be necessary to fully understand the impact of different SN-host relationships on cosmological parameter measurements.

Although the galaxy targeted nature of the CSP sample has made measuring the mass step in this work difficult (nearly all CSP galaxies are in the ``high-mass'' regime), we do find evidence suggesting that the mass step in the NIR is about the size of the mass step in the optical.  We note that our results are not in tension with other recent NIR mass step results, including those of \citet{Johansson21}, who saw that the mass step was insignificant in the $JH$ bands and less significant ($\sim$2$\sigma$) in the $Y$ band.  Our distances primarily probe the rest-frame $Y$ band, albeit with some $i$ and $J$-band overlap, where \citet{Johansson21} find a mass step of approximately $0.07 \pm 0.03$~mag.  However, our results do not favor the conclusion of \citet{Brout21} that the mass step is caused by variations in dust properties, as the effect of variation in $R_V$ should decrease by a factor of $\sim$3 between e.g., rest-frame $V$ and rest-frame $Y$.  Further study of the origin of the mass step is clearly needed and larger infrared samples may prove useful.

Future measurements of this and other host galaxy dependencies as a function of wavelength may be necessary to separate intrinsic color, dust, other SN\,Ia physics, or even multiple SN\,Ia progenitor models.  For example, when measuring the mass step, we did not fit for individual $R_V$ values as was done in  \citet{Johansson21}, or simultaneously constrain the population distribution of $R_V$ values as was done in \citet{Thorp21}; these methods can help to disentangle the role of dust in the mass step's size.  Although SNe\,Ia are used as an empirical tool for measuring distance, understanding the objects and their connection to stellar populations is also valuable in itself. The evolution of stellar populations and of SN\,Ia progenitors across the long span of cosmic time that {\it Roman} will provide gives an opportunity to understand the nature of SN progenitors and their environments and to use them for more precise cosmological measurements. 

\section{Conclusions}
\label{sec:conclusions}

We present photometric measurements, a Hubble diagram, and NIR-only cosmological constraints from the RAISIN survey.  SN candidates were observed in cycles 20 and 23 by {\it HST} and combined with low-$z$, NIR-observed SNe from CSP to yield a NIR Hubble diagram and a better understanding of cosmological parameter measurements in the NIR.  From CSP data we build a new, optical-to-NIR dispersion model for SNe\,Ia to predict distance-dependent biases and measure the independence of SN\,Ia distances in the optical and NIR.  We find that NIR-derived distances are well correlated with distances derived from optical data, but offer some independent information.
The dispersion of our NIR-only distances is 0.18~mag, $\sim$25\% larger than the Pantheon sample at 0.141~mag but not standardized using shape or color information.  When measuring the host galaxy mass step after applying shape corrections to our distances, in order to match the procedures used in optical analyses, we measure a NIR mass step of \massstepstretch~mag at a step location of 10~dex and $0.123 \pm 0.034$~mag at a step location of 10.44~dex.

In combination with CMB constraints from \citet{Planck18} and assuming a flat $w$CDM model, we measure $1+w =$~\onepw\ (stat$+$sys), consistent with the $\Lambda$CDM expectation of $w = -1$.
This NIR measurement of $w$ may embed unknown systematic uncertainties that will become apparent when future, larger NIR samples are analyzed.  The dominant systematic uncertainties in this analysis stem from bias corrections, which will be improved by future SN standardization models that extend to the NIR; building these models will be facilitated by larger NIR sample sizes on the way from CSP-II, SIRAH, VISTA, and UKIRT. 
We find that combined cosmological parameter measurements from optical$+$NIR data appear to have lower systematic uncertainties than when using optical data alone.  We also see shifts in $w$ of up to 0.1 between optical, optical$+$NIR and NIR data alone, pointing to the possibility of inconsistency in the optical versus NIR SN standardization models.

We also measure ${\rm H_0} = 75.9 \pm 2.2~{\rm km~s^{-1}~Mpc^{-1}}$ from a local distance ladder approach tied to 8 NIR-observed SNe\,Ia with Cepheid or TRGB distances versus ${\rm H_0} = 71.2 \pm 3.8~{\rm km~s^{-1}~Mpc^{-1}}$ using a so-called ``inverse distance ladder" approach anchored to CMB measurements of the angular size of the sound horizon.  
Future NIR measurements can help to constrain whether the present dark energy density or systematic uncertainties due to SN\,Ia dust could play any role in the H$_0$ tension.

Finally, these data will help to better understand the behavior and standardization of SNe\,Ia in the NIR and assist the community in preparing for data from the {\it Roman Space Telescope}.  Improved dispersion models will help improve distances by using an optical$+$NIR combination, and wavelength-dependent measurements of the dependence of SN\,Ia distance measurements on host galaxy properties will break the degeneracies with dust properties.  When combined with future datasets from SIRAH and improved SN\,Ia standardization models that can be trained with this dataset, we hope the data published here will have legacy value for the SN\,Ia community in the {\it Roman Space Telescope} era\footnote{The data used in this analysis are published at \url{https://doi.org/10.5281/zenodo.6349657}.}.

\acknowledgements
We would like to acknowledge that much of the data and analysis in this manuscript were made possible through the work of our friend and colleague Dr.~Andrew S.~Friedman, who passed away during the final stages of this project.  We would also like to thank G.~Narayan and N.~Drakos for useful discussions and their assistance with $HST$ photometry and $K$-corrections.  We would like to thank R.~Kessler for implementing the latest SNooPy model in SNANA and C.~Rojas-Bravo for assistance with the spectra.  We would also like to thank the anonymous referee for many helpful suggestions.

Support for this work was provided by a Gordon and Betty Moore Foundation postdoctoral fellowship to D.O.J. at the University of
California, Santa Cruz and by NASA through the NASA Hubble Fellowship grant HF2-51462.001 awarded by the Space Telescope Science Institute, which is operated by the Association of Universities for Research in Astronomy, Inc., for NASA, under contract NAS5-26555. The Cambridge University team acknowledges support from  the European Research Council under the European Union’s Horizon 2020 research and innovation programme (ERC Grant Agreement No. 101002652), the ASTROSTAT-II collaboration, enabled by the Horizon 2020, EU Grant Agreement No. 873089, and the Cambridge Centre for Doctoral Training in Data-Intensive Science funded by the UK Science and Technology Facilities Council (STFC).
L.G.~acknowledges financial support from the Spanish Ministerio de Ciencia e Innovaci\'on (MCIN), the Agencia Estatal de Investigaci\'on (AEI) 10.13039/501100011033, and the European Social Fund (ESF) ``Investing in your future" under the 2019 Ram\'on y Cajal program RYC2019-027683-I and the PID2020-115253GA-I00 HOSTFLOWS project, from Centro Superior de Investigaciones Cient\'ificas (CSIC) under the PIE project 20215AT016, and the program Unidad de Excelencia María de Maeztu CEX2020-001058-M.
M.R.S. is supported by the NSF Graduate Research Fellowship Program Under grant 1842400.
The UCSC team is supported in part by NASA grants 14-WPS14-0048, NNG16PJ34C, NNG17PX03C,
NSF grants AST-1518052 and AST-1815935, NASA through grant number AR-14296 from the Space
Telescope Science Institute, which is operated by AURA, Inc., under NASA contract NAS 5-26555,
the Gordon and Betty Moore Foundation, the Heising-Simons Foundation, and by fellowships from
the Alfred P.\ Sloan Foundation and the David and Lucile Packard Foundation to R.J.F.
Based on observations with the NASA/ESA Hubble Space Telescope obtained from the Data Archive at the Space Telescope Science Institute, which is operated by the Association of Universities for Research in Astronomy, Incorporated, under NASA contract NAS5-26555. Support for Program numbers 13046 and 14216 was provided through grants from the STScI under NASA contract NAS5-26555.  The Carnegie Supernova Project has been supported by the National Science Foundation under grants AST0306969, AST0607438, AST1008343, AST1613426, AST1613455, and AST1613472.  L.K. thanks the UKRI Future Leaders Fellowship for support through the grant MR/T01881X/1.  This research has made use of the NASA/IPAC Extragalactic Database (NED), which is funded by the National Aeronautics and Space Administration and operated by the California Institute of Technology.

This paper includes data gathered with the 6.5 meter Magellan Telescopes located at Las Campanas Observatory, Chile.  Some of the observations reported here were also obtained at the MMT Observatory, a joint facility of the University of Arizona and the Smithsonian Institution, and the W.M. Keck Observatory, which is operated as a scientific partnership among the California Institute of Technology, the University of California and the National Aeronautics and Space Administration. The Observatory was made possible by the generous financial support of the W.M. Keck Foundation.  The authors also wish to recognize and acknowledge the very significant cultural role and reverence that the summit of Maunakea has always had within the indigenous Hawaiian community.  We are most fortunate to have the opportunity to conduct observations from this mountain.  This paper also uses observations obtained at the international Gemini Observatory, a program of NSF’s NOIRLab, which is managed by the Association of Universities for Research in Astronomy (AURA) under a cooperative agreement with the National Science Foundation on behalf of the Gemini Observatory partnership: the National Science Foundation (United States), National Research Council (Canada), Agencia Nacional de Investigación y Desarrollo (Chile), Ministerio de Ciencia, Tecnología e Innovación (Argentina), Ministério da Ciência, Tecnologia, Inovações e Comunicações (Brazil), and Korea Astronomy and Space Science Institute (Republic of Korea).  Finally, we include data acquired at the Anglo-Australian Telescope. We acknowledge the traditional owners of the land on which the AAT stands, the Gamilaroi people, and pay our respects to elders past and present.

\software{PythonPhot \citep{Jones15}, hstphot (\url{https://github.com/srodney/hstphot}), TinyTim \citep{Krist11}, HOTPANTS \citep{Becker15}, ZOGY \citep{Zackay16}, corner \citep{Foreman-Mackey16}, SNANA \citep{Kessler10}, LePHARE \citep{Arnouts11}, SNooPy \citep{Burns11,Burns14}, RVSAO \citep{Kurtz98}, Marz \citep{Hinton16}}

\appendix

\section{Photometric Measurements from {\it HST}}
\label{app:hstphot}

In this appendix, we describe the photometric measurement of RAISIN SNe from {\it HST}.  The photometric measurements themselves are given in Appendix \ref{sec:app_phot}.

 First, we used AstroDrizzle to combine RAISIN FLT images downloaded from the Mikulski Archive for Space Telescopes (MAST) and create geometric distortion-corrected images for each epoch of each SN.  For a given SN, each epoch of observations is drizzled, then the output, drizzled images are aligned to a reference image via {\tt tweakreg} so that SN coordinates are consistent in all epochs, and then the data from each epoch are re-drizzled. Finally, we subtract the drizzled template image from each image containing the SN.  We make use of {\tt sndrizpipe}\footnote{\url{https://github.com/srodney/sndrizpipe}}, a pipeline for {\it HST} drizzling and image subtraction originally written for the CANDELS and CLASH {\it HST} Multi-Cycle Treasury programs \citep{Grogin11,Postman12} that has continued being actively developed in the last few years.  Images were drizzled to 0.11\arcsec~pix$^{-1}$, slightly smaller than the native pixel scale of {\it WFC3}, to improve the resolution of the {\it HST} PSF (we found minimal differences between 0.09-0.13\arcsec~pix$^{-1}$ choices).  The photometric zeropoints that we use are derived from the latest {\it HST} calibrations of \citet{Bohlin20}.

SN positions from discovery data or optical imaging are known to approximately 1\arcsec\ precision or better and are given in Table \ref{table:coords}.  We use centroiding algorithms from the {\tt hstphot}\footnote{\url{https://github.com/srodney/hstphot}} and {\tt PythonPhot} \citep{Jones15} packages to refine those positions.  The best centroid was given by the weighted average of coordinates measured from {\it F125W} images for RAISIN1 and {\it F160W} images for RAISIN2.  Because SN\,Ia are fainter in redder bands and host galaxies are often brighter, coordinates from {\it F125W} images tend to be more reliable when available.  

Using the measured SN coordinates, we performed aperture photometry with a fixed 0.4\arcsec\ radius using the zeropoints and aperture corrections determined from standard star observations\footnote{\url{https://www.stsci.edu/hst/instrumentation/wfc3/data-analysis/photometric-calibration/ir-photometric-calibration}}.  We found that due to the undersampled and somewhat variable {\it HST} PSF, aperture photometry was more reliable than PSF photometry for these images.  There are few stars in a typical RAISIN frame from which to determine the PSF on an epoch-by-epoch basis and, likely due to the breathing of {\it HST} and the undersampled {\it WFC3} PSF, we found the PSF was too variable for a PSF model based on archival data to be sufficient.  By performing artificial star tests we found we were unable to perform PSF photometry without introducing mmag-level biases.

As the drizzling process produces pixels with correlated noise, uncertainties were estimated by planting 1000 fake stars per epoch at random locations.  Fake stars were generated using a P330E PSF model (again created using {\tt hstphot}) and each fake star had a magnitude equal to the measured SN mag in a given epoch.  The dispersion in these fake star magnitudes was added in quadrature to the Poisson noise from the SN to give the total uncertainty due to sky background variation, correlated pixels, and Poisson noise.  Given that our final pixel scale is near the native pixel size of {\it WFC3-IR}, the noise in the recovered stars is likely dominated by Poisson noise of the sky background.

The expected bias in the photometry given the centroid uncertainties derived above is derived by \citet{Rest14}, their Appendix D.  This requires an estimate of how much the magnitude changes when the coordinate is incorrect by 1 pixel.  P330E images show that this is approximately 0.022~mag for a 0.11~\arcsec\ pixel scale, but typical centroid uncertainties are closer to $\sim$0.02 pix.  We corrected for these biases, but in practice find that they are typically $<$1~mmag.

\subsection{Host Galaxy Noise}
\label{sec:hostnoise}

SNe\,Ia on top of bright host galaxies can have biased measurements or underestimated errors.  The additional noise introduced by a bright host galaxy tends to create significant subtraction residuals by increasing the effect of small PSF changes between the template and SN images.  We occasionally see significant dipole-like subtraction artifacts.

To model this effect we used {\tt TinyTim} to create up to 25 fake stars near the {\it same host galaxy} as each SN.  We require that the fake stars are at least 0.7 arcsec from the SN position and from each other, which is $>$4 times the full width at half maximum of the {\it F160W} PSF.  We cannot measure the noise at the exact location of the SN, but we can gain a statistical sense of the noise as a function of host galaxy surface brightness from these fake stars.

\begin{figure}
  \includegraphics[width=7in]{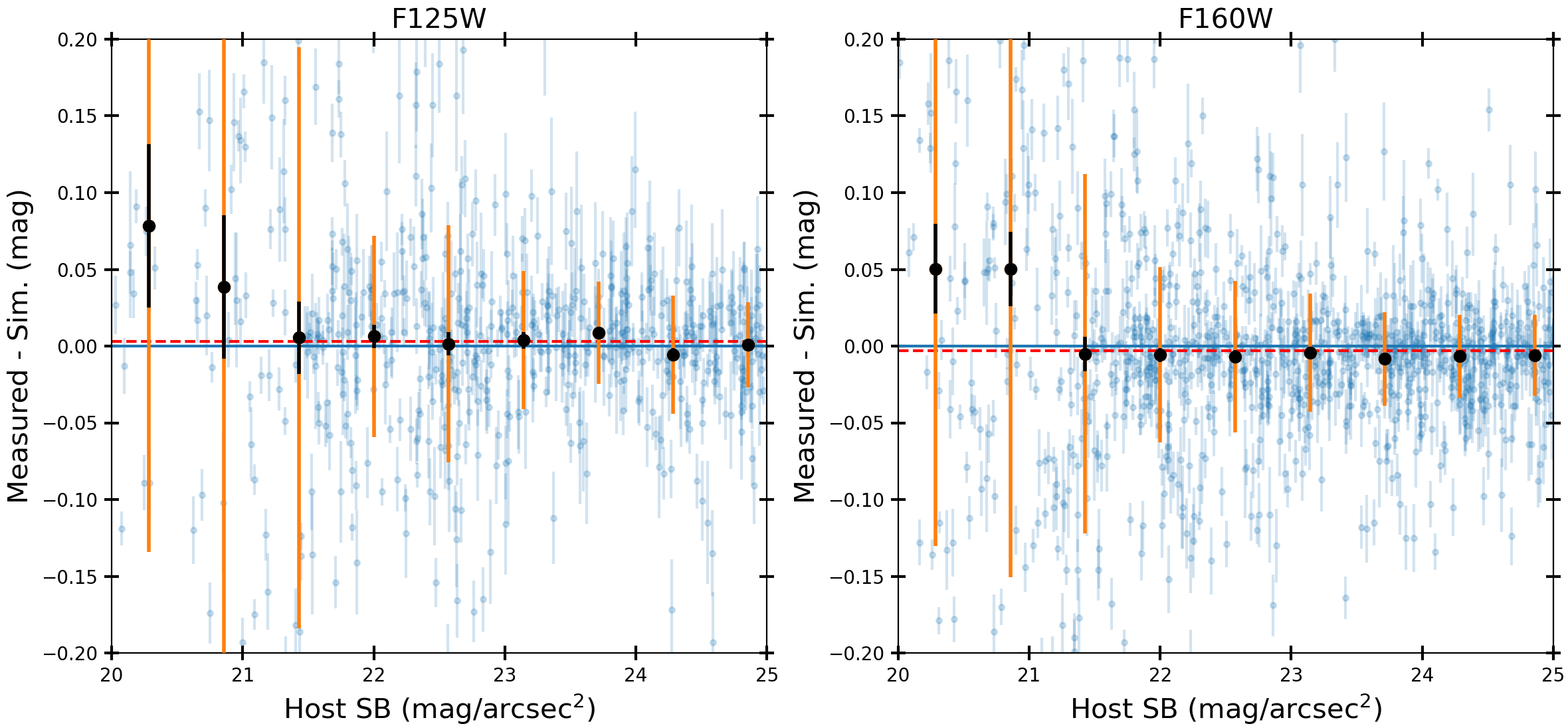}
  \caption{Dispersion of fake star magnitudes as a function of host galaxy surface brightness at the SN location for $F125W$ (left) and $F160W$ (right).  Large error errors (orange) show the dispersion while small errors (green) show the error on the mean.  We restrict to the sample of SNe with predicted host galaxy error less than 0.1~mag and verify with visual inspections that the subtracted images for these SNe appear reliable.}
  \label{fig:sbbias}
  \end{figure}

The results of this test are shown in Figure \ref{fig:sbbias}.  Small error bars give the uncertainty on the average magnitude in each surface brightness bin, and large error bars give the dispersion of the fake magnitudes in each bin.  To compute more realistic uncertainties, we then subtract the measured magnitude uncertainties in quadrature from the total magnitude dispersion of the fake stars.  These quantities give the extra uncertainty that should be added in quadrature to each SN magnitude as a function of the brightness of the host galaxy underneath the SN.  SNe with estimated host galaxy uncertainties of greater than 0.1~mag are removed from the sample; we chose this cut based on visual inspection, which shows that such SNe typically have subtraction artifacts likely due to misalignments or ``breathing'' of {\it HST} and are unlikely to provide reliable, unbiased photometry.  Our team attempted applying convolutional subtraction algorithms to the data including HOTPANTS \citep{Becker15} and ZOGY \citep{Zackay16} but did not achieve improved results for these bright hosts.  Five SNe were removed from the final analysis based on excessive host galaxy noise.

\subsection{Residual SN flux in Template Imaging}
\label{sec:templates}

The RAISIN programs took {\it HST} template imaging between 129 to 285~days after the estimated time of maximum light for subtracting from the {\it HST} images with SN light.  To test that our time interval between SN images and template images was sufficient, we used early- and late-time observations of SN~2012fr, which has $YJ$ imaging at $\sim$150~days after maximum light from \citet{Contreras18}.  With these data, we can extrapolate to later times to estimate the residual flux of RAISIN SNe at the time the template was taken.

For all RAISIN SNe, we use this estimated template flux to correct the magnitudes of RAISIN SN observations for which the closest rest-frame band is $Y$.  Due to the faster decline of SNe in the $J$ band, the $J$-band correction is expected to be negligible.  To account for variability in the typical decline rate of SNe at late times in the NIR, we created a version of our analysis in which we conservatively increase the predicted late-time template flux by 0.5~mag for each RAISIN SN and correct the photometry for this revised template flux prediction.  We include this analysis variant in our systematic error budget (\S\ref{sec:analysis_systematics}).  Unfortunately, SN~2012fr is the only SN for which high-quality NIR data exist at $\sim$six months after maximum light.  We predict that 36 of 47 RAISIN SNe would have declined by $<5$~mag, giving expected magnitude corrections that are more than 1\% for these SNe with a maximum predicted correction of $\sim$3.7\%.  We note that this source of uncertainty could be eliminated by re-visiting the sites of RAISIN SNe to obtain templates with zero flux from the SNe.

\subsection{End-to-end Fake Star Tests}
\label{sec:fakestar}

Finally, to verify both our photometric measurement
technique and our drizzling procedure, we used
{\tt TinyTim} to simulate fake, geometrically-distorted
stars in the FLT images.  Then, we drizzled
those images, subtracted them by a template, and
measured output photometry on the sources detected
in those images.

\begin{figure}
  \includegraphics[width=6.5in]{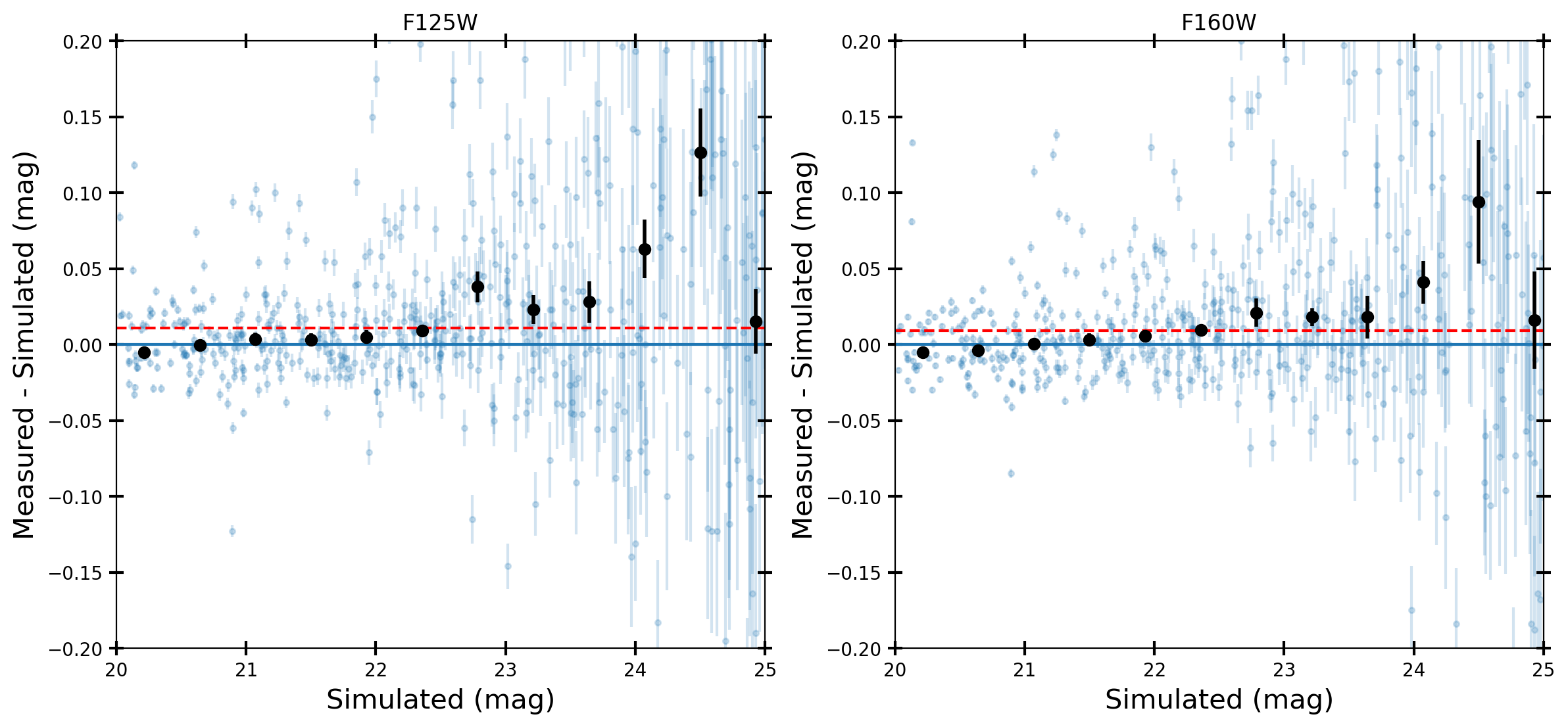}
  \caption{Measured fake star magnitudes minus simulated
    magnitudes as a function of mag for $F125W$ (left) and $F160W$ (right) with the median bias shown in red.  We find just 1~mmag biases for sources brighter than 22~mag and $\sim$1\% median biases overall likely due to centroiding errors that are corrected in the real data thanks to multiple observational epochs.}
  \label{fig:fakecomp}
\end{figure}

The results of these tests are shown in Figure \ref{fig:fakecomp}.
The median {\it F125W} and {\it F160W} biases are $\le$1~mmag for bright sources, and $\sim$1\% for all sources.  However, because these tests use only a single epoch of data, this 1\% bias may be due to inaccurate fake star centroids.  In the real RAISIN data this bias is likely not present because we use multiple epochs to determine the precise PSF centroid.

\section{Coordinates and Photometric Measurements of RAISIN SNe}
\label{sec:app_phot}

This appendix contains coordinates and {\it HST} photometry for the RAISIN sample.  Optical light curves and spectra are provided in the online data release accompanying this paper at \url{https://doi.org/10.5281/zenodo.6349657}.

\begin{table}[H]
    \centering
    \caption{RAISIN Coordinates, Redshifts, and Discovery Dates}
    \begin{tabular}{lrrrrrr}
    \hline \hline
         ID & $\alpha$ & $\delta$ & $z_{\rm Helio}$&$z_{\rm CMB}$ & $z_{\rm source}^{\rm a}$&MJD$_{\rm disc}$ \\
         \hline
PS1-480464&09:59:12.147& 03:13:17.555&0.220&0.221&host&55243\\
PS1-450082&23:33:25.446& 00:18:37.609&0.250&0.249&SN&56233\\
PS1-540087&22:21:30.369& 00:07:46.110&0.275&0.275&host&56505\\
PS1-520188&14:22:05.831& 52:32:40.679&0.280&0.280&host&56469\\
DES16E2cxw&00:40:42.701&-43:55:18.948&0.293&0.293&host&57686\\
PS1-440005&03:31:57.424&-28:53:00.481&0.306&0.306&host&56206\\
PS1-520062&16:17:48.345& 54:37:59.502&0.308&0.308&SN&56442\\
PS1-500100&08:46:33.237& 45:21:21.978&0.310&0.311&host&56383\\
PS1-500301&14:22:11.455& 52:37:54.908&0.325&0.325&host&56394\\
PS1-470041&08:45:21.034& 44:45:17.194&0.331&0.332&host&56290\\
PS1-480794&09:58:10.055& 01:21:50.792&0.334&0.334&host&56332\\
PS1-490521&09:55:27.619& 01:34:30.925&0.340&0.341&SN&56358\\
PS1-470110&03:28:16.989&-28:37:23.988&0.346&0.346&host&56299\\
DES16E2clk&00:36:48.876&-44:08:23.460&0.367&0.367&SN&57680\\
DES16C2cva&03:34:44.035&-29:19:02.028&0.403&0.403&host&57686\\
DES15X2kvt&02:21:46.262&-05:38:23.640&0.404&0.404&SN&57298\\
PS1-450339&02:25:09.004&-03:21:12.182&0.410&0.410&host&56234\\
DES15E2nlz&00:35:51.350&-44:40:36.444&0.410&0.410&SN&57356\\
PS1-530251&22:18:32.238& 01:06:53.316&0.413&0.412&host&56477\\
DES15C1nhv&03:39:00.948&-27:50:35.520&0.421&0.421&host&57355\\
PS1-550202&23:27:11.964&-00:46:59.074&0.422&0.421&host&56558\\
PS1-490037&09:59:49.727& 00:48:59.738&0.422&0.423&host&56354\\
DES16E2cqq&00:39:50.076&-43:33:53.388&0.426&0.426&host&57682\\
PS1-440236&02:28:24.442&-04:25:28.243&0.430&0.429&host&56197\\
PS1-470240&08:42:53.508& 45:54:04.910&0.430&0.431&SN&56302\\
DES16X1cpf&02:16:49.512&-04:13:06.096&0.436&0.436&host&57681\\
DES15E2mhy&00:41:17.134&-43:53:10.104&0.439&0.439&host&57329\\
PS1-560027&02:23:24.170&-03:05:02.026&0.440&0.439&SN&56167\\
DES16E1dcx&00:35:52.440&-43:21:29.232&0.453&0.453&host&57694\\
DES15X2nkz&02:22:13.210&-05:52:18.228&0.469&0.469&host&57356\\
DES16S1bno&02:50:03.823&-00:01:17.724&0.470&0.470&SN&57653\\
PS1-540118&23:27:17.446&-00:02:53.063&0.477&0.477&host&56508\\
PS1-560054&23:33:10.992&-00:20:11.692&0.482&0.480&host&56562\\
DES16S2afz&02:46:24.329&-01:13:31.044&0.483&0.483&host&57626\\
DES16E2rd&00:32:21.334&-43:50:46.032&0.494&0.494&host&57629\\
DES16X3zd&02:26:56.465&-04:25:09.804&0.495&0.495&host&57325\\
PS1-510457&12:25:33.373& 48:08:11.868&0.502&0.503&host&56420\\
DES16S1agd&02:52:20.062&-00:54:40.284&0.504&0.504&host&57633\\
DES15C3odz&03:27:55.027&-28:33:52.092&0.508&0.508&host&57378\\
PS1-520107&12:16:38.631& 47:48:57.881&0.519&0.520&SN&56444\\
DES16C1cim&03:39:30.362&-26:38:57.048&0.531&0.531&host&57667\\
DES16C3cmy&03:27:13.440&-27:29:20.724&0.556&0.556&host&57680\\
DES15E2uc&00:37:53.419&-43:18:51.552&0.566&0.566&host&57252\\
DES15X2mey&02:20:30.785&-06:26:47.472&0.608&0.608&host&57327\\
DES16X3cry&02:24:10.548&-03:49:51.996&0.612&0.612&host&57686\\
         \hline\\*[-2ex]
    \multicolumn{7}{l}{
    \begin{minipage}{4in}
    {\bf Note:} Coordinates, redshifts, and approximate discovery MJD (the first ${\rm S/N} > 5$ detection) for RAISIN SNe.
    $^{\rm a} - $SN redshifts are accurate to $\sigma_z \simeq 0.01$ while host redshifts are accurate to better than $\sigma_z = 0.001$.
    \end{minipage}}
    \end{tabular}
    \label{table:coords}
\end{table}

\clearpage
\begin{longtable*}[H]{lrrrrrrrrr}
  \caption{RAISIN Photometry }\\
\hline \hline

SNID&MJD&filter&$m_{\rm SN}$&$\sigma_{\rm SN}$&$m_{\rm host}$&$\sigma_{\rm host}$&$\sigma_{c}$&$\Delta_m^{\mathrm{tmpl}}$&$\sigma_{\mathrm{tmpl}}$\\
&&&(mag)&(mag)&(mag)&(mag)&(\arcsec)&(mag)&(mag)\\
\hline
\endfirsthead

\multicolumn{10}{c}%
{{\tablename\ \thetable{} -- continued from previous page}} \\
\hline \hline
SNID&MJD&filter&$m_{\rm SN}$&$\sigma_{\rm SN}$&$m_{\rm host}$&$\sigma_{\rm host}$&$\sigma_{c}$&$\Delta_m^{\mathrm{tmpl}}$&$\sigma_{\mathrm{tmpl}}$\\
&&&(mag)&(mag)&(mag)&(mag)&(\arcsec)&(mag)&(mag)\\
\hline
\endhead

\hline \multicolumn{10}{r}{{Continued on next page}}
\endfoot

\endlastfoot

PS1-480464&56370&$F160W$&$24.017\pm0.080$&0.073&23.304&0.033&0.006&0.000&0.000\\
PS1-480464&56370&$F125W$&$22.437\pm0.039$&0.022&23.459&0.033&0.006&0.009&0.005\\
PS1-480464&56362&$F160W$&$24.333\pm0.114$&0.109&23.304&0.033&0.006&0.000&0.000\\
PS1-480464&56362&$F125W$&$22.726\pm0.047$&0.033&23.459&0.033&0.006&0.011&0.007\\
PS1-480464&56357&$F160W$&$24.253\pm0.122$&0.117&23.304&0.035&0.006&0.000&0.000\\
PS1-480464&56357&$F125W$&$22.921\pm0.049$&0.035&23.459&0.034&0.006&0.013&0.008\\
PS1-450082&56265&$F125W$&$22.834\pm0.238$&0.030&20.889&0.236&0.003&0.014&0.008\\
PS1-450082&56265&$F160W$&$24.573\pm0.250$&0.150&20.564&0.200&0.003&0.000&0.000\\
PS1-450082&56272&$F125W$&$22.505\pm0.237$&0.022&20.889&0.236&0.003&0.010&0.006\\
PS1-450082&56278&$F160W$&$24.469\pm0.235$&0.124&20.564&0.200&0.003&0.000&0.000\\
PS1-450082&56278&$F125W$&$22.339\pm0.237$&0.022&20.889&0.236&0.003&0.009&0.005\\
PS1-450082&56272&$F160W$&$24.606\pm0.239$&0.132&20.564&0.199&0.003&0.000&0.000\\
PS1-540087&56528&$F125W$&$22.873\pm0.078$&0.043&21.797&0.065&0.002&0.011&0.006\\
PS1-540087&56536&$F160W$&$24.962\pm0.303$&0.294&21.539&0.073&0.002&0.000&0.000\\
PS1-540087&56536&$F125W$&$23.002\pm0.078$&0.045&21.797&0.064&0.002&0.012&0.007\\
PS1-540087&56528&$F160W$&$23.748\pm0.114$&0.088&21.539&0.072&0.002&0.000&0.000\\
PS1-540087&56523&$F160W$&$23.394\pm0.103$&0.072&21.539&0.074&0.002&0.000&0.000\\
PS1-540087&56523&$F125W$&$22.657\pm0.074$&0.036&21.797&0.065&0.002&0.009&0.005\\
PS1-520188&56488&$F125W$&$22.534\pm0.035$&0.027&26.192&0.021&0.001&0.022&0.013\\
PS1-520188&56501&$F160W$&$24.140\pm0.075$&0.071&26.190&0.024&0.001&0.000&0.000\\
PS1-520188&56501&$F125W$&$22.573\pm0.034$&0.026&26.192&0.022&0.001&0.023&0.014\\
PS1-520188&56494&$F160W$&$23.815\pm0.048$&0.041&26.190&0.025&0.001&0.000&0.000\\
PS1-520188&56494&$F125W$&$22.689\pm0.034$&0.026&26.192&0.021&0.001&0.026&0.015\\
PS1-520188&56488&$F160W$&$23.133\pm0.037$&0.028&26.190&0.024&0.001&0.000&0.000\\
DES16E2cxw&57715&$F125W$&$22.771\pm0.275$&0.021&20.420&0.274&0.002&0.008&0.005\\
DES16E2cxw&57715&$F160W$&$23.874\pm0.200$&0.072&20.203&0.187&0.002&0.000&0.000\\
DES16E2cxw&57723&$F125W$&$22.691\pm0.275$&0.020&20.420&0.274&0.002&0.007&0.004\\
DES16E2cxw&57723&$F160W$&$23.902\pm0.204$&0.082&20.203&0.187&0.002&0.000&0.000\\
DES16E2cxw&57730&$F125W$&$22.455\pm0.274$&0.018&20.420&0.273&0.002&0.006&0.003\\
DES16E2cxw&57730&$F160W$&$24.314\pm0.213$&0.102&20.203&0.187&0.002&0.000&0.000\\
PS1-440005&56229&$F125W$&$22.965\pm0.071$&0.033&21.715&0.063&0.007&0.014&0.008\\
PS1-440005&56229&$F160W$&$24.079\pm0.127$&0.090&21.387&0.090&0.007&0.000&0.000\\
PS1-440005&56235&$F160W$&$24.502\pm0.159$&0.131&21.387&0.090&0.007&0.000&0.000\\
PS1-440005&56242&$F125W$&$22.734\pm0.070$&0.031&21.715&0.063&0.007&0.011&0.007\\
PS1-440005&56242&$F160W$&$24.137\pm0.119$&0.078&21.387&0.090&0.007&0.000&0.000\\
PS1-440005&56235&$F125W$&$22.998\pm0.075$&0.042&21.715&0.063&0.007&0.014&0.008\\
PS1-520062&56473&$F125W$&$23.082\pm0.077$&0.033&21.979&0.069&0.002&0.016&0.009\\
PS1-520062&56473&$F160W$&$25.973\pm0.315$&0.310&21.769&0.056&0.002&0.000&0.000\\
PS1-520062&56482&$F125W$&$22.894\pm0.074$&0.030&21.979&0.068&0.002&0.013&0.008\\
PS1-520062&56482&$F160W$&$25.101\pm0.197$&0.189&21.769&0.056&0.002&0.000&0.000\\
PS1-520062&56468&$F125W$&$23.087\pm0.080$&0.040&21.979&0.069&0.002&0.016&0.009\\
PS1-520062&56468&$F160W$&$24.669\pm0.100$&0.084&21.769&0.054&0.002&0.000&0.000\\
PS1-500100&56424&$F160W$&$24.034\pm0.076$&0.072&25.917&0.024&0.002&0.000&0.000\\
PS1-500100&56424&$F125W$&$22.806\pm0.042$&0.036&26.200&0.021&0.002&0.015&0.009\\
PS1-500100&56416&$F160W$&$24.521\pm0.116$&0.114&25.917&0.021&0.002&0.000&0.000\\
PS1-500100&56411&$F125W$&$23.141\pm0.041$&0.035&26.200&0.021&0.002&0.020&0.012\\
PS1-500100&56411&$F160W$&$24.920\pm0.134$&0.133&25.917&0.016&0.002&0.000&0.000\\
PS1-500100&56416&$F125W$&$22.959\pm0.047$&0.042&26.200&0.021&0.002&0.017&0.010\\
PS1-500301&56418&$F125W$&$23.058\pm0.166$&0.038&21.263&0.161&0.002&0.027&0.016\\
PS1-500301&56431&$F160W$&$23.823\pm0.134$&0.046&21.080&0.126&0.002&0.000&0.000\\
PS1-500301&56431&$F125W$&$22.813\pm0.164$&0.030&21.263&0.162&0.002&0.021&0.012\\
PS1-500301&56423&$F160W$&$23.995\pm0.140$&0.060&21.080&0.126&0.002&0.000&0.000\\
PS1-500301&56423&$F125W$&$22.956\pm0.166$&0.037&21.263&0.161&0.002&0.025&0.014\\
PS1-500301&56418&$F160W$&$24.027\pm0.147$&0.077&21.080&0.125&0.002&0.000&0.000\\
PS1-470041&56333&$F160W$&$23.018\pm0.193$&0.026&20.335&0.191&0.018&0.000&0.000\\
PS1-470041&56333&$F125W$&$22.488\pm0.289$&0.023&20.518&0.288&0.018&0.006&0.004\\
PS1-470041&56325&$F160W$&$23.109\pm0.193$&0.019&20.335&0.192&0.018&0.000&0.000\\
PS1-470041&56325&$F125W$&$22.632\pm0.289$&0.022&20.518&0.288&0.018&0.007&0.004\\
PS1-470041&56320&$F160W$&$23.964\pm0.205$&0.074&20.335&0.191&0.018&0.000&0.000\\
PS1-470041&56320&$F125W$&$23.121\pm0.291$&0.040&20.518&0.288&0.018&0.012&0.007\\
PS1-470041&56395&$F125W$&$24.080\pm0.302$&0.091&20.518&0.288&0.018&0.028&0.016\\
PS1-470041&56395&$F160W$&$23.642\pm0.202$&0.062&20.335&0.192&0.018&0.000&0.000\\
PS1-480794&56368&$F125W$&$23.284\pm0.053$&0.034&23.014&0.041&0.002&0.013&0.007\\
PS1-480794&56361&$F160W$&$24.189\pm0.087$&0.080&22.905&0.034&0.002&0.000&0.000\\
PS1-480794&56361&$F125W$&$23.064\pm0.058$&0.040&23.014&0.042&0.002&0.010&0.006\\
PS1-480794&56355&$F160W$&$23.786\pm0.060$&0.049&22.905&0.035&0.002&0.000&0.000\\
PS1-480794&56368&$F160W$&$24.635\pm0.165$&0.161&22.905&0.036&0.002&0.000&0.000\\
PS1-480794&56355&$F125W$&$22.887\pm0.054$&0.034&23.014&0.042&0.002&0.009&0.005\\
PS1-490521&56396&$F160W$&$23.974\pm0.084$&0.066&22.011&0.052&0.001&0.000&0.000\\
PS1-490521&56396&$F125W$&$22.979\pm0.080$&0.032&22.226&0.073&0.001&0.005&0.003\\
PS1-490521&56389&$F160W$&$24.295\pm0.097$&0.082&22.011&0.052&0.001&0.000&0.000\\
PS1-490521&56389&$F125W$&$23.089\pm0.079$&0.027&22.226&0.074&0.001&0.006&0.003\\
PS1-490521&56383&$F125W$&$23.197\pm0.077$&0.021&22.226&0.074&0.001&0.007&0.004\\
PS1-490521&56383&$F160W$&$24.300\pm0.081$&0.062&22.011&0.052&0.001&0.000&0.000\\
PS1-470110&56326&$F160W$&$24.419\pm0.079$&0.074&23.854&0.027&0.007&0.038&0.022\\
PS1-470110&56321&$F125W$&$22.968\pm0.054$&0.038&23.948&0.038&0.007&0.022&0.013\\
PS1-470110&56321&$F160W$&$23.979\pm0.072$&0.067&23.854&0.026&0.007&0.025&0.014\\
PS1-470110&56326&$F125W$&$23.133\pm0.058$&0.045&23.948&0.037&0.007&0.026&0.015\\
PS1-470110&56334&$F160W$&$24.592\pm0.108$&0.105&23.854&0.025&0.007&0.044&0.026\\
PS1-470110&56334&$F125W$&$23.198\pm0.053$&0.038&23.948&0.037&0.007&0.027&0.016\\
DES16E2clk&57707&$F125W$&$23.375\pm0.058$&0.036&22.835&0.046&0.001&0.010&0.006\\
DES16E2clk&57712&$F160W$&$24.640\pm0.147$&0.141&22.709&0.042&0.001&0.014&0.008\\
DES16E2clk&57719&$F125W$&$23.310\pm0.054$&0.028&22.835&0.046&0.001&0.009&0.005\\
DES16E2clk&57719&$F160W$&$24.514\pm0.117$&0.111&22.709&0.037&0.001&0.012&0.007\\
DES16E2clk&57707&$F160W$&$24.873\pm0.195$&0.191&22.709&0.039&0.001&0.017&0.010\\
DES16E2clk&57712&$F125W$&$23.364\pm0.053$&0.028&22.835&0.046&0.001&0.010&0.006\\
DES16C2cva&57729&$F160W$&$24.126\pm0.074$&0.054&22.106&0.051&0.015&0.007&0.004\\
DES16C2cva&57721&$F160W$&$24.586\pm0.081$&0.063&22.106&0.051&0.015&0.010&0.006\\
DES16C2cva&57714&$F160W$&$24.798\pm0.090$&0.074&22.106&0.051&0.015&0.012&0.007\\
DES15X2kvt&57321&$F160W$&$24.926\pm0.166$&0.162&23.081&0.036&0.002&0.010&0.006\\
DES15X2kvt&57327&$F160W$&$24.722\pm0.131$&0.126&23.081&0.036&0.002&0.008&0.005\\
DES15X2kvt&57337&$F160W$&$24.081\pm0.102$&0.096&23.081&0.034&0.002&0.005&0.003\\
PS1-450339&56282&$F125W$&$23.432\pm0.070$&0.063&24.447&0.031&0.001&0.010&0.006\\
PS1-450339&56266&$F125W$&$23.527\pm0.060$&0.052&24.447&0.030&0.001&0.011&0.007\\
PS1-450339&56266&$F160W$&$24.604\pm0.112$&0.109&24.387&0.026&0.001&0.008&0.005\\
PS1-450339&56272&$F125W$&$23.541\pm0.059$&0.051&24.447&0.030&0.001&0.011&0.007\\
PS1-450339&56282&$F160W$&$24.293\pm0.098$&0.094&24.387&0.028&0.001&0.006&0.004\\
PS1-450339&56272&$F160W$&$24.665\pm0.140$&0.137&24.387&0.029&0.001&0.008&0.005\\
DES15E2nlz&57404&$F160W$&$24.366\pm0.058$&0.053&25.538&0.023&0.007&0.025&0.014\\
DES15E2nlz&57396&$F160W$&$24.641\pm0.065$&0.061&25.538&0.022&0.007&0.032&0.018\\
DES15E2nlz&57390&$F160W$&$24.604\pm0.086$&0.083&25.538&0.022&0.007&0.031&0.018\\
PS1-530251&56515&$F125W$&$23.766\pm0.092$&0.087&25.038&0.030&0.002&0.015&0.009\\
PS1-530251&56508&$F160W$&$24.884\pm0.185$&0.183&25.120&0.027&0.002&0.012&0.007\\
PS1-530251&56508&$F125W$&$23.830\pm0.100$&0.096&25.038&0.028&0.002&0.016&0.009\\
PS1-530251&56502&$F160W$&$24.541\pm0.141$&0.139&25.120&0.024&0.002&0.009&0.005\\
PS1-530251&56502&$F125W$&$23.613\pm0.068$&0.062&25.038&0.030&0.002&0.013&0.008\\
PS1-530251&56515&$F160W$&$24.662\pm0.150$&0.148&25.120&0.024&0.002&0.010&0.006\\
PS1-550202&56600&$F160W$&$24.425\pm0.070$&0.057&22.584&0.042&0.002&0.014&0.008\\
PS1-550202&56586&$F160W$&$24.761\pm0.113$&0.105&22.584&0.042&0.002&0.018&0.011\\
PS1-550202&56593&$F125W$&$23.616\pm0.069$&0.043&22.692&0.054&0.002&0.023&0.013\\
PS1-550202&56593&$F160W$&$24.661\pm0.086$&0.075&22.584&0.042&0.002&0.017&0.010\\
PS1-550202&56610&$F160W$&$23.898\pm0.052$&0.030&22.584&0.042&0.002&0.008&0.005\\
PS1-550202&56586&$F125W$&$23.520\pm0.064$&0.035&22.692&0.054&0.002&0.021&0.012\\
DES15C1nhv&57401&$F160W$&$23.655\pm0.045$&0.036&24.039&0.026&0.001&0.025&0.015\\
DES15C1nhv&57387&$F160W$&$24.035\pm0.051$&0.043&24.039&0.027&0.001&0.036&0.021\\
DES15C1nhv&57394&$F160W$&$23.955\pm0.045$&0.036&24.039&0.026&0.001&0.034&0.020\\
PS1-490037&56390&$F125W$&$23.568\pm0.119$&0.095&22.154&0.073&0.000&0.028&0.017\\
PS1-490037&56376&$F125W$&$23.161\pm0.092$&0.057&22.154&0.072&0.000&0.019&0.011\\
PS1-490037&56376&$F160W$&$24.209\pm0.153$&0.144&21.858&0.052&0.000&0.014&0.008\\
PS1-490037&56381&$F125W$&$23.358\pm0.092$&0.057&22.154&0.072&0.000&0.023&0.013\\
PS1-490037&56390&$F160W$&$24.769\pm0.231$&0.225&21.858&0.052&0.000&0.025&0.014\\
PS1-490037&56381&$F160W$&$24.450\pm0.167$&0.159&21.858&0.051&0.000&0.018&0.011\\
DES16E2cqq&57731&$F160W$&$24.139\pm0.050$&0.037&23.247&0.034&0.002&0.006&0.003\\
DES16E2cqq&57721&$F160W$&$24.552\pm0.051$&0.039&23.247&0.033&0.002&0.008&0.005\\
DES16E2cqq&57714&$F160W$&$24.897\pm0.082$&0.075&23.247&0.033&0.002&0.011&0.007\\
PS1-440236&56242&$F160W$&$24.150\pm0.065$&0.062&28.020&0.020&0.000&0.009&0.005\\
PS1-440236&56234&$F125W$&$23.580\pm0.042$&0.037&29.261&0.021&0.000&0.019&0.011\\
PS1-440236&56234&$F160W$&$24.640\pm0.094$&0.092&28.020&0.019&0.000&0.014&0.008\\
PS1-440236&56242&$F125W$&$23.471\pm0.036$&0.030&29.261&0.021&0.000&0.018&0.010\\
PS1-440236&56229&$F160W$&$24.872\pm0.129$&0.127&28.020&0.023&0.000&0.017&0.010\\
PS1-440236&56229&$F125W$&$23.659\pm0.041$&0.034&29.261&0.022&0.000&0.021&0.012\\
PS1-470240&56333&$F160W$&$24.507\pm0.076$&0.076&\nodata&\nodata&0.001&\nodata&\nodata\\
PS1-470240&56333&$F125W$&$23.638\pm0.039$&0.039&\nodata&\nodata&0.001&\nodata&\nodata\\
PS1-470240&56350&$F125W$&$23.469\pm0.040$&0.040&\nodata&\nodata&0.001&\nodata&\nodata\\
PS1-470240&56350&$F160W$&$23.779\pm0.045$&0.045&\nodata&\nodata&0.001&\nodata&\nodata\\
PS1-470240&56325&$F160W$&$24.662\pm0.047$&0.047&\nodata&\nodata&0.001&\nodata&\nodata\\
PS1-470240&56325&$F125W$&$23.659\pm0.055$&0.055&\nodata&\nodata&0.001&\nodata&\nodata\\
PS1-470240&56320&$F160W$&$24.561\pm0.066$&0.066&\nodata&\nodata&0.001&\nodata&\nodata\\
PS1-470240&56320&$F125W$&$23.522\pm0.037$&0.037&\nodata&\nodata&0.001&\nodata&\nodata\\
DES16X1cpf&57708&$F160W$&$23.703\pm0.201$&0.034&20.531&0.198&0.004&0.010&0.006\\
DES16X1cpf&57716&$F160W$&$23.592\pm0.200$&0.025&20.531&0.199&0.004&0.009&0.005\\
DES16X1cpf&57723&$F160W$&$23.424\pm0.200$&0.026&20.531&0.198&0.004&0.008&0.005\\
DES15E2mhy&57357&$F160W$&$24.070\pm0.072$&0.049&21.922&0.053&0.002&0.027&0.016\\
DES15E2mhy&57373&$F160W$&$24.293\pm0.078$&0.057&21.922&0.053&0.002&0.033&0.020\\
DES15E2mhy&57363&$F160W$&$24.378\pm0.078$&0.058&21.922&0.052&0.002&0.036&0.021\\
PS1-560027&56601&$F160W$&$24.147\pm0.082$&0.079&25.655&0.022&0.001&0.007&0.004\\
PS1-560027&56593&$F125W$&$23.494\pm0.041$&0.036&26.393&0.019&0.001&0.014&0.008\\
PS1-560027&56593&$F160W$&$24.495\pm0.090$&0.087&25.655&0.023&0.001&0.010&0.006\\
PS1-560027&56611&$F160W$&$23.800\pm0.049$&0.044&25.655&0.022&0.001&0.005&0.003\\
PS1-560027&56587&$F125W$&$23.509\pm0.043$&0.038&26.393&0.020&0.001&0.015&0.009\\
PS1-560027&56587&$F160W$&$24.784\pm0.124$&0.122&25.655&0.022&0.001&0.013&0.008\\
DES16E1dcx&57715&$F160W$&$24.177\pm0.135$&0.124&21.997&0.053&0.002&0.004&0.002\\
DES16E1dcx&57730&$F160W$&$24.218\pm0.076$&0.055&21.997&0.052&0.002&0.004&0.003\\
DES16E1dcx&57730&$F125W$&$23.525\pm0.083$&0.047&22.416&0.068&0.002&0.007&0.004\\
DES16E1dcx&57722&$F160W$&$24.594\pm0.143$&0.133&21.997&0.053&0.002&0.006&0.004\\
DES16E1dcx&57722&$F125W$&$23.733\pm0.084$&0.049&22.416&0.068&0.002&0.009&0.005\\
DES16E1dcx&57715&$F125W$&$23.487\pm0.083$&0.048&22.416&0.068&0.002&0.007&0.004\\
DES15X2nkz&57384&$F160W$&$24.529\pm0.074$&0.069&23.892&0.029&0.003&0.032&0.019\\
DES15X2nkz&57391&$F160W$&$24.853\pm0.108$&0.104&23.892&0.029&0.003&0.042&0.025\\
DES15X2nkz&57399&$F160W$&$24.630\pm0.082$&0.078&23.892&0.028&0.003&0.035&0.020\\
DES16S1bno&57686&$F160W$&$24.823\pm0.084$&0.080&26.342&0.026&0.001&0.009&0.006\\
DES16S1bno&57693&$F160W$&$24.409\pm0.074$&0.070&26.342&0.024&0.001&0.007&0.004\\
DES16S1bno&57701&$F160W$&$24.134\pm0.054$&0.048&26.342&0.025&0.001&0.005&0.003\\
PS1-540118&56528&$F160W$&$24.572\pm0.097$&0.091&23.302&0.033&0.001&0.017&0.010\\
PS1-540118&56536&$F160W$&$24.721\pm0.099$&0.093&23.302&0.034&0.001&0.020&0.012\\
PS1-540118&56523&$F160W$&$24.262\pm0.077$&0.070&23.302&0.032&0.001&0.013&0.008\\
PS1-560054&56603&$F125W$&$23.667\pm0.073$&0.063&24.039&0.038&0.001&0.024&0.014\\
PS1-560054&56587&$F160W$&$24.522\pm0.109$&0.106&23.871&0.025&0.001&0.017&0.010\\
PS1-560054&56587&$F125W$&$23.444\pm0.066$&0.055&24.039&0.036&0.001&0.019&0.011\\
PS1-560054&56592&$F160W$&$24.831\pm0.192$&0.190&23.871&0.028&0.001&0.022&0.013\\
PS1-560054&56603&$F160W$&$24.432\pm0.116$&0.112&23.871&0.030&0.001&0.015&0.009\\
PS1-560054&56592&$F125W$&$23.726\pm0.102$&0.095&24.039&0.037&0.001&0.025&0.015\\
DES16S2afz&57657&$F160W$&$24.244\pm0.065$&0.052&22.778&0.039&0.016&0.005&0.003\\
DES16S2afz&57663&$F160W$&$24.420\pm0.085$&0.076&22.778&0.038&0.016&0.007&0.004\\
DES16S2afz&57671&$F160W$&$24.296\pm0.072$&0.061&22.778&0.038&0.016&0.006&0.003\\
DES16E2rd&57657&$F160W$&$23.980\pm0.195$&0.074&19.559&0.180&0.007&0.030&0.018\\
DES16E2rd&57663&$F160W$&$24.049\pm0.200$&0.087&19.559&0.180&0.007&0.032&0.019\\
DES16E2rd&57671&$F160W$&$23.632\pm0.191$&0.064&19.559&0.180&0.007&0.021&0.012\\
DES16X3zd&57657&$F160W$&$24.558\pm0.124$&0.120&23.247&0.031&0.001&0.018&0.010\\
DES16X3zd&57663&$F160W$&$24.546\pm0.102$&0.098&23.247&0.032&0.001&0.018&0.010\\
DES16X3zd&57672&$F160W$&$24.240\pm0.113$&0.108&23.247&0.033&0.001&0.013&0.008\\
PS1-510457&56457&$F160W$&$25.043\pm0.112$&0.098&21.964&0.054&0.002&0.033&0.019\\
PS1-510457&56444&$F160W$&$25.221\pm0.121$&0.108&21.964&0.054&0.002&0.039&0.023\\
PS1-510457&56439&$F160W$&$25.086\pm0.088$&0.071&21.964&0.053&0.002&0.034&0.020\\
DES16S1agd&57672&$F160W$&$24.285\pm0.063$&0.051&22.833&0.037&0.003&0.009&0.005\\
DES16S1agd&57657&$F160W$&$24.565\pm0.085$&0.076&22.833&0.038&0.003&0.011&0.007\\
DES16S1agd&57663&$F160W$&$24.501\pm0.104$&0.098&22.833&0.035&0.003&0.011&0.006\\
DES15C3odz&57429&$F160W$&$24.508\pm0.065$&0.061&25.733&0.022&0.007&0.021&0.012\\
DES15C3odz&57421&$F160W$&$25.022\pm0.094$&0.091&25.733&0.023&0.007&0.033&0.019\\
DES15C3odz&57414&$F160W$&$25.291\pm0.136$&0.134&25.733&0.023&0.007&0.042&0.025\\
PS1-520107&56473&$F160W$&$24.132\pm0.055$&0.048&23.998&0.027&0.001&0.036&0.021\\
PS1-520107&56482&$F160W$&$24.114\pm0.058$&0.051&23.998&0.027&0.001&0.035&0.020\\
PS1-520107&56468&$F160W$&$24.064\pm0.049$&0.041&23.998&0.027&0.001&0.034&0.020\\
DES16C1cim&57721&$F160W$&$24.318\pm0.060$&0.050&23.086&0.033&0.002&0.007&0.004\\
DES16C1cim&57712&$F160W$&$24.527\pm0.063$&0.053&23.086&0.034&0.002&0.008&0.005\\
DES16C1cim&57706&$F160W$&$24.772\pm0.095$&0.088&23.086&0.036&0.002&0.010&0.006\\
DES16C3cmy&57706&$F160W$&$24.541\pm0.068$&0.064&25.646&0.023&0.001&0.010&0.006\\
DES16C3cmy&57720&$F160W$&$24.492\pm0.061$&0.057&25.646&0.022&0.001&0.010&0.006\\
DES16C3cmy&57712&$F160W$&$24.743\pm0.075$&0.071&25.646&0.024&0.001&0.012&0.007\\
DES15E2uc&57293&$F160W$&$24.889\pm0.081$&0.075&23.465&0.030&0.016&0.029&0.017\\
DES15E2uc&57299&$F160W$&$24.614\pm0.082$&0.075&23.465&0.033&0.016&0.022&0.013\\
DES15E2uc&57309&$F160W$&$24.396\pm0.060$&0.051&23.465&0.031&0.016&0.018&0.011\\
DES15X2mey&57357&$F160W$&$24.556\pm0.092$&0.086&23.068&0.032&0.001&0.029&0.017\\
DES15X2mey&57365&$F160W$&$24.722\pm0.092$&0.086&23.068&0.032&0.001&0.034&0.020\\
DES15X2mey&57374&$F160W$&$24.525\pm0.083$&0.075&23.068&0.035&0.001&0.029&0.017\\
DES16X3cry&57713&$F125W$&$24.019\pm0.043$&0.026&24.218&0.034&0.001&0.007&0.004\\
DES16X3cry&57720&$F125W$&$24.500\pm0.051$&0.037&24.218&0.035&0.001&0.011&0.006\\
DES16X3cry&57720&$F160W$&$24.626\pm0.054$&0.046&24.015&0.028&0.001&0.007&0.004\\
DES16X3cry&57728&$F125W$&$24.455\pm0.064$&0.054&24.218&0.034&0.001&0.011&0.006\\
DES16X3cry&57728&$F160W$&$24.349\pm0.080$&0.076&24.015&0.025&0.001&0.006&0.003\\

\hline\\
\multicolumn{10}{l}{
  \begin{minipage}{6in}
    {\bf Note:} The variable $m_{SN}$ is the magnitude of the SN
    and $m_{host}$ is the mag of the host {\it at the SN location}.
    $\sigma_{SN}$ is the combination of the 
    SN Poisson uncertainties and the dispersion in
    magnitudes measured from fake star tests.  $\sigma_{host}$
    is the uncertainty on the SN magnitude due to the noise
    of the host galaxy (computed using fake star tests).
    $\sigma_{c}$ is the uncertainty of
    the SN centroid, in arcseconds.
\end{minipage}}
\end{longtable*}
\clearpage

\section{Simulating the RAISIN Sample}
\label{app:sims}

Here, we describe our simulations of the RAISIN sample.  These simulations rely on an optical selection function that was determined for the PS1 and DES analyses \citep{Scolnic18,Kessler19}.  We modeled the sample under the assumption that the RAISIN SNe were selected in an unbiased way from the subset of Pan-STARRS and DES SNe with spectroscopic classifications, and the typical S/N of the RAISIN optical light curves compared to the full PS1/DES spectroscopic samples shows this to be a reasonable approximation.  Though RAISIN targeted SNe in specific redshift ranges, those redshift ranges --- $0.25 < z < 0.45$ for RAISIN1 and $0.4 < z < 0.6$ for RAISIN2 --- were within one standard deviation of the mean redshifts of the PS1 and DES SN spectroscopic surveys.  Therefore, much of the necessary work for building these simulations for RAISIN was already undertaken by the MDS and DES teams.  The main difference for RAISIN is the use of a new SN model, SNooPy, with the new optical-to-NIR dispersion model discussed in Section \ref{sec:dispmodel}.  We implemented the SNooPy model in SNANA by using the {\tt Python}-based SNooPy code to generate a grid of model realizations as a function of $s_{BV}$ and $A_V$, which SNANA may then use for both simulations and light curve fitting (Figure \ref{fig:snoopy}).  We then added a module to the SNANA simulation engine containing the SNooPy dispersion model, which generates correlated-random band-to-band offsets to be applied to the simulated magnitudes; this results in simulated distance-dependent biases.  Both the SNooPy models and the SNooPy dispersion model are publicly available in SNANA.

For the low-$z$ simulations, we do not have sky noise and zeropoint information for the original NIR observations.  However, SNANA is able to estimate these quantities by using the magnitudes and uncertainties in the data themselves to create a ``simulation library''.  Other low-$z$ survey characteristics and selection effects were modeled in the Pantheon analysis \citep{Scolnic18} and we use the magnitude-based SN detection efficiency determined in that work to simulate the low-$z$ CSP sample here.  For the high-$z$ simulations, our simulation library is built from observations of the RAISIN SNe themselves, which allows us to simulate SNe at the specific redshifts of the RAISIN objects and control for possible differences between the mean survey observing conditions and the conditions during which the RAISIN SNe were observed.

Finally, we require that simulated distributions of $x_1$ and $c$ parameters for SALT2 are replaced by distributions of the stretch parameter $s_{BV}$ and the extinction $A_V$.  Even though the light curves in NIR bands are less sensitive to $s_{BV}$ and $A_V$ than those in optical bands, a NIR-only approach with the RAISIN data does not allow shape/color to be corrected for and therefore the results will be very sensitive to sample-to-sample differences in these parameters and the optical selection effects that will bias the sample towards, e.g., high $s_{BV}$ and low $A_V$.  We therefore adapt the method of \citet{Scolnic16} to determine the intrinsic $s_{BV}$ distribution for the low-$z$ and high-$z$ samples.  The distribution of the $s_{BV}$ parameter is treated as an asymmetric Gaussian following \citet{Scolnic16} and we estimate the $s_{BV}$ distribution by fitting $s_{BV}$ with our NIR data.

\begin{figure*}
  \includegraphics[width=7in]{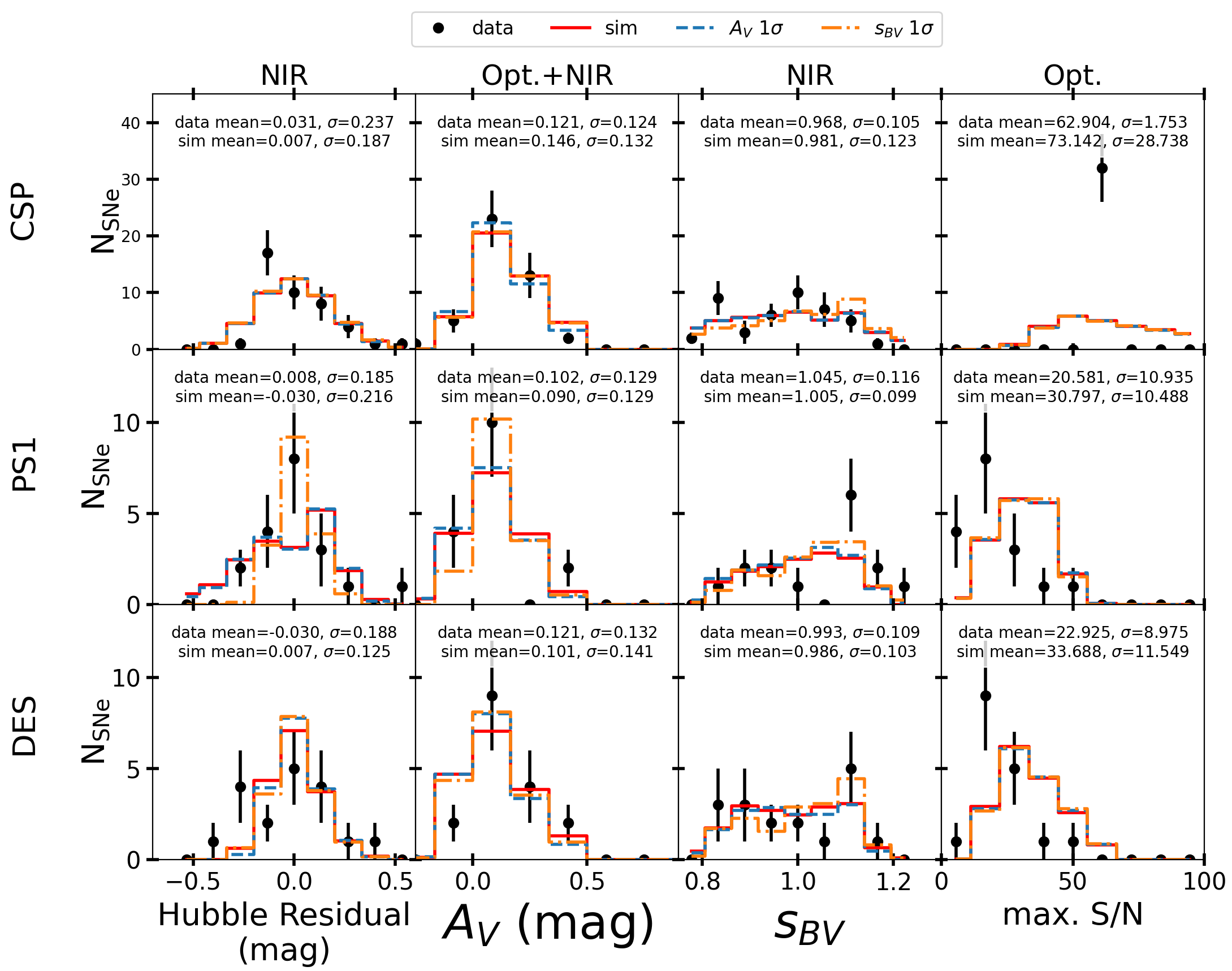}
  \caption{RAISIN simulations (red) compared to the data (black). Histograms of Hubble residuals, $A_V$, $s_{BV}$ and maximum S/N are shown for the CSP sample (top), the RAISIN1 sample (middle) and the RAISIN2 sample (bottom). Various combinations of data are used for comparison purposes as indicated above each column.  Simulations after applying a 1$\sigma$ shift to parameters of the $A_V$ and $s_{BV}$ distributions are shown in blue and orange, respectively.  Some discrepancy in maximum S/N may be due to statistical fluctuation due to the small sample sizes in this analysis or uncertainty in the chosen SNooPy model.}
  \label{fig:sim}
\end{figure*}

For $A_V$, our baseline analysis assumes the mean intrinsic dust extinction is independent of redshift, but we vary this assumption in our systematic error budget.  We simulate an exponential dust distribution with $\tau_{A_V} = 0.2$~mag, which we find is a good approximate match to the optical$+$NIR data in Figure \ref{fig:sim} and matches nominal dust distributions from the SN rates analysis of \citep{Rodney14}.  We also find that varying $\tau$ globally has a minimal impact on the redshift-dependence of our predicted bias corrections.  Our two $A_V$-related systematic uncertainty analysis variants reduce the $A_V$ scale length first by 0.05 for both samples (a small additional uncertainty) and second by 0.05 for the low-$z$ sample only.

The low-$z$ sample from CSP used a galaxy targeted approach to find SNe, which gives a sample of predominantly massive host galaxies and, as a result, increases the fraction of fast-declining SNe in the sample \citep{Childress13}.  For this reason we assume {\it a priori} that the low-$z$ and RAISIN samples have different intrinsic stretch and color distributions.  However, because the DES and MDS samples were selected in much the same way and to avoid statistical noise, we assume that the intrinsic distributions of shape and color in these SN samples is the same.  This is an approximation that is necessary due to limited statistics, but the \citet{Scolnic16} measurements of intrinsic population parameters in high-$z$ samples show that this may be a good approximation.  However, some evolution in population parameters \citep{Nicolas20} is expected, and larger samples both at high- and low-$z$ in future analyses will allow constraints on this evolution as well as a more robust determination of the stretch and color distributions themselves.

The resulting stretch and color distributions are shown in Figure \ref{fig:sim}.  We find that both the $s_{BV}$ and $A_V$ distributions are statistically consistent between low- and high-$z$ samples, albeit with relatively large uncertainties.

Finally, throughout this analysis we assume the default SNooPy total-to-selective extinction ratio of $R_V = 1.52$ from \citet[their Table 8;][]{Folatelli10}, which reduces dispersion about the Hubble diagram compared to choosing the Milky Way value of $R_V = 3.1$ (Figure \ref{fig:optical}).  The lower-than-expected value of $R_V$ may be due to the lack of an intrinsic color variation component of the SNooPy model \citep{Mandel22,Thorp21} and because extinction is allowed to be (unphysically) negative in this analysis.  By using the CSP third data release and allowing SNooPy to fit for $R_V$, we find that SNooPy gives a median $R_V \simeq 2$, though unfortunately we find the high-$z$ RAISIN data are insufficient to constrain $R_V$ with SNooPy.  

Because data at NIR wavelengths are a factor of $\sim$4-5 less sensitive to dust than the optical, distance error due to uncertainty in the value of $R_V$ is a sub-percent level effect and will not be a significant component of the error budget in this analysis.  We note that in the NIR fitting itself, we treat $A_V$ as a constant, and therefore this effect will only change the distances in the bias correction stage.

The maximum S/N distributions are shown on the right side of Figure \ref{fig:sim}.  Generally, the simulations expect slightly higher S/N than is observed in the data for high-$z$.  Because the larger PS1 and DES samples have a similar S/N near maximum light compared to the subset with RAISIN observations $-$ nearly identical for PS1 and marginally higher S/N in the DES RAISIN subset $-$ we choose not to modify the simulations here.  The low-$z$ S/N, on the other hand, has an unusual shape that we were unable to reproduce, perhaps due to small sample sizes; in the full SNooPy sample, before applying our selection cuts, we do not see the same concentration around a narrow range of S/N.  In future, larger NIR samples it will be much easier to understand such artifacts in the data.  We are unsure whether the NIR sample was selected in the same way as the rest of the CSP sample, but we assume that for this analysis; the simulations developed for past analyses are therefore sufficient given systematic uncertainties on the $s_{BV}$ and $A_V$ distributions.  A low-$z$ sample that selects NIR-observed SNe in an unbiased way is a key potential improvement for future cosmological analyses.

\subsection{Distance Measurement Methodologies}
\label{app:simdist}

Using the RAISIN simulations, we test the assumption that NIR-only distance measurements are most precise when $s_{BV}$ and $A_V$ are fixed to a constant value.  From simulations, we measure a Hubble residual RMS of 0.16~mag, slightly lower than the 0.19 mag measured from the real RAISIN data, when $s_{BV}$ and $A_V$ are fixed to a constant.  When we attempt to use the NIR data alone to fit for $A_V$, the Hubble residual RMS increases to 0.28~mag, and when we fit for both $s_{BV}$ and $A_V$ we measure an RMS of 0.25~mag.  However, when we use the NIR data alone to fit for $s_{BV}$ we find that the RMS is unchanged from the $s_{BV}$ fixed case; this may be because the simulated SNooPy model does not include the increase in scatter at later phases that we observe and discuss in Section \ref{sec:latetimenir}.  However, despite this modest discrepancy, both the simulations and real RAISIN data appear to show no benefit from fitting to $A_V$ and/or $s_{BV}$ with NIR data alone.

\section{Host Mass Measurement}
\label{app:hostmass}

Host galaxy masses were reported by the CSP, DES, and MDS teams \citep{Krisciunas17,Smith20,Jones18}.  However, to ensure consistency between all methods, we estimated the host galaxy masses of SNe in the RAISIN and low-$z$ samples ourselves.  For the low-$z$ sample we used data from {\it GALEX} \citep{Martin05}, SDSS DR16 \citep{Ahumada19}, PS1 DR2 \citep{Flewelling16}, 2MASS \citep{Skrutskie06}, while for the high-$z$ sample we used only SDSS and PS1 as the other catalogs do not have sufficient depth for the faint, high-$z$ RAISIN galaxies.  For RAISIN2, we also used SN-free DES photometry from \citep{Wiseman20}.  Finally, because PS1 3$\pi$ images may be contaminated by SN light, we used MDS single-season template images for RAISIN1 SNe.  The stacked images make it possible to detect ${\rm log(M_{\ast}/M_{\odot})} = 10$ SN host galaxies with sufficient depth to measure masses on either side of the typical mass step location.

To find the host galaxy for each SN, we use SExtractor to determine the ``directional light radius" (DLR) of between potential host galaxies and each SN \citep{Sullivan06,Gupta16}, a method that incorporates the size and orientation of each galaxy to determine which galaxy is the most likely host.  Each most probable host was confirmed by eye and, thanks to {\it HST} imaging, we were able to determine which galaxies were the host without significant ambiguity.  We then used SExtractor to measure the elliptical parameters of each galaxy in the $r$ band, and then used elliptical aperture photometry to measure the magnitudes of each galaxy in each available bandpass.  The aperture size was chosen to extend slightly beyond the isophotal radius determined by SExtractor, and was also extended to account for the increased PSF sizes of 2MASS and {\it WISE}.  Any contamination by foreground stars was removed by using SExtractor to identify possible contaminants and masking those objects by setting the value of the pixels SExtractor deemed as belonging to those objects to the median value of the nearby pixels.

Once aperture photometry was measured, we used LePHARE \citep{Arnouts11} with \citet{Bruzual03} spectral templates and a Chabrier initial mass function \citep{Chabrier03} to determine the stellar masses of each host galaxy.  The templates include 9 exponentially decreasing star formation histories in three metallicity bins, and we allow $E(B-V)$ to vary from 0 to 0.4 in steps of 0.1~mag with a range of extinction laws.  Uncertainties on these masses were estimated by Monte Carlo sampling of the photometry using the photometric uncertainties for each band and assuming a 1\% error floor for bright galaxies.  We note that mass estimation requires determining an absolute magnitude, making these mass estimates dependent on an assumed cosmology and the SN brightness residual.  However, none of our mass estimates would change from $<$10~dex to $>$10~dex with modest adjustments in cosmological parameters; a 20\% shift in $w$ at a redshift of 0.6 would result in a systematic shift of just 0.03~dex relative to a low-$z$ mass.

We compare these masses to estimates from \citet{Roman18} for the low-$z$ sample and \citet{Smith20} for the high-$z$ sample.  Out of the 15 SNe with host masses in \citet{Roman18}, only one SN, SN~2004ey, disagrees with our high versus low-mass designations.  We do see a large median offset of $-0.5$~dex when subtracting the \citet{Roman18} masses from ours, but find that this is due to our addition of {\it GALEX} and 2MASS data; running LePHARE on our optical-only measurements gives a marginally significant median difference of $+0.21\pm0.08$~dex for these SNe.  For DES masses, we find a median difference of just 0.06~dex between our masses and those of \citet{Smith20}, with no disagreement between our high versus low-mass designations.  Figure \ref{fig:mass_hist} shows histograms of the masses for CSP, RAISIN1 and RAISIN2 SNe; due to the targeted nature of the low-$z$ CSP data, the CSP SNe are found in significantly more massive host galaxies.

For a given mass step location, with log$(M_{\ast}/M_{\odot}) = 10$ as the default, we assume the uncertainties on the mass estimates are Gaussian and use these uncertainties to compute the probability that each SN has a mass greater than or less than the value at the step location.  We then built a maximum likelihood, Gaussian-mixture model that includes free parameters of the intrinsic dispersions for CSP, RAISIN1, and RAISIN2 (RAISIN2 has higher dispersion as it uses one filter instead of two for most SNe), median Hubble residuals at low, medium, and high redshifts to remove sensitivity to the cosmological model, and a single mass step parameter.  This procedure largely follows \citet{Jones18b}.  We correct for the measured mass step in our data and apply two systematic uncertainties based on the value and location of the mass step as discussed in Section \ref{sec:massstepsys}.

\section{Distance Measurements for RAISIN and Low-$z$ SNe}
\label{sec:app_dist}

This appendix contains NIR distances and optical+NIR stretch ($s_{BV}$) and $A_V$
measurements for RAISIN, computed using SNooPy and assuming an R$_V$ of 1.52.

\begin{table}
    \centering
    \caption{RAISIN Distances and Cuts}
    \begin{tabular}{lrrrrrrr}
    \hline \hline
        ID & $z_{\rm Helio}$ & Raw Distance & Rest-frame Bands&Bias Corr. & $\sigma_{\rm tmax}$ & Avg.\ $\sigma_{\rm phot}$ & Cuts \\
        &&(mag)&&(mag)&(days)&(mag)&\\
        \hline

PS1-480464&0.220&40.188 $\pm$ 0.033&$Y,J$&0.040&0.120&0.073&\nodata\\
PS1-450082&0.250&\nodata&\nodata&\nodata&0.310&0.234&bad host sub.\\
PS1-540087&0.275&40.621 $\pm$ 0.043&$Y,J$&0.073&0.350&0.145&\nodata\\
PS1-520188&0.280&40.350 $\pm$ 0.025&$Y,J$&0.073&0.050&0.040&\nodata\\
DES16E2cxw&0.293&\nodata&\nodata&\nodata&0.160&0.237&bad host sub.\\
PS1-440005&0.306&40.771 $\pm$ 0.044&$Y,J$&0.049&0.410&0.102&\nodata\\
PS1-520062&0.308&41.168 $\pm$ 0.047&$Y,J$&0.049&0.290&0.128&\nodata\\
PS1-500100&0.310&40.991 $\pm$ 0.032&$Y,J$&0.046&0.390&0.069&\nodata\\
PS1-500301&0.325&40.727 $\pm$ 0.068&$Y,J$&0.038&0.420&0.145&\nodata\\
PS1-470041&0.331&\nodata&\nodata&\nodata&0.120&0.242&bad host sub.\\
PS1-480794&0.334&\nodata&\nodata&\nodata&0.280&0.077&high $s_{BV}$\\
PS1-490521&0.340&40.982 $\pm$ 0.042&$Y,J$&0.033&0.430&0.082&\nodata\\
PS1-470110&0.346&41.128 $\pm$ 0.033&$Y,J$&0.035&0.190&0.074&\nodata\\
DES16E2clk&0.367&41.340 $\pm$ 0.039&$Y,J$&0.041&0.240&0.103&\nodata\\
DES16C2cva&0.403&41.699 $\pm$ 0.055&$Y$&0.019&0.330&0.080&\nodata\\
DES15X2kvt&0.404&41.741 $\pm$ 0.078&$Y$&0.019&0.710&0.132&\nodata\\
PS1-450339&0.410&41.577 $\pm$ 0.034&$i,Y$&0.022&0.450&0.082&\nodata\\
PS1-530251&0.413&41.720 $\pm$ 0.045&$i,Y$&0.024&0.500&0.120&\nodata\\
DES15E2nlz&0.410&41.715 $\pm$ 0.047&$Y$&0.021&0.390&0.066&\nodata\\
DES15C1nhv&0.421&\nodata&\nodata&\nodata&0.370&0.043&Chauvenet\\
PS1-550202&0.422&41.698 $\pm$ 0.033&$i,Y$&0.025&0.800&0.073&\nodata\\
PS1-490037&0.422&41.500 $\pm$ 0.051&$i,Y$&0.025&0.370&0.147&\nodata\\
DES16E2cqq&0.426&41.905 $\pm$ 0.042&$Y$&0.025&0.380&0.063&\nodata\\
PS1-470240&0.430&41.712 $\pm$ 0.024&$i,Y$&0.025&0.410&0.065&\nodata\\
DES16X1cpf&0.436&\nodata&\nodata&\nodata&0.380&0.201&bad host sub.\\
PS1-440236&0.430&41.732 $\pm$ 0.024&$i,Y$&0.025&0.340&0.070&\nodata\\
DES15E2mhy&0.439&41.528 $\pm$ 0.051&$Y$&0.029&0.340&0.079&\nodata\\
PS1-560027&0.440&41.690 $\pm$ 0.027&$i,Y$&0.024&0.210&0.075&\nodata\\
DES16E1dcx&0.453&41.679 $\pm$ 0.040&$i,Y$&0.029&0.340&0.105&\nodata\\
DES15X2nkz&0.469&42.025 $\pm$ 0.056&$Y$&0.037&0.560&0.095&\nodata\\
DES16S1bno&0.470&42.094 $\pm$ 0.046&$Y$&0.037&0.510&0.076&\nodata\\
PS1-540118&0.477&41.938 $\pm$ 0.058&$Y$&0.004&0.490&0.084&\nodata\\
PS1-560054&0.482&41.813 $\pm$ 0.040&$i,Y$&0.004&0.250&0.112&\nodata\\
DES16S2afz&0.483&41.706 $\pm$ 0.051&$Y$&0.033&0.300&0.068&\nodata\\
DES16E2rd&0.494&\nodata&\nodata&\nodata&0.210&0.194&bad host sub.\\
DES16X3zd&0.495&41.905 $\pm$ 0.071&$Y$&0.030&0.220&0.103&\nodata\\
PS1-510457&0.502&42.625 $\pm$ 0.065&$Y$&0.004&0.890&0.106&\nodata\\
DES16S1agd&0.504&41.925 $\pm$ 0.055&$Y$&0.031&0.490&0.086&\nodata\\
DES15C3odz&0.508&42.538 $\pm$ 0.058&$Y$&0.031&0.320&0.094&\nodata\\
PS1-520107&0.519&\nodata&\nodata&\nodata&1.960&0.049&possible non-Ia\\
DES16C1cim&0.531&42.282 $\pm$ 0.049&$Y$&0.024&0.570&0.070&\nodata\\
DES16C3cmy&0.556&42.326 $\pm$ 0.048&$Y$&0.017&0.250&0.066&\nodata\\
DES15E2uc&0.566&42.710 $\pm$ 0.048&$Y$&0.019&0.640&0.070&\nodata\\
DES15X2mey&0.608&42.700 $\pm$ 0.058&$Y$&0.025&0.280&0.093&\nodata\\
DES16X3cry&0.612&42.567 $\pm$ 0.028&$i,Y$&0.025&0.340&0.062&\nodata\\
\hline
\\*[-5pt]
    \multicolumn{8}{c}{
  \begin{minipage}{7in}
    {\bf Note.} Distance moduli from RAISIN SNe, calibrated to our best-fit H$_0$ of $75.4~{\rm km~s^{-1}~Mpc^{-1}}$.  The ``Raw Distance" (column 3) does not include the bias correction (column 4), which is added to the raw distance prior to cosmological parameter fitting.  The ``Rest-frame Bands" column indicates the rest-frame SNooPy templates that were used to fit the RAISIN observations.
\end{minipage}}
    \end{tabular}
    \label{table:raisin_distance}
\end{table}

\clearpage
\bibliographystyle{apj}
\bibliography{main}

\suppressAffiliationsfalse
\allauthors

\end{document}